\documentclass[aps,prx,10pt,longbibliography,superscriptaddress,twocolumn,showpacs,oneside]{revtex4-2}

\usepackage[utf8]{inputenc}

\usepackage[american,british]{babel}
\usepackage[T1]{fontenc}
\usepackage{lmodern}
\usepackage{xcolor}

\usepackage{wrapfig}
\usepackage{graphicx}
\graphicspath{{old_figs/}}
\usepackage[caption=false]{subfig}

\usepackage{amsmath,amssymb,amsthm,mathrsfs,mathtools,amsfonts,physics,bm,bbm,booktabs}
\usepackage{mathdots}
\usepackage{dsfont}
\usepackage[version=4]{mhchem}

\usepackage{bbold}

\usepackage[colorlinks=true, linktocpage=true]{hyperref} 
\hypersetup{allcolors = {black}}

\usepackage{tikz}
\usetikzlibrary{shapes.geometric, arrows}

\usepackage{algorithm}
\usepackage{algpseudocode}
\usepackage{setspace}

\tikzstyle{Atensor} = [rectangle, rounded corners, minimum width=1cm, minimum height=1cm, text centered, draw=black]
\tikzstyle{Gtensor} = [circle, minimum width=1cm, minimum height=1cm, text centered, draw=black]
\tikzstyle{arrow} = [thick,->]

\usepackage{comment}
\usepackage[normalem]{ulem}  

\newtheorem{lemma}{Lemma}[section]
\newtheorem{definition}{Definition}[section]

\begin{document}

\title{Temporal entanglement transition in chaotic quantum many-body dynamics}

\author{Ilya Vilkoviskiy}
\affiliation{Department of Physics, Princeton University, Princeton NJ 08544, USA}
\affiliation{Department of Theoretical Physics,
University of Geneva, Quai Ernest-Ansermet 30,
1205 Geneva, Switzerland}

\author{Michael Sonner}
\affiliation{Department of Theoretical Physics,
University of Geneva, Quai Ernest-Ansermet 30,
1205 Geneva, Switzerland}
\affiliation{Max Planck Institute for the Physics of Complex Systems, 01187 Dresden, Germany}
\affiliation{Department of Physics, University of California, Berkeley CA 94720, USA}

\author{Qi Camm Huang}
\affiliation{Department of Physics, National University of Singapore, Singapore 117551}
\author{Wen~Wei Ho}
\affiliation{Department of Physics, National University of Singapore, Singapore 117551}
\affiliation{Centre for Quantum Technologies, National University of Singapore,  Singapore 117543}

\author{Alessio Lerose}
\affiliation{Department of Theoretical Physics,
University of Geneva, Quai Ernest-Ansermet 30,
1205 Geneva, Switzerland}
\affiliation{Rudolf Peierls Centre for Theoretical Physics, University of Oxford, Oxford OX1 3PU, United Kingdom}
\affiliation{Institute for Theoretical Physics, KU Leuven, Celestijnenlaan 200D, 3001 Leuven, Belgium}

\author{Dmitry A. Abanin}
\affiliation{Department of Physics, Princeton University, Princeton NJ 08544, USA}
\affiliation{\'Ecole Polytechnique F\'ed\'erale de Lausanne (EPFL), 1015 Lausanne, Switzerland}
\affiliation{Google Research, Brandschenkestrasse 150, 8002 Zürich, Switzerland}
\date{\today}

\begin{abstract}

Temporal entanglement (TE) of an influence matrix (IM) has been proposed as a measure of complexity of simulating dynamics of local observables in many-body systems. Foligno {\it et al.}~\cite{FolignoBertiniPRX2023} recently argued that the TE in chaotic 1d quantum circuits obeys linear (volume-law) scaling with evolution time. To reconcile this apparent high complexity of IM with the rapid thermalization of local observables, here we study the relation between TE, non-Markovianity, and local temporal correlations for chaotic quantum baths. By exactly solving a random-unitary bath model, and bounding distillable entanglement between future and past degrees of freedom, we identify a regime where TE is extensive and reflects genuine non-Markovianity.
This memory, however, is entirely contained in highly complex temporal correlations, and its effect on few-point temporal correlators is negligible. An IM coarse-graining procedure, reducing the allowed frequency of measurements of the probe system, results in a transition from volume- to {subvolume- (area or logarithmic)} law TE scaling. We demonstrate the generality of this TE transition in 1d circuits by analyzing the kicked Ising model analytically at dual-unitary points, as well as numerically away from them. This finding indicates that dynamics of local observables are fully captured by a 
subvolume-law IM. We provide evidence that the compact IM MPS 
obtained via standard compression algorithms 
accurately describes local~evolution.

\end{abstract}
\maketitle

\section{Introduction}

Quantum chaos and thermalization play a foundational role in many-body physics. Fueled by progress
in quantum simulation and computing, which opened the door to their experimental studies, these phenomena continue to attract much interest and generate new lines of research. Theoretically, two classes of quantum circuits --- random unitary~\cite{RUCreview} and dual-unitary (DU)~\cite{Guhr1,Bertini2019,bertini2025exactly} circuits --- have been identified as analytical testbeds to study
  entanglement dynamics and operator growth under generic quantum dynamics, as they generally do not harbor any conservation laws~\cite{nahum2017quantum,nahum2018operator,BertiniSFF,BertiniPRXEntanglementDU}.
  Intuitively, under time evolution, a probe system interacting
with such a chaotic system quickly becomes maximally entangled with it,
and its reduced density matrix quickly evolves to a featureless mixed state (i.e., it thermalizes to an infinte-temperature state).
This process is accompanied by a rapid delocalization of the information
about {the} initial state of the probe system into the chaotic bath, which can no longer be retrieved by acting locally on the probe. 

The dynamics of local observables in quantum many-body dynamics can be viewed through the
lens of open quantum systems: one considers a small local subsystem and views it as a probe
 that interacts repeatedly with the rest of the system, viewed as a bath. At each step, the
probe system, which we will assume to be a qudit, can be acted on by arbitrary physical operations, including preparation in a fresh
state and measurement in any basis. This description gives rise to a multi-time quantum channel~\cite{pollock2018NonMarkovian}
(or a many-body temporal {\it state}, via Choi
duality~\cite{choi1975completely}) which is closely related to the Feynman-Vernon
influence functional~\cite{feynman2000Theory} of a quantum bath. In the context of quantum information theory, this object is known as a process tensor~\cite{pollock2018NonMarkovian}, while in the context of many-body dynamics, it has been dubbed the influence matrix (IM)~\cite{lerose2021Influence}. 
 
The entanglement properties of the IM, 
which we refer to as \textit{temporal entanglement} (TE),
provide insights into non-Markovianity
\cite{pollock2018NonMarkovian,pollock2018operational,LuchnikovPRL2019,figueroa2019almost} and
into the complexity of simulating dynamics of local observables in many-body
systems in different regimes~\cite{banuls2009Matrix,lerose2021Scaling,sonner2021Influence,Chan21, sonner2022Characterizinga,carignano2024temporal}, leading to new computational algorithms~\cite{hastings2015Connecting,FriasBanulsPRB2022,lerose2023overcoming,carignano2025overcoming}. Scaling of TE
with evolution time is of central importance: in particular, an area-law or logarithmic scaling
with the evolution time $T$ strongly suggest that the IM may be efficiently
represented as a matrix-product state
(MPS)~\cite{banuls2009Matrix,lerose2021Influence, sonner2021Influence}.
Physically, compressing the IM of a large bath is to be expected 
 if only a limited
amount of information backflows from the bath to affect the subsequent dynamics of the probe
system. In integrable, and especially in non-interacting many-body baths, information is carried away by ballistic quasiparticle excitations, leading to a favorable temporal entanglement scaling~\cite{muller-hermes2012Tensor,lerose2021Scaling,klobas20exact,giudice2022Temporal,lerose2023overcoming,thoenniss2023Nonequilibrium}.
In this case efficient methods have been
developed to compress the IM to MPS form, providing an accurate
simulation of the dynamics of local observables~\cite{TEMPO,thoenniss2023Efficient,ng2023real,sonner2025semi},
supported by rigorous justification~\cite{VilkovPRB2024,Thoenniss2024PseudoModes}.

In contrast, for chaotic quantum many-body systems, a precise connection between the entanglement properties of the IM and the possibility of efficiently approximating temporal correlators of local observables 
remains largely elusive. A recent study of TE in 1d chaotic quantum unitary circuits with brickwork architecture by Foligno et al.~\cite{FolignoBertiniPRX2023} established a volume-law
scaling of TE with evolution time for dual-unitary circuits, in agreement with numerical results for random brickwork circuits,  suggesting that for these chaotic quantum baths the memory required to describe the IM increases  exponentially with evolution time. 
On the other hand, 
these systems undergo rapid thermalization, characterized by a finite relaxation timescale $\tau_{\rm th}$ for local observables. This intuition agrees with the claim of Markovianization of the process tensor in chaotic systems \cite{figueroa2019almost,o2025diagnosing}.
Naively, one may thus expect that the object
controlling the dynamics of local observables -- the IM -- should be
well approximated using a MPS with a finite range of correlations, of the order of~$\tau_{\rm th}$~\cite{lerose2021Influence}.
However, this expectation appears to be at odds with
the conclusion of Ref.~\cite{FolignoBertiniPRX2023} that the IM has exponential complexity. {The fact that generic 1d quantum circuits  exhibit volume-law scaling of TE 
raises a question of the physical origin of this scaling: 
does it correspond to an operational protocol that probes the information recoverability between past and future degrees of freedom?
%
Even with the presence of the such extensive temporal entanglement scaling, does it pose a fundamental obstruction to an efficient numerical simulation of the IM, if we restrict attention to only practically relevant observables, such as few-point temporal correlations?}

The goal of this paper is to investigate the connection between volume-law TE of 
quantum chaotic systems' IMs, non-Markovianity, and temporal correlators of local
observables. 
{Focusing on 1d local quantum circuits}, 
we argue that, while volume-law TE can reflect genuine
non-Markovianity of dynamics, this high complexity is only manifest in temporal
correlation functions involving a high frequency of measurements of the probe system. Instead, temporal
correlation functions involving a sufficiently low (but finite) frequency of measurements of the probe system, including all the physically relevant few-point correlation
functions, can be captured by a modified IM with {subvolume-}law scaling of TE. 
This IM is generated by a coarse-graining procedure, which effectively
projects out high-frequency correlators 
while retaining an exact description of low-frequency correlators.
{In particular}, 
we establish that
there is a \textit{finite} critical
 coarse-graining density $0<n_{\rm cg}^\star<1$ at which volume-law TE scaling  {of the IM} sharply transitions to a {subvolume-}law TE scaling. 

\begin{figure}
    \centering
    \includegraphics[width=\columnwidth]{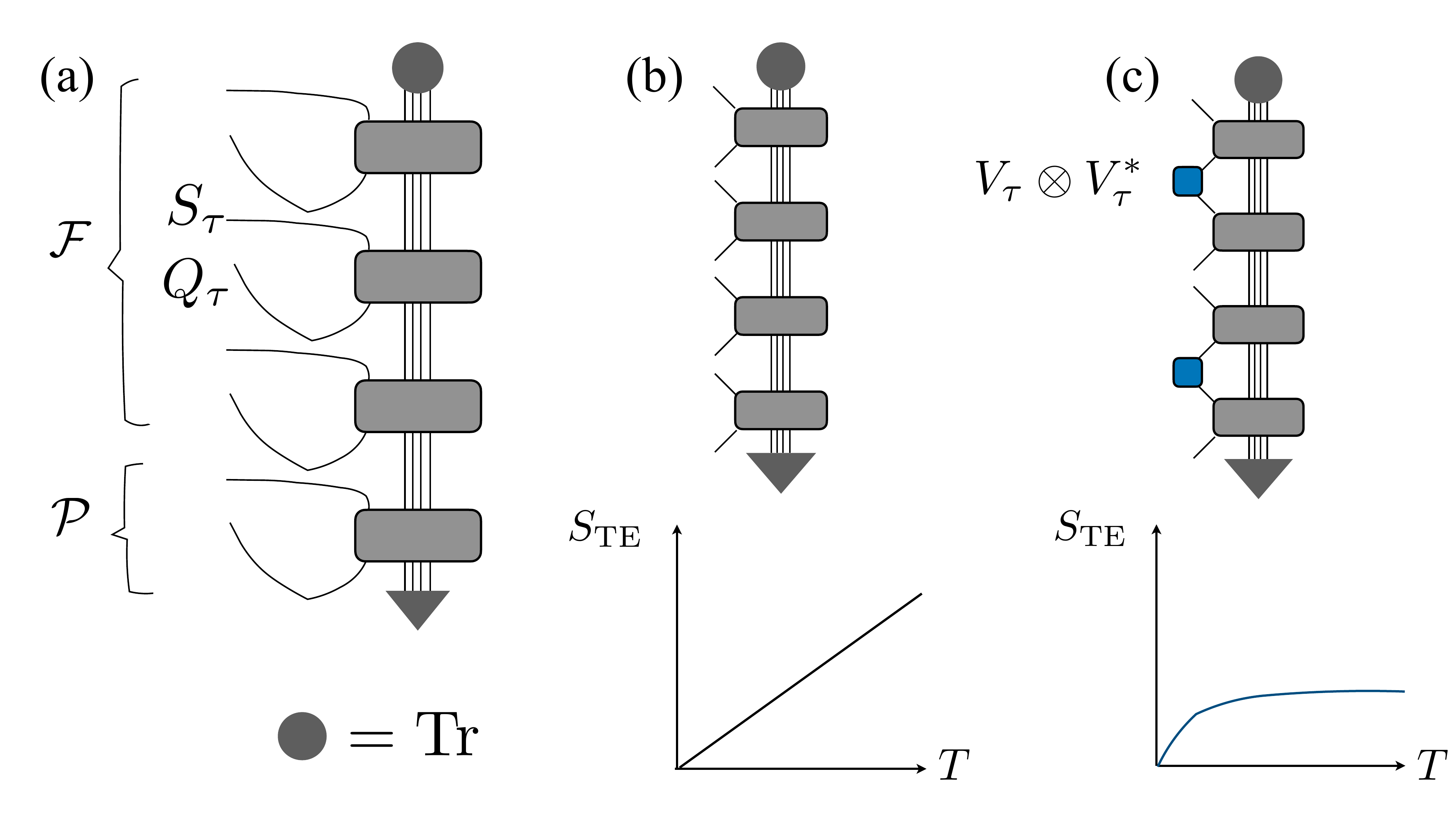}
    \caption{Schematic of a general influence matrix (IM) and temporal entanglement (TE) transition. (a) The IM state is obtained by preparing $T$ qudit pairs (each comprising a ``system'' qudit $S_\tau$ and a ``reference'' qudit $Q_\tau$) in maximally entangled Bell states, and then sequentially bringing the qudits $S_1$, $S_2$, \dots, $S_T$ to interact with a given bath. The IM state is the state of these $2T$ qudits at the end of the procedure (i.e., the state obtained by tracing over the bath degrees of freedom). (b,c) IM before (b) and after (c) the coarse-graining operation, which amounts to inserting unitary evolution of the probe qudit at a fraction of the time steps. As we show, beyond a critical coarse-graining density, TE undergoes a transition from volume-law to {subvolume}-law (see main text for details).}
    \label{fig1:schematic}
\end{figure}

\subsection{Main results}

Before describing our main results,
we recall some basic properties of IMs, with a more detailed discussion provided in
Section~\ref{sec:IM}. Physically, an IM can be identified with a quantum state of $T$ pairs of qudits, with local Hilbert space dimension $d$: Starting with a product state of $T$ maximally entangled qudit pairs (Bell pairs), we let one qudit of each pair sequentially interact with the bath (see Fig.~\ref{fig1:schematic}). The final state of the $2T$ qudits is, up to a
factor of $d^{-T}$, identical to the IM. 

{Knowledge of the IM allows one to describe an experiment where a probe system, prepared in some state, interacts with the bath and is measured after each interaction step, i.e., probe a {\it multi-time correlation function}. 
In particular, one can consider correlations between the first $p$ qudit pairs (``past'' subsystem,
$\mathcal{P}$), and the remaining $f=T-p$ pairs (the ``future'' subsystem,
$\mathcal{F}$); 
the size of correlations determine the degree of Markovianity of the bath, with small(large) correlations corresponding to the Markovian(non-Markovian) regimes. 
The operator-space entanglement entropy of the associated density matrix of the IM, called the {\it temporal entanglement} (TE), has been proposed to characterize the Markovian property of a bath.
}
%

 However, {in practice}, one is often interested in a more restricted setup
of {\it few-point} temporal correlation functions, with a given unitary evolution of the probe between interaction steps with the bath. Such
correlators are obtained by projecting the IM onto a hyperplane determined by the local evolution of the probe qudit, as well as the chosen observables (see below for
details). A central question addressed in this work is to elucidate how TE and Markovianity properties of chaotic quantum baths depend on the allowed frequency of observations of the probe system.

{Our analysis proceeds as follows.} 
First, to develop intuition, we study a tractable toy model of
thermalization where a probe system of Hilbert dimension $d$ (taken to be $2$ in most of the paper) is repeatedly coupled to a large structureless
bath with Hilbert space dimension $D_{\mathcal{B}}$, modeled via the repeated application of a random unitary acting on the $dD_{\mathcal{B}}$-dimensional
space [see Fig.~\ref{fig1:schematic}(a)].
We compute TE, which is a function of the ratio of
evolution time $T$ and bath size $b=\log_d D_{\mathcal{B}}$,  
\begin{equation}\label{eq:r}
    r=\frac{T}{b},
\end{equation}
and find two opposite regimes: In the limit $T\ll b$, or equivalently $r\ll 1$, the IM is close to a product state, 
{implying that} 
no experiment performed on the future system ($\tau>p$) can retrieve information about the past $\tau\leq p$. Thus, as previously found in Ref.~\cite{figueroa2019almost}, this limit is genuinely Markovian.
{Conversely}, in the opposite limit $r\gg 1$, {we instead find the TE saturates the upper boundset by the finite bath size, $S_\text{TE} \sim 2b$ (i.e., TE is extensive in $b$ or $T$),
suggesting that this limit is non-Markovian}. 
{These two regimes are separated by a} 
sharp transition at $r=1/2$.
A similar transition is found in a dynamical version of the {toy} model, where the number of bath degrees of freedom is taken to grow linearly in time as $b(\tau)=2v_B \tau$, mimicking the locality constraints in 1d unitary circuits. Here, the TE is found to saturate for $v_B>1$ and to grow extensively for $0<v_B<1$. 
 

{We then make a connection}
to the black-hole {quantum} information retrieval problem~\cite{Page1993BlackHole,hayden2007black}. We apply quantum-information techniques, such as decoupling theorems~\cite{hayden2008decoupling}, to show that for $r\gg 1$, one can distill $b$ Bell pairs between $\mathcal{P}$ and $\mathcal{F}$. {Since} the number of distilled Bell pairs puts a lower bound on the \textit{distillable entanglement}~\cite{bennett1996purification,bennett1996mixed} between the two parties, {this 
shows the $r \gg 1$ limit is indeed associated with dynamics being genuinely non-Markovian:} in particular, both TE and the number of distillable Bell pairs scale extensively here.  
{At the same time, we argue that} this non-Markovianity is encoded in highly nonlocal temporal correlations involving measurements at an extensive number of times. By contrast, few-point correlation functions decay rapidly to random fluctuations of order $1/D_{\mathcal{B}}$ and therefore do not exhibit any discernible signature of non-Markovianity.


{Our analysis of the toy model suggests a simple {picture explaining why}  volume-law scaling of TE {is seen ubiquitously in more physical extended}  1d brickwork circuits:} due to locality of interactions, a chaotic bath prepared in an initial short-range-correlated state (e.g., fully mixed state or a product state), has an effective Hilbert space dimension seen by a probe which grows in time as $D_B(t)\sim 2^{2v_B t}$, $v_B$ being the butterfly velocity. A rough estimate of the numbers of temporal and bath degrees of freedom thus places the system in the non-Markovian regime,
{$
r\approx\frac{1}{2v_B}>\frac{1}{2}$,
where the inequality follows from the constraint $v_B<1$ imposed by the assumption that the brickwork circuit is composed of two-local gates (as most past works assume \cite{nahum2018operator,von2018operator,chan2018solution,Bertini2019}). In such a setting, we therefore expect TE to grow linearly with $T$.}

An interesting possibility {to curb such extensive growth of TE}, further suggested by this analysis, is to consider the effect of a restricted setup by reducing the allowed frequency of operations that we can perform on the probe system.
Mathematically, preventing access to the probe at a given time step is implemented by a local projection onto the unitary evolution of the probe; we refer to this projection as temporal \textit{coarse-graining} of the IM state introduced earlier in Ref.~\cite{figueroa2019almost}. 
Suppose, in particular, that the probe system's  evolution at each step is set by the unitary operator $V_\tau$, see Fig.~\ref{fig1:schematic}(b,c). Then we can 
impose this evolution on the probe system at a fraction $1-n_{\rm cg}$ of time steps, 
effectively restricting the frequency of measurements we can access. This corresponds to projecting the IM on the remaining $n_{\rm cg}T$ probes, producing a new, coarse-grained IM state with a reduced temporal Hilbert space dimension $D_T^{\rm cg}=d^{2n_{\rm{cg}}T}$. By construction, coarse-graining preserves the values of few-point temporal correlations exactly. 
{The role of coarse-graining has been studied in Ref.~\cite{figueroa2019almost} for the toy random unitary bath model considered in our Section~\ref{sec:quantum_dot}, and also in Ref.~\cite{o2025diagnosing} where the authors considered the process tensor of a Hamiltonian system, with probe interventions at discrete times with tunable separation $\delta t$ playing the role of $n_{cg}^{-1}$.}

The central result of the present paper is {our claim} that coarse-graining can in fact lead {dramatically} to a {\it TE transition} of the IM from volume-law to {subvolume-}law scaling at a finite coarse-graining density $n_{\rm cg}<n_{\rm cg}^\star$, where the threshold value $n_{\rm cg}^\star$ is system-specific, {i.e., a {\it complexity transition} in the influence of the bath on the local system, see Fig.~\ref{fig1:schematic}(b,c). Intuitively, coarse-graining effectively renormalizes $r\to r n_{\rm cg}$, thereby eliminating the volume-law contribution whenever $n_{\rm cg}<n^\star_{\rm cg}\simeq 1/(2r)$. We demonstrate this transition rigorously in the random-unitary toy model and argue that an analogous transition occurs in generic quantum circuits in 1d.}


{Namely}, 
we investigate spatially extended quantum bath models {in 1d} with local interactions, exemplified by the Floquet kicked Ising model. Following Ref.~\cite{FolignoBertiniPRX2023}, we first consider dual-unitary points of this model with non-solvable initial conditions. We {can analytically} demonstrate that the TE transitions {from a volume to area-law} at an arbitrarily weak coarse-graining (i.e., $n_{\rm cg}^\star = 1$). We also show that, as expected, local observables $\langle O_\tau\rangle$ rapidly decay to zero, with irregular fluctuations decaying in magnitude of order $2^{-T/2}$. These fluctuations are very sensitive to the initial state and cannot be efficiently captured for arbitrarily long times by an MPS representation of the IM with a moderate bond dimension. However, owing to the rapid decay of these fluctuations, observables can be approximated to absolute accuracy $\epsilon$ using a truncated bond dimension that grows only polynomially in $1/\epsilon$. 


{Turning to more generic circuits, we numerically study the kicked Ising model away from the dual-unitary points and observe a TE transition upon coarse-graining. In contrast to random-unitary and dual-unitary circuits, below the transition the TE obeys a log-law rather than an area-law. This model also exhibits a much richer phenomenology: in particular, certain parameter regimes host prethermal, nearly conserved edge operators with parametrically slow temporal decay~\cite{FendleyPRXPreth}. We show that the behavior of TE under coarse-graining is sensitive to these prethermal effects and differs qualitatively between the corresponding regimes.}

{Lastly, the subvolume-law scaling of TE below the transition point suggests the existence of an effective low-bond-dimension IM that accurately reproduces few-point correlation functions.} To test this expectation in a concrete setting, we study the convergence of these correlation functions with increasing bond dimension in the kicked Ising model and find that relatively modest bond dimensions are indeed sufficient to faithfully capture the thermalization dynamics.


{All together, our work provides a unifying framework to understand and relate the concepts of temporal entanglement within the influence matrix, non-Markovianity of a bath, and temporal correlations. It also potentially provides a practical route to reducing the complexity needed to simulate a quantum chaotic bath, through the idea of coarse-graining. 
}

The rest of the paper is organized as follows: in Section~\ref{sec:IM}, we review the properties of the IM. In Section~\ref{sec:quantum_dot}, we analytically study TE and distillable entanglement for the case of a structureless random unitary bath, and discuss TE transition upon coarse-graining. Further, in Section~\ref{sec:Dual-Unitary}, we consider TE transitions in dual-unitary circuits with non-solvable initial conditions. After that, we turn to a more generic, non-dual-unitary kicked Ising model, discovering a phenomenology consistent with predictions from the structureless random unitary bath model. We argue that the IM can be significantly compressed into a compact MPS form, still capturing dynamics of local observables accurately. Finally, in Section~\ref{sec:discussion} we summarize the main results and outline directions for future research.

\section{Setting the stage 
}\label{sec:IM}

\subsection{IM formalism}\label{sec:im_formalism}

\begin{figure}
    \centering
    \includegraphics[width=0.95\linewidth]{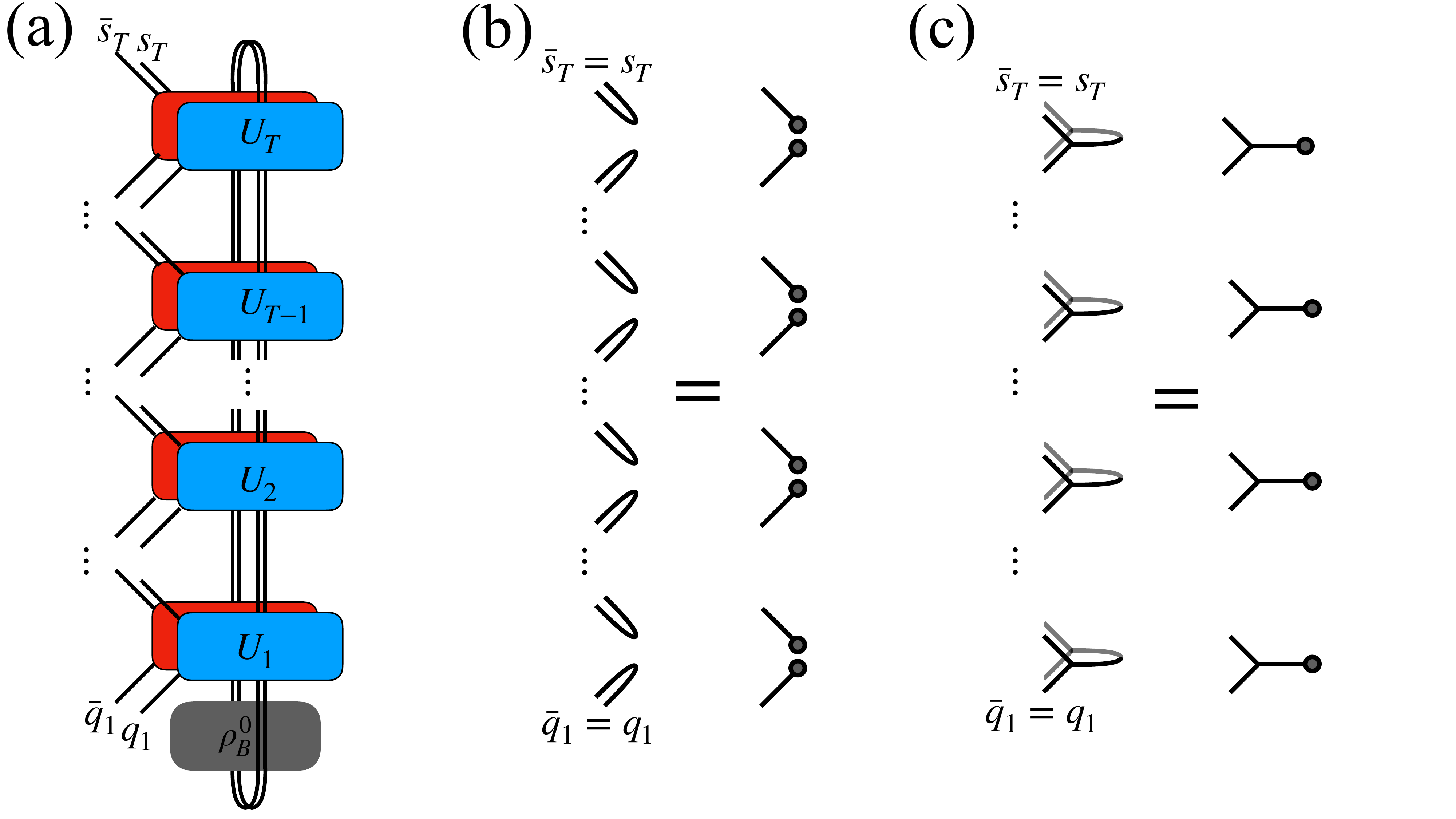}
    \caption{{(a) Tensor network representation of the IM: grey box at the bottom corresponds to an initial density matrix $\rho_{\mathcal{B}}^0$, blue gates correspond to the forward action of unitary gates $U_i$, while red gates correspond to the corresponding backward action. This is identical to Fig.~\ref{fig1:schematic} except that there all forwards and backwards diagrams (gates, legs, etc.) are overlaid (``folded representation'') while here all diagrams are explicitly shown (``unfolded representation'').
    In this paper, all diagrams drawn will predominantly be in the folded representation. 
    (b,c) Two illustrative perfectly Markovian IMs. (b) Perfect depolarizer solution in the unfolded (left) and folded (right) representation. 
    (c) Perfect dephaser solution in unfolded (left) and folded (right) representation. Both IMs exhibit area-law scaling of their TE. 
    }}
    \label{fig:IM_def}
\end{figure}

We start by {reminding the reader of the} definition of the IM and reviewing its key properties~\cite{pollock2018NonMarkovian,lerose2021Influence} [see Fig.~\ref{fig1:schematic}(a)]. An IM can be viewed as a multi-time generalization of a quantum channel: Given a many-body bath $\mathcal{B}$ with Hilbert-space dimension $D_\mathcal{B}$, at each time step $\tau=1,\dots,T$ a new probe qudit interacts with 
$\mathcal{B}$ via a unitary transformation ${U}_\tau$ acting on the composite $dD_\mathcal{B}$-dimensional space. This unitary contains both the bath's internal dynamics 
and qudit-bath interactions. At the end, bath degrees of freedom are traced out. The resulting object, which we denote by $\mathcal{I}$, is a quantum channel with $T$ inputs and $T$ outputs, causally related in a temporal sequence, see Fig.~\ref{fig:IM_def}(a). 
%

{Explicitly}, the IM elements are given by 
\begin{multline}\label{eq:IM_definition}
\mathcal{I}_{q_1,s_1\dots q_T,s_T}^{\bar{q}_1,\bar{s}_1\dots \bar{q}_T,\bar{s}_T}=\\ =\text{tr}_{\mathcal{B}}\left(({U}_{T})^{s_T}_{q_T}\dots({U}_{1})^{s_1}_{q_1}\, \rho^{\text{0}}_{\mathcal{B}} \,({U}^{\dagger}_{1})^{\bar{s}_1}_{\bar{q}_1}\dots ({U}^{\dagger}_{T})^{\bar{s}_T}_{\bar{q}_T}\right), 
\end{multline}
with indices $q_\tau,s_\tau$ [$ \bar{q}_\tau,\bar{s}_\tau$] associated with the input and output state of the $\tau$-th probe qudit on the  forward [backward] branch of time evolution, 
respectively. In this equation, we have defined the conditioned operators $\left(U_\tau\right)_{q_\tau}^{s_\tau}=\langle s_\tau |U_\tau| q_\tau \rangle$, acting on $\mathcal{B}$ only.

As illustrated in Fig.~\ref{fig1:schematic}a, the multi-time quantum channel $\mathcal{I}$ can be mapped to a $2T$-qudit (mixed) quantum state via Choi's channel-state  duality~\cite{choi1975completely}. 
At each step $\tau$, we prepare a maximally entangled (Bell pair) state of two qudits -- a ``reference'' qudit $Q_\tau$ and ``system'' qudit $S_\tau$, and we let the latter interact with the bath. This results in a state $\rho_\mathcal{I}$ of $2T$ qudits, which is dual to the IM channel: The density matrix associated with $\rho_\mathcal{I}$ has exactly the same elements as the IM $\mathcal{I}$ up to normalization,
\begin{equation}
\left(\rho_{\mathcal{I}}\right)_{q_1,s_1\dots q_T,s_T}^{\bar{q}_1,\bar{s}_1\dots \bar{q}_T,\bar{s}_T}=\frac{1}{d^T}\mathcal{I}_{q_1,s_1\dots q_T,s_T}^{\bar{q}_1,\bar{s}_1\dots \bar{q}_T,\bar{s}_T}.
\end{equation}
%

When studying circuits in Sections~\ref{sec:Dual-Unitary} and~\ref{sec:generic}, we will instead consider a constrained type of IM arising from probe-bath interactions described by a (time-independent) product operator, i.e., $U=e^{-i H_{\rm probe} \otimes H_{\rm bath}}$.
In this case, working in a basis where $H_{\rm probe}$ is diagonal, $H_{\rm probe} |s\rangle= h_s |s\rangle$, the IM elements simplify as
\begin{equation}\label{eq:Ising_constraint}
\langle s_\tau |U| q_\tau \rangle=\delta_{q_\tau,s_{\tau}} e^{-i h_{s_\tau} H_{\rm bath}}  = \delta_{q_\tau,s_{\tau}} U_{s_{\tau}},
\end{equation}
where the conditioned bath operator $U_s$ is here unitary,
 similarly to the original Feynman-Vernon influence functional~\cite{feynman2000Theory}. Then, we can identify $q_\tau=s_\tau$ and $\bar q_\tau=\bar s_\tau$ in Eq.~\eqref{eq:IM_definition}, which yields 
\begin{equation}\label{eq:IM_kicked_ising}
\mathcal{I}_{s_1\dots s_T}^{\bar{s}_1\dots \bar{s}_T}=\text{tr}_{\mathcal{B}}\left(U_{s_T}\dots U_{s_1}\rho^0_{\mathcal{B}} U_{\bar{s}_1}^\dagger\dots U_{\bar{s}_T}^\dagger\right).
\end{equation} 
In tensor notation, the general IM pictures in Fig.~\ref{fig1:schematic}b,c  simplify into the corresponding ones in Fig.~\ref{fig:CG_Ising}a,b. {Such an IM naturally arises in applications to the kicked Ising model (see Ref.~\cite{lerose2021Influence} for earlier work), and we therefore refer to it as an ``Ising-type'' IM.}

\subsection{Entanglement properties of the IM: Temporal entanglement}\label{sec:entanglement_properties}

{The IM provides a complete description of the influence of a many-body bath on a probe system.  Intuitively, the IM  encompasses the flow of information from earlier times (the ``past'') to  later times (the ``future'') via the bath, see Fig.~\ref{fig1:schematic}(a). A central question is to characterize this information flow, and to find out whether the IM admits compression -- in other words, whether the large unitary bath can be replaced by a smaller, potentially dissipative, effective bath, which still accurately approximates the probe system's dynamics. 
If the bath is Markovian (correlations between between past and future are small), then it is expected that compression can be performed well; otherwise if the bath is non-Markovian (correlations are large), then it is expected that such compression is not possible.  
Different measures of Markovianity based on the IM (or process tensor) have been introduced~\cite{pollock2018operational,LuchnikovPRL2019}.
}

{As mentioned before, temporal entanglement (TE) of the IM will be the key measure of complexity studied in this paper, following Refs.~\cite{banuls2009Matrix,muller-hermes2012Tensor,lerose2021Influence,lerose2023overcoming,ye2021constructing}.
We view the IM (or its associated density matrix $\rho_\mathcal{I}$)}
as a \textit{vector} $|\mathcal{I}\rangle$ in $d^{4T}$-dimensional space, and {study} the entanglement properties of this vector, in particular its von Neumann entanglement entropy with a past-future bi-partition $(\mathcal{P},\mathcal{F})$, 
{Explicitly, this is given by} 
\begin{equation}
\label{eq:defTE}
    S_{\rm TE} = - \Tr_\mathcal{P} \left( \rho_\mathcal{P} \log_d \rho_\mathcal{P}\right), \quad \rho_\mathcal{P} = \Tr_{\mathcal{F}} \left(\frac{|\mathcal{I}\rangle\langle\mathcal{I}|}{\langle\mathcal{I}|\mathcal{I}\rangle}\right)\, ,
\end{equation}
in contrast with other notions of IM entanglement that we investigate. 
{In particular, the {\it scaling} of TE with $T$ is an important aspect we are interested in (namely, how the amount of entanglement scales with the number of temporal degrees of freedom).}

Indeed, a particularly simple {illustrative} example of
an IM {describing a perfectly Markovian (no correlations between past and future) bath} is the \textit{perfect depolarizer}~\cite{lerose2021Influence}, 
\begin{equation}\label{eq:PD_definition}
    \left(\mathcal{I}_{\rm depol}\right)_{q_1,s_1\dots q_T,s_T}^{\bar{q}_1,\bar{s}_1\dots \bar{q}_T,\bar{s}_T}= \frac 1 {d^T} \prod_{\tau=1}^T \delta_{q_\tau \bar{q}_\tau}  \delta_{s_\tau \bar{s}_\tau},
\end{equation}
{which can easily be seen to exhibit an {\it area-law} scaling of TE}. {
Physically, this corresponds to a bath that transforms any incoming impurity degree of freedom into a maximally mixed state, thereby completely erasing the memory of its prior history.
}
Despite its simplicity, this state arises as an exact IM in a class of chaotic circuits known as dual-unitary circuits, for a family of solvable initial conditions~\cite{piroli2020exact}.
{A pictorial illustration of this state is provided in Fig.~\ref{fig:IM_def}(b) both in unfolded (left) and folded  (right) representation.} The many-body state associated with this  IM is the maximally mixed state $\rho_{\rm depol}=  \mathbb{1}_{d^{2T}}/d^{2T}$. {A relative of the perfect depolarizer solution, specific to the Ising-type IM is the \textit{perfect dephaser} solution:
\begin{equation}\label{eq:IPD_KIM}
\mathcal{I}_{T,\rm deph}=\prod\limits_{i=1}^T\delta_{s_i \bar{s}_i},
\end{equation}
illustrated in Fig.~\ref{fig:IM_def} (c). Physically, this IM maps any incoming density matrix to its diagonal part, eliminating all off-diagonal coherences --- hence the name ``perfect dephaser.'' It arises, for example, as an exact IM of the kicked Ising model at a dual-unitary point and also exhibits area-law scaling of TE.
}
{Area-law scaling of TE was further found in some integrable systems
~\cite{lerose2021Influence,klobas20exact,giudice2022Temporal,lerose2023overcoming,thoenniss2023Nonequilibrium}.}

{In contrast,} numerical simulations for  {generic} chaotic non-integrable baths  suggested a linearly increasing TE; this volume-law scaling was  recently proven by Foligno et al.~\cite{FolignoBertiniPRX2023} for chaotic 1d dual-unitary circuits {upon using non-solvable initial states, indicating a surprising fragility of the area-law depolarizer IM Eq.~\eqref{eq:PD_definition} solution of the perfectly Markovian quantum bath~\footnote{These authors also found that the IM vector for 1d chaotic baths has a high (i.e., only polynomially suppressed in $T$) overlap with a product state, resulting in area-law at the dual-unitary point or logarithmic scaling away from this point for all higher Rényi TEs}}.


\subsection{Temporal coarse-graining}
The IM is a very general object, which contains significantly more information than the standard few-point temporal correlation functions, $\langle  O_N(\tau_N) \dots O_1(\tau_1) \rangle$, where $O_j$ is an operator acting on the probe qudit and {the number of probes is assumed few:} $N=O(T^0)$.
Generally, such correlators can be computed by contracting the IM with quantum channels acting on the probe qudit only and inserting the operators $O_1,\dots,O_N$ at the appropriate time steps. Assuming probe system's internal dynamics is given by a time-independent unitary $V_\tau=V$,~\footnote{The extension to dissipative and time-dependent probe system's dynamics is straightforward.} 
the temporal correlation function is explicitly given by
\begin{multline}\label{eq:corr_function}
    \langle  O_N(\tau_N) \dots O_1(\tau_1) \rangle = \\ =\sum_{\{ q_i,\bar{q}_i, s_i, \bar{s}_i\}} \mathcal{I}_{q_1,s_1\dots q_T,s_T}^{\bar{q}_1,\bar{s}_1\dots \bar{q}_T,\bar{s}_T} \prod\limits_{k\ne \{\tau_i\}}V_{s_{k},q_{k+1}} V^*_{\bar{s}_{k},\bar{q}_{k+1}} \times \\ \times
    \prod\limits_{k=1}^{N} \left[O_k(\tau_k)V\right]_{s_{\tau_k},q_{\tau_k+1}} V^*_{\bar{s}_{\tau_k},\bar{q}_{\tau_k+1}} \, . 
\end{multline}

\begin{figure}
    \centering
    \includegraphics[width=0.95\linewidth]{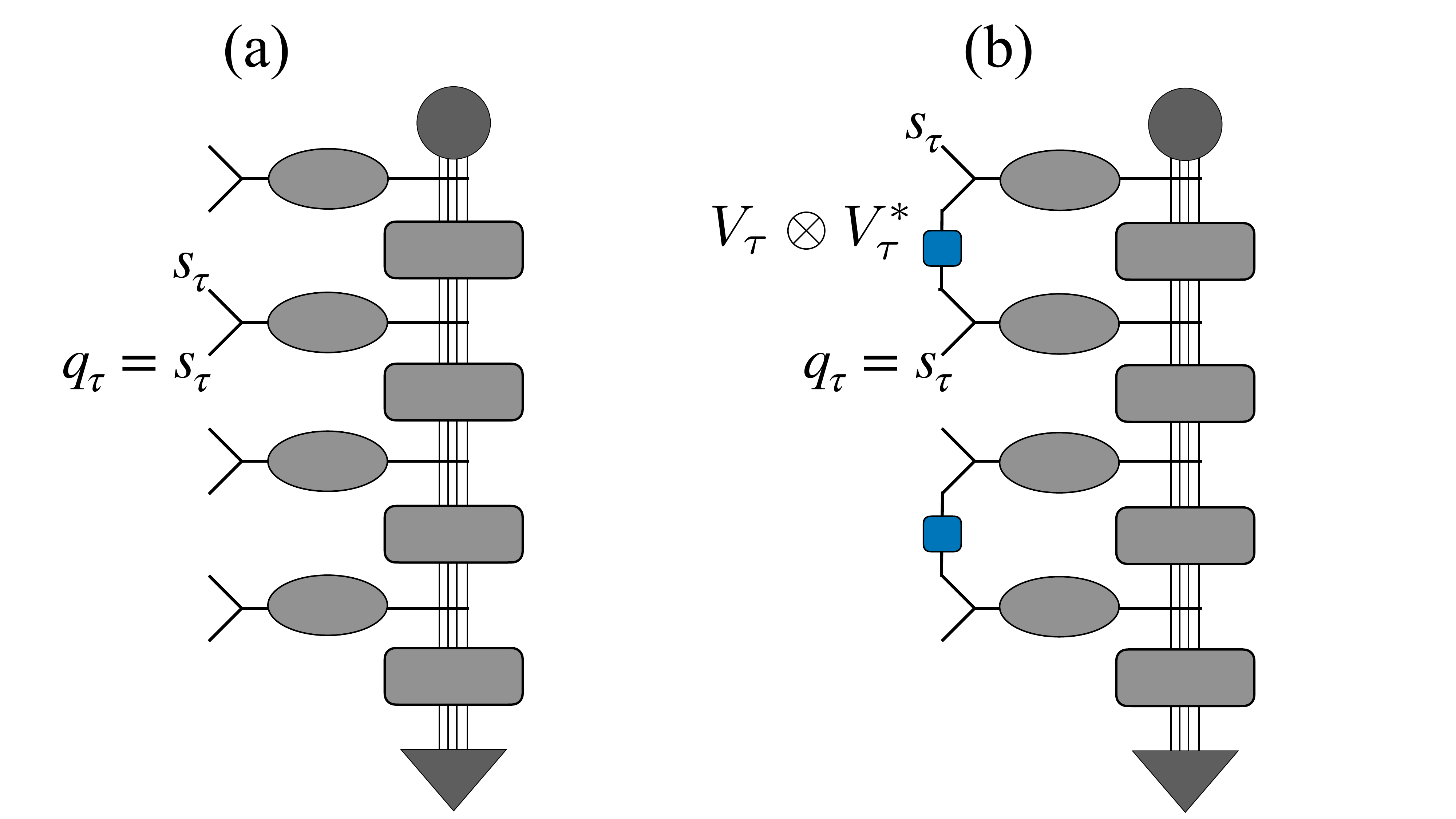}
    \caption{{(a)} IM of the constrained type, arising from probe-bath interactions in product operator form $U=e^{-i H_{\rm probe} \otimes H_{\rm bath}}$ {represented by oval gates}. We use here the diagonal tensor notation as in Ref.~\cite{lerose2021Influence}. {(b)} The IM after coarse-graining procedure with $n_{\rm cg}=1/2$.}
    \label{fig:CG_Ising}
\end{figure}

In this light, besides the task of globally approximating the IM channel $\mathcal{I}$ (or the dual state $\rho_\mathcal{I}$), another simpler task emerges: approximating the IM to the extent that a given relevant subset temporal correlation functions are accurately captured. On physical grounds, we are interested in describing processes where the frequency of ``interrogations'' of the probe is low; for instance, we could be interested in temporal correlations with a specified minimum temporal spacing between consecutive measurements of the probe. 
As we will show below, this task is significantly less demanding. To that end, we will consider a coarse-graining procedure for the IM, which amounts to inserting undisturbed unitary evolutions $V_\tau$ 
at a uniformly distributed fraction $(1-n_{\rm cg})$ of time steps~$\tau$. This way the number of IM degrees of freedom is reduced by a factor $n_{\rm cg}$. (For a vectorized IM, coarse-graining is equivalent to projection onto a specific hyperplane set by the choice of $V_\tau$.) This procedure is shown in Fig.~\ref{fig1:schematic}c for a general IM and in Fig.~\ref{fig:CG_Ising}b for the constrained-type IM of later relevance for the kicked Ising model. (In the latter case, for simplicity, we choose to restrict the operators $V_{\tau}$ to be diagonal, so as to preserve the constraint from Eq.~\eqref{eq:Ising_constraint}.) 
By construction, a coarse-grained IM still allows to exactly compute low-frequency correlators. Crucially, we will show that below a critical coarse-graining threshold $n^\star_{\rm cg}$
the IM complexity is greatly reduced, as TE collapses to a {subvolume-}law scaling. Such a coarse-grained IM can be effectively approximated by an MPS.  

\section{A toy random unitary bath model}\label{sec:quantum_dot}

\begin{figure}
\vspace{-1.3cm}
  \centering
\includegraphics[width=\linewidth]{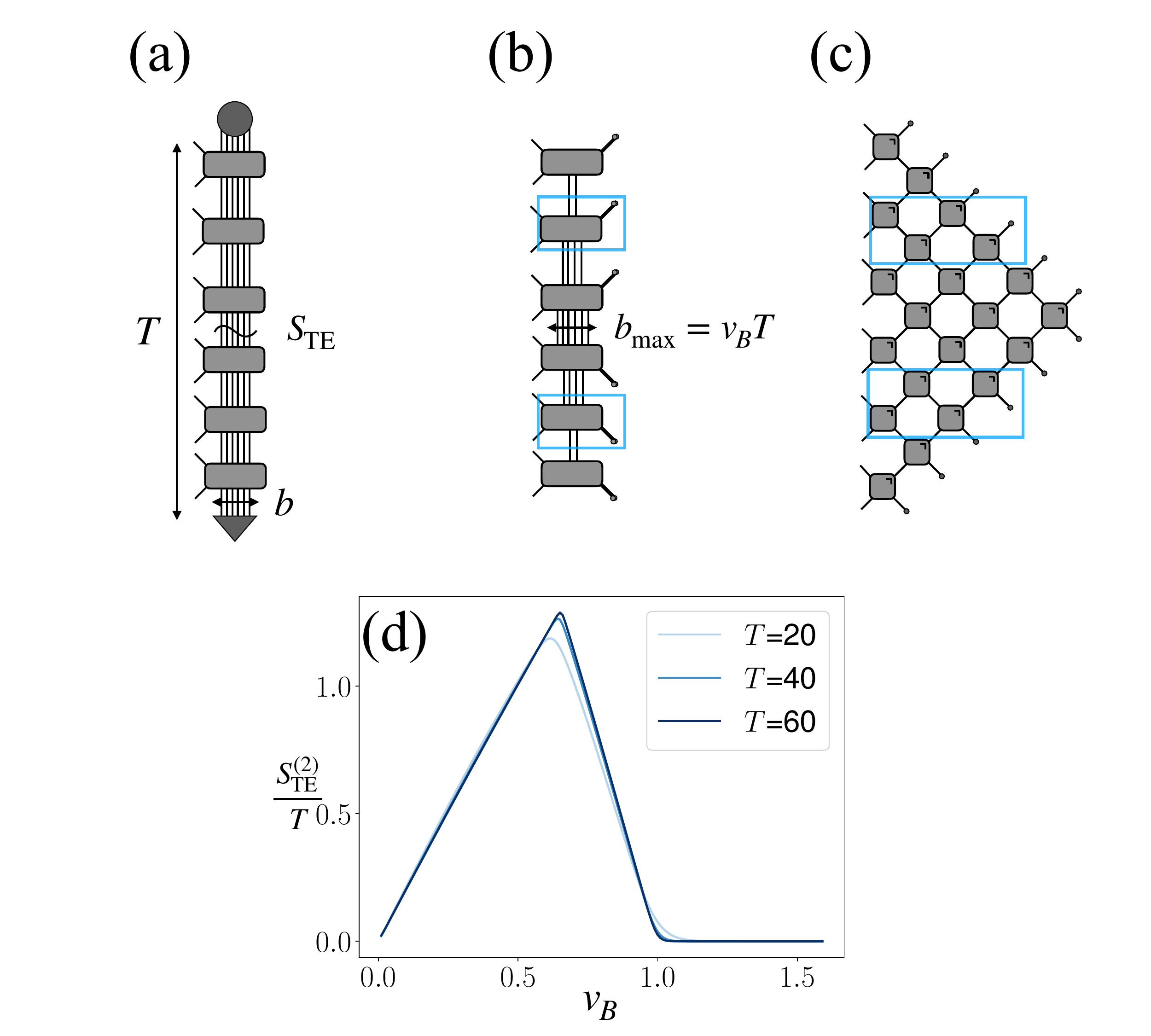}
  \caption{Illustration of the IMs of the static (a) and dynamically growing (b) bath toy models, for $T=6$ time steps and $b=6$ bath  qudits (a) or  $b_{\text{max}}=6$ maximal number of bath qudits (b), with $v_B=1$. {For illustration, we also show the corresponding quantum circuit for an infinite-temperature initial state (c). The random-bath approximation consists of replacing the structured gate blocks, highlighted by blue boxes, with unstructured random unitary matrices.}
 Panel (d): Slope of TE scaling as a function of $v_B$ for several values of $T$, showing the convergence to a limiting slope as $T\to\infty$ and a transition from volume-law to {subvolume-}law TE scaling at the critical value $v_B^\star=1$.}
  \label{fig:growing_bath}
\end{figure}

{Although our ultimate goal is to understand spatially extended systems with a nontrivial locality structure, much of the intuition for their dynamics comes from the spreading and scrambling of initially local quantum information \cite{nahum2017quantum,nahum2018operator,von2018operator}. Once this information has spread over a sufficiently large region and become strongly scrambled, the detailed spatial structure within that region may become secondary to its collective role as an effective bath. Motivated by this picture, we begin by considering a toy model of a completely structureless random bath. This model isolates the dynamics of an already-scrambled region while abstracting away its detailed spatial organization, allowing us to study the relation between thermalization, temporal entanglement (TE), and coarse-graining in a particularly simple setting. It also provides a useful baseline for the local circuit models considered below, in which the effective bath emerges dynamically through information spreading.

We consider two variants of the random-bath model, which we call the `static'' and `dynamical'' models. In both variants, at each time step $\tau$, a probe qudit interacts with a structureless bath of Hilbert-space dimension $D_{\mathcal{B}}(\tau)=d^{b(\tau)}$, where $b(\tau)$ denotes the effective number of bath qudits. The joint probe--bath evolution is generated by an independently drawn Haar-random unitary $U_\tau$ of dimension $dD_{\mathcal{B}}(\tau)\times dD_{\mathcal{B}}(\tau)$.

In the static model, introduced in Ref.~\cite{figueroa2019almost} and further studied in Ref.~\cite{Ippoliti2022solvablemodelofdeep}, the bath size is fixed throughout the evolution, $b(\tau)=b$. In the dynamical extension introduced here, we instead take $b(\tau)=2v_B\min(\tau,T-\tau)$, so that the bath grows linearly until $\tau=T/2$ and then shrinks at the same rate. The parameter $v_B$ plays the role of a butterfly velocity: it controls the rate at which new bath degrees of freedom become accessible, thereby mimicking the light-cone constraint on information propagation in a one-dimensional unitary circuit while leaving the bath itself structureless.
More explicitly, for $\tau<T/2$, the bath is enlarged at each time step by adding $2v_B$ maximally mixed qudits. The probe and the enlarged bath are then acted upon by a Haar-random unitary $U_\tau$ of dimension $d^{2v_B\tau+1}\times d^{2v_B\tau+1}$. For $\tau>T/2$, the bath is correspondingly reduced by tracing out $2v_B$ qudits at each time step. The IMs for static and dynamical random-bath models, together with the local quantum-circuit models considered below, are illustrated in Fig.~\ref{fig:growing_bath}(a,b).
}

\subsection{Overview of results}

{We give a high-level overview of the results contained within this section.} 
First, {in Sec.~\ref{Sec:TE_structureless_bath}}, we  analyze TE [Eq.~\eqref{eq:defTE}] in these toy models using Weingarten calculus~\cite{brouwer1996diagrammatic,collins2022weingarten}. In the static version, its behavior is controlled by the ratio $r=T/b$ between the number of degrees of freedom of the probe system and of the bath, both assumed large, $T,b \gg 1$  [see Eq.~(\ref{eq:r})]. We show that TE transitions from an exponentially suppressed behavior for $r<1/2$ to an extensive scaling (with $T$) for $r>1/2$, reaching 
the maximum possible value $2b$ 
at $r=1$.
In the dynamical version, a similar behavior is found, with the role of $r$ taken up by $1/({2v_B})$.
%
{Further, we explain how} the IM coarse-graining procedure {alluded to in the introduction} simply amounts to reducing the parameter $T\to n_{\rm cg} T$, and hence $r\to n_{\rm cg} r$ (static) or $v_B\to v_B / n_{\rm cg} $ (dynamical). This 
{results in the claimed} TE transition from volume-law to {subvolume-}law scaling.  

{In Sec.~\ref{sec:Non-Markovianity}} we investigate the relation between TE -- which, by itself, is not a physical quantity -- and an operationally well-defined notion of (non-)Markovianity of the probe system's dynamics, throughout the parameter space of this model. Focusing for definiteness on the static bath model, 
we first show that in the regime $r < 1/2$, where TE is exponentially suppressed, one can establish an upper bound on any possible connected correlation function between operators supported on $\mathcal{P}$ and $\mathcal{F}$, which ought to be smaller than $\epsilon = d^{-b(1/2-r)}$. 
This bound follows from the observation that 
the IM is $\epsilon$-close in trace distance to the Markovian perfect depolarizer IM in Eq.~\eqref{eq:PD_definition}. This fact was previously pointed out in Ref.~\cite{figueroa2019almost}, where the authors further bounded the probability of significant deviations from Markovianity. 

To go beyond this Markovian regime, we study the relation between TE and non-Markovianity for $r>1/2$.  In particular, we provide intuition for the physical origin and meaning of the extensive scaling of TE in this regime, 
by 
demonstrating the concomitant existence of an extensive amount of genuine quantum correlations between $\mathcal{P}$ and $\mathcal{F}$ in the form of shared Bell pairs, i.e.,  \textit{distillable entanglement}~\cite{bennett1996purification,bennett1996mixed}.
More precisely, 
we show that 
$\mathcal{F}$ can retrieve at least $ N_{\mathcal{P}\mathcal{F}} = T \min\left(\frac{2r-1}{2r},\frac 1 r\right)$ qudits of quantum information injected in the bath by $\mathcal{P}$. This result shows that the IM viewed as a many-body quantum state has extensive genuine quantum entanglement, and that this entanglement has a transparent physical  interpretation as non-Markovian memory of the bath.   
For $r>3/2$, in particular, the capacity of the bath to transmit quantum information from the past to the future is saturated, as $N_{\mathcal{P}\mathcal{F}}=b$. 
Our argument may be viewed as a multi-time extension of the Hayden-Preskill protocol for the problem of information retrieval from evaporating black holes~\cite{hayden2007black,yoshida2017efficient}, as it reduces to it in the limit of a single time step. 


\subsection{Temporal entanglement}\label{Sec:TE_structureless_bath}

We start by analyzing the Rényi TE, defined as bipartite Rényi entanglement entropy of the IM vector, ${S^{(\alpha)}_{\rm TE}}= \frac 1 {1-\alpha} \log_d \Tr \rho_{\mathcal{P}}^\alpha$, in the static bath model with constant $b$. We compute its behavior as a function of $d$ and $D_\mathcal{B}$ using Weingarten calculus~\cite{brouwer1996diagrammatic,collins2022weingarten} for a pure initial state of the bath. The details of the calculation, along with the analysis of a fully mixed initial state of the bath, are provided in Appendix \ref{app:Renyi}. The result for $\alpha=2$ and even bipartition $p=f=T/2$ is the following asymptotic TE formula:
\begin{equation}\label{eq:Renyi_2_pure}
{S^{(2)}_{\rm TE}}= -\log_d\frac{1+D_\mathcal{B}^{4(r-1)}+\epsilon(D_\mathcal{B},r)}{(1+D_\mathcal{B}^{2r-1})^2}+o\left(\frac{1}{D_\mathcal{B}}\right). 
\end{equation} 
{Although the second Rényi entropy provides only a lower bound on the von Neumann TE [see  Eq.~\eqref{eq:defTE}], it is sufficient to identify the regime in which TE grows. Conversely, in the regime of small TE, we derive an independent upper bound, thereby obtaining a clear picture of the overall behavior of TE.
}

The term $\epsilon(D_\mathcal{B},r)$ in this equation is negligible for all $r$ except in the regime $r\approx 1$, where it induces appreciable corrections of order $b^{-1} = (\log_d D_\mathcal{B})^{-1}$. This term can be computed exactly:
\begin{equation}\label{eq:Renyi_epsilon}
\epsilon(D_\mathcal{B},r)=2D_\mathcal{B}^{3(r-1)}+\left(\frac{3}{4} r \log_d D_\mathcal{B}+2 \right)D_\mathcal{B}^{2(r-1)}+2D_\mathcal{B}^{(r-1)}.
\end{equation} 

\begin{figure}[t]
\centering
\includegraphics[width=0.99\linewidth]{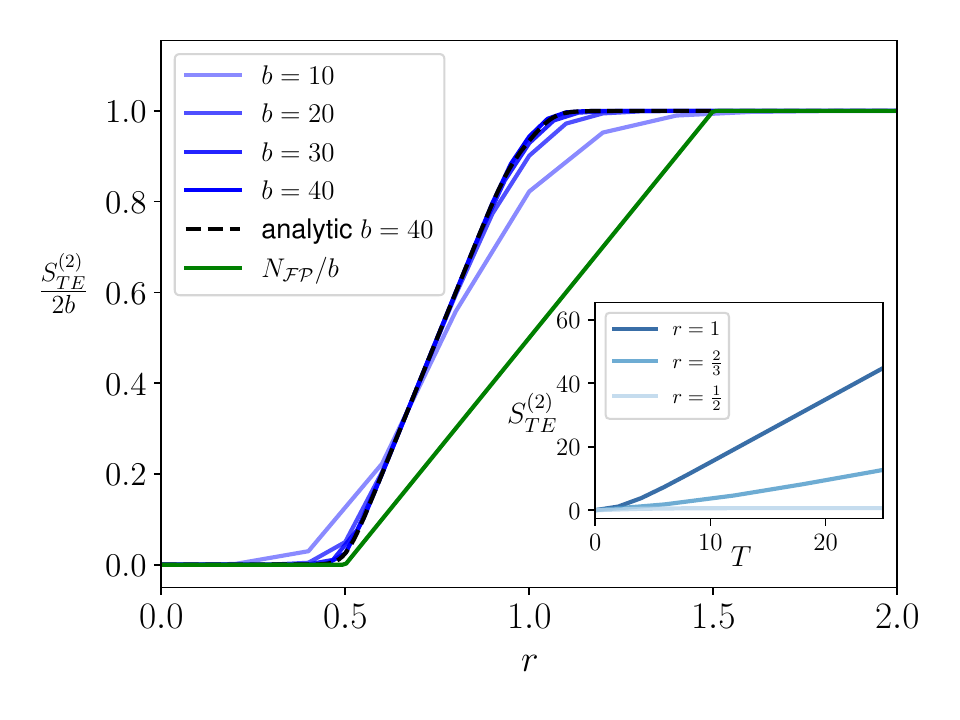}
    \caption{Maximum  temporal Rényi-2 TE for the structureless random unitary bath model, plotted as a function of $r$ for different bath sizes $b=\log_2 \mathcal{D}_B$, fixed probe system dimension $d=2$, and pure bath initial state. The dashed line illustrates the analytical prediction, see Eq.~(\ref{eq:Renyi_2_pure}). The green curve gives a lower bound on distillable entanglement measured in units of $b$. 
        {\it Inset:} Rényi-2 TE for $r=1$ and coarse-graining parameters $rn_{\rm cg}=1, 2/3, 1/2$. We observe a transition from volume-law to {area-}law scaling for $r^\star=1/2$. 
    } 
    \label{fig5:TE_quantum_dot}
\end{figure}

Below we specialize to the case of qubits, $d=2$. The exact behavior of ${S^{(2)}_{\rm TE}}$ as a function of $r$ is illustrated in Fig.~\ref{fig5:TE_quantum_dot} for different values of $b=\log_2 D_\mathcal{B}$, illustrating the approach to the asymptotic limit for large $b$. Notably, this approach is relatively slow near $r\approx 1$, due to the previously mentioned $\mathcal{O}(1/b)$ 
corrections.

The above result indicates that there are three regimes of Rényi TE behavior, also visible in Fig.~\ref{fig5:TE_quantum_dot}: 
\begin{itemize}
    \item For $r< 1/2$, TE is decaying to zero exponentially in $b$, $S^{(2)}_{\rm TE}\sim \frac{2d^{-b(1-2r)}}{\ln d}$;
    \item For $1/2< r\leq 1$, 
    TE approaches the limiting behavior $S^{(2)}_{\rm TE}\sim  2(2r-1) b $ for large $b$; 
    \item For $r> 1$, TE saturates at the maximum possible value $2b$. 
\end{itemize}
We note that extensivity of higher Rényi TEs implies extensivity of the regular von Neumann TE, due to monotonicity with the Rényi index $\alpha$.

In the regime $r<1/2$, the probe system's dynamics is nearly Markovian~\cite{figueroa2019almost}, and the IM is well approximated by a {perfect depolarizer IM, such that the corresponding density matrix $\rho_{\mathcal{I}}$ is close to a maximally mixed state}:
\begin{equation}\label{eq:rho_PD}
\rho_{\rm depol}=\frac{1}{d^{2T}}\mathbb{1}_{d^{2T}}. 
\end{equation}
In Appendix~\ref{app:replicas} [see Eq.~(\ref{eq:norm_pure_MM}) and Ref.~\cite{figueroa2019almost}], we estimate the trace distance from the maximally mixed state, obtaining the bound 
\begin{equation}\label{eq:Pure_purity_bound}
\|\rho_{\mathcal{I}}-\rho_{\rm depol}\|_1< d^{-b(1/2-r)}.
\end{equation}
This indicates that all correlations of the probe system are suppressed in the limit $b\to\infty$. In contrast, for $r>1/2$, where TE is proportional to $T$, the dynamics is non-Markovian, as discussed in the next Subsection.

Furthermore, the  random unitary bath model allows to illustrate the effects of coarse-graining by elementary means, as previously noted in Ref.~\cite{figueroa2019almost}. Since the product of Haar random unitaries is also a Haar-random unitary, coarse-graining the temporal axis (with an arbitrary $V_\tau$) simply amounts to  reducing the number of time-steps $T\to n_{\rm{cg}} T$, leading to a rescaling of the parameter $r$:
\begin{equation}
    r\to n_{\rm cg} r. 
\end{equation}
Thus, if the initial value $r=r_0$ satisfies the inequality $r_0>1/2$, such that the IM is initially an extensively entangled vector, coarse-graining beyond the critical value
\begin{equation}
    n_{\rm cg}^*=(2r_0)^{-1} 
\end{equation}
brings the IM down to the regime $r<1/2$, with exponentially small TE, exponentially small temporal correlators, and close to the Markovian perfect depolarizer `fixed-point' in trace distance. The bound~\eqref{eq:Pure_purity_bound} also 
guarantees that distilling even a single Bell pair between subsystems $\mathcal{P}$ and $\mathcal{F}$ is impossible.



{In order to capture the local structure of the quantum circuits, we now turn to the dynamically growing random unitary bath model. }
Figure~\ref{fig:growing_bath}(d) shows the dependence of the asymptotic slope of TE scaling  on the parameter $v_B$. We observe a transition from volume-law to area-law TE scaling at the critical value $v_B^\star = 1$. 
Coarse-graining reduces the number of time steps, effectively rescaling $v_B\to {v_B}/{n_{\rm cg}}$ (see Appendix \ref{app:growing_bath}), hence leading to a transition from volume-law to area-law TE at the critical value
\begin{eqnarray} \label{eq:n_CG}
n_{\rm{cg}}^{\star}=v_B.
\end{eqnarray}


{This result in the dynamical version of the toy model leads to an intriguing proposition for the extended 1d quantum circuits it is supposed to mimic: that coarse-graining of their IMs (which reduces $v_B$ too) may similarly lead to a transition in their temporal entanglement scaling. We shall explore this idea in Section \ref{sec:Dual-Unitary} and \ref{sec:generic}.}

\subsection{Non-Markovianity and transmission of quantum information through the bath}\label{sec:Non-Markovianity}

{Before exploring the effect of coarse-graining in extended 1d quantum circuits, we explore here within the toy model the relationship between
}
(non-)Markovianity of the probe system's dynamics and its IM's TE  --- {the latter} quantity we remind the reader does not have a straightforward operational meaning.
{The aim of this subsection is to ground the expectation that TE indeed serves as a good proxy for the degree of correlations between the past and future.}
Again, we divide the temporal degrees of freedom into `past' $\mathcal{P}$ and `future' $\mathcal{F}$, as in Fig.~\ref{fig1:schematic}(a), containing $2p$ and $2f$ qudits, respectively, such that
\begin{equation}\label{eq:past_future_steps}
    p+f=T. 
\end{equation}
Analogously to the previous subsection, we will assume that $b$ and $T$, as well as $p$, $f$ are large, and the ratio $r$ is kept fixed.

Correlations {between disjoint subsystems} can be rigorously estimated using quantum-information techniques, in particular decoupling theorems for noisy quantum channels~\cite{hayden2008decoupling}. Physically, the problem we consider is reminiscent of the problem of information retrieval from evaporating black holes~\cite{Page1993BlackHole, hayden2007black}, with the difference that here we are interested in information transfer from past to the future via the bath. At the technical level, we will look for an optimal temporal partition that maximizes distillable entanglement --- that is, the number of shared Bell pairs --- between $\mathcal{P}$ and $\mathcal{F}$. 
Below we provide an overview of the results along with intuitive explanations; details can be found in Appendix~\ref{app:information_recovery}.

\begin{figure}
    \centering
    \includegraphics[width=\columnwidth]{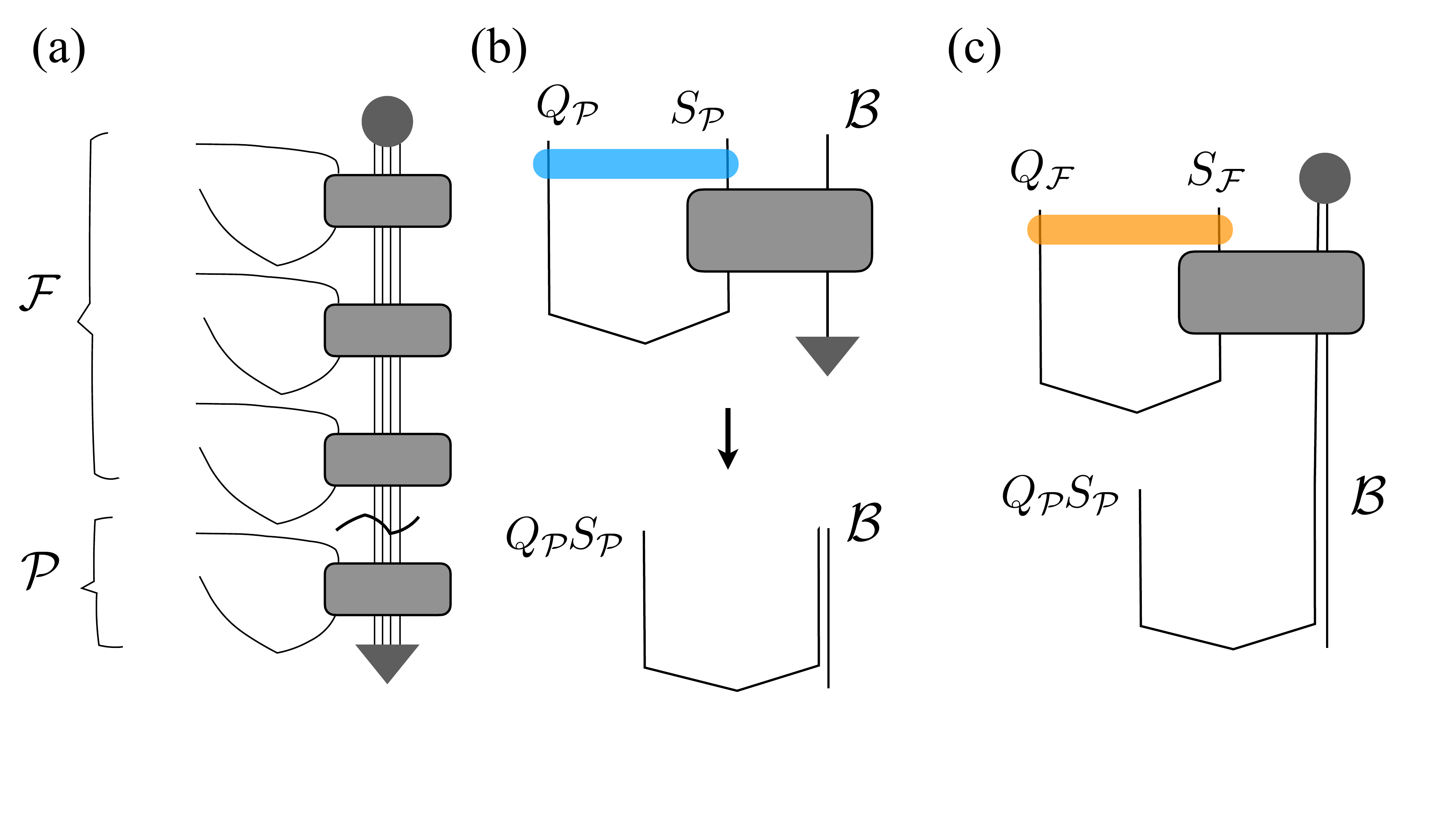}
    \caption{Illustration of the estimate of distillable entanglement. (a) We divide the temporal system into past $\mathcal{P}$ and future $\mathcal{F}$ subsystems, and think of evolution as consisting of two steps. (b) (top) In the first step, the bath gets entangled with the $\mathcal{P} $ qubits. (bottom) There exists a distillation protocol, acting only on the $\mathcal{P}$ degrees of freedom (blue box), which distills $N_{\mathcal BP}$ Bell pairs, provided the bath is sufficiently large, $b>2p$. (c) In the second step, the Bell-pair partners 
    in $\mathcal{B}$ get entangled with $\mathcal{F}$ qubits. As explained in the text, provided $\mathcal{F}$ is large enough, this allows to distill Bell pairs between $\mathcal{P} $ and $\mathcal{F}$ via a distillation protocol acting on $\mathcal F$ only (orange box).}
    \label{fig6:NonMarkovian}
\end{figure}

As discussed in Section~\ref{sec:im_formalism}, it is convenient to {divide} the degrees of freedom in subsystem $\mathcal{P}$ into a single ``reference'' set $Q_\mathcal{P}$ of $p$ qudits, and a single ``system'' set $S_\mathcal{P}$. The two sets are initially in a maximally entangled state. An analogous grouping into $Q_\mathcal{F}$ and $S_\mathcal{F}$ is defined for subsystem $\mathcal{F}$, as illustrated in Fig.~\ref{fig6:NonMarkovian}.  
Our goal is to estimate the number of Bell pairs $N_{\mathcal{PF}}$ shared between $\mathcal{P}$ and $\mathcal{F}$. In order to do so, we break down our problem into two steps: 
(i) First, $S_\mathcal{P}$ gets entangled with the bath $\mathcal{B}$ by a sequence of random unitaries, and (ii) $\mathcal{B}$ gets entangled with $S_\mathcal{F}$ by another sequence of random unitaries. Here we will focus on the case where the bath is initially in a pure state; the case of a mixed initial state is briefly considered in Appendix~\ref{app:information_recovery}.

Our strategy is to first estimate the number $N_{\mathcal{BP}}$ of shared Bell pairs between $\mathcal{B}$ and $\mathcal{P}$, and then estimate how much of that information can be retrieved by $\mathcal{F}$. The basic piece of intuition that guides us, adapted from Refs.~\cite{Page1993Entropy,hayden2007black}, is that the large random unitaries acting on bipartite unentangled states yield random pure states whose bipartite entanglement is close to maximal, leaving the smaller subsystem in a maximally mixed state containing no information. Here, we adapt this intuition to the multi-time process that generates an IM. 

In the first step, the bath gets entangled with $\mathcal{P}$, and the number of shared Bell pairs is determined by the smaller Hilbert space dimension: $N_{\mathcal{BP}}=\min (b,2p)$. In the second step, $\mathcal{F}$ is entangled with the $\mathcal{PB}$ system that can now be thought of as $N_{\mathcal{BP}}$ shared Bell pairs and the remaining $\max(b-2p,0)$ bath qudits, which 
are in a pure state. 

In the subsequent step, as in the black-hole information retrieval problem, the information about the state of $\mathcal{P}$ is contained mostly in $\mathcal{F}$, provided the size of $\mathcal{F}$ is large enough. 
More precisely, for $b>2p$, information can be retrieved by $\mathcal{F}$ provided $2f>b+2p$ (this condition simply states that the size of $\mathcal{F}$ exceeds the size of $\mathcal{B}$ and $\mathcal{P}$ combined); since $f=T-p$, this implies an inequality $4p<2T-b$, which can only be satisfied for $r=T/b>1/2$. The optimal choice of $p$, yielding a maximum transfer of information from past to future, is given by $p_{\rm opt}=\frac{2T-b}{4}$, and therefore we can transmit 
\begin{equation}
 N_{\mathcal{PF}}=2p_{\rm opt}=\frac{2T-b}{2}=\frac{(2r-1)b}{2} 
\end{equation}
qudits from the past to the future. This expression holds as long as $N_\mathcal{PF}$ does not exceed $b$: indeed, at most $b$ Bell pairs can be transmitted via the bath. This is saturated at $T=3b/2$, or, equivalently, $r=3/2$. 

To sum up, the number of Bell pairs between $\mathcal{P}$ and $\mathcal{F}$, that is, the distillable entanglement, is lower bounded by:  
\begin{equation}\label{eq:N_shared}
N_\mathcal{PF} \gtrsim\begin{cases} 0 ,& \text{ for}\  r<1/2\, , \\
\frac{(2r-1)b}{2}, &\text{ for}\  1/2<r<3/2 \, ,\\
b , &\text{ for} \ r>3/2 \, .
\end{cases}
\end{equation}

{Comparing this result with the estimate of TE from Eq.~\eqref{eq:Renyi_2_pure} and Fig.~\ref{fig5:TE_quantum_dot}, we conclude that, although the bound suggests a slope for the number of shared Bell pairs that is smaller by a factor of two, both temporal and distillable entanglement begin to grow {\it concomittantly} at $r=1/2$. This suggests that, although TE does not explicitly measure any physical quantity, it still provides a useful measure diagnosing both distillable entanglement and quantum non-Markovianity. 
} 

{Now,} the fact that past and future temporal subsystems share Bell pairs indicates {the existence of} strong quantum correlations. {At the same time  however}, as pointed out in Ref.~\cite{figueroa2019almost}, such correlations are highly non-local {(involving observables on many degrees of freedom in $\mathcal{P}$ and $\mathcal{F}$ at the same time)} and detecting them would generally require measuring complex, multi-time operators. As a result, the impact of these correlations on {\it few-point} temporal correlators of local observables is expected to be negligible. This expectation is
{indeed consistent with the coarse-graining procedure described previously, which  brings the IM close to the perfect depolarizer ``fixed-point'', and suggests that coarse-graining is a useful tool to eliminate complicated (and hence physically irrelevant) correlators and hence effectively reduce the complexity of the IM, as we will explore next}.

{We close this subsection by outlining}
 the technical steps leading to the result (\ref{eq:N_shared}). The decoupling theorems in Ref.~\cite{hayden2008decoupling}, colloquially, state that Bell pairs can be transmitted by a quantum channel, provided the state of the reference system (in the standard quantum-information terminology) is decoupled from environment, 
where the quality of the decoupling is measured by the trace distance from a product state. We adapt those results to our setting as follows: 
\begin{itemize}
    \item (i) The first step, as discussed above, is to show that $\mathcal{P}$ forms $\text{min}(b,2p)$ Bell pairs with the bath~$\mathcal{B}$. This can be demonstrated by bounding the trace distance between the reduced density matrix of the smaller of subsystems $\mathcal{P}$, $\mathcal{B}$ and the maximally mixed state. The calculation proceeds by first computing the Haar-averaged Frobenius distance using Weingarten calculus, and then applying the following inequality to relate it to the trace distance:
\begin{equation}\label{eq:trace_Frobenius_ineq}
    \|\rho\|_1\le \sqrt{\text{Dim}(\rho)}\|\rho\|_2,
\end{equation}
where $\text{Dim}(\rho)$ is the dimension of the Hilbert space on which $\rho$ acts.

\item (ii) When this distance is smaller than $\epsilon$, Ulhman's theorem imples that the subsystems $\mathcal{P}$ and $\mathcal{B}$ are in a maximally entangled state with fidelity $\mathit{F}\ge 1-\epsilon$.

\item (iii) Finally, to demonstrate that the information contained in the $\mathcal{BP}$ system is transmitted to system $\mathcal{F}$, we use the decoupling lemma from Ref.~\cite{hayden2008decoupling}, which states that a sufficient condition for distilling Bell pairs between ${\mathcal P, \mathcal F}$ is provided by the decoupling of the degrees of freedom in $\mathcal{P}$ and the ``outgoing'' bath degrees of freedom following the interaction with $\mathcal{F}$. The decoupling is demonstrated by estimating the trace distance between the bipartite state on ${\mathcal{BP}}$ (obtained by tracing out ${\mathcal{F}}$) and the maximally mixed product state, with the same strategy as above. 
\end{itemize}

For $r<1/2$, step (i) above suffices to obtain the bound in Eq.~\eqref{eq:Pure_purity_bound}. For $r>1/2$, we show that at least $N_{\mathcal{PF}}-s$ Bell pairs are shared between $\mathcal{F}$ and $\mathcal{P}$ with fidelity $|1-\it{F}|<d^{-s}$; thus, in the limit where both the bath $\mathcal{B}$ and the probe 
system $\mathcal{P}\mathcal{F}$ are large, Eq.~(\ref{eq:N_shared}) holds asymptotically. When $r>\frac{3}{2}$, the maximum possible number of  qudits, $b$, is shared between $\mathcal{F}$ and $\mathcal{P}$ through $\mathcal{B}$. 

Summarizing this subsection, for a structureless random unitary bath model we showed that extensive TE does signal extensive distillable entanglement between the probe system's past and future, and hence genuine non-Markovian memory~\footnote{Distillable entanglement provides a lower bound for the TE, but the opposite is not true in general. Nothing excludes the scenario of high TE and low distillable entanglement. For example this could happen in the system where correlations are almost classical but non-Markovian.}. Yet, the coarse-graining argument indicates that arbitrary temporal correlation functions with up to $n_{\rm cg}^*T$ operator insertions are extremely small, as indicated by the bounds in Eq.~(\ref{eq:Pure_purity_bound}). Thus, the extensive IM entanglement of the random unitary bath in the regime $r>1/2$ stems entirely from highly complex multi-time temporal correlations involving more than $n_{\rm cg}^*T$ interrogations of the probe. 

We emphasize that the `fixed-point' Markovian description applies -- with exponentially small corrections -- to a large family of temporal correlators that are still highly complex, associated with an extensive number $N=n T$ of operator insertions, with $0\le n<n_{\rm cg}^\star$, all the way down to few-point correlators with $N=\mathcal{O}(1)$.

\section{Temporal entanglement transition in dual-unitary circuits}\label{sec:Dual-Unitary}

Equipped with the intuition {of how coarse-graining can dramatically change the complexity structure of the IM in} the  random unitary bath models above, next we focus on {coarse-graining's effect on the} TE properties of {IMs of} 1d Floquet circuits. 
In this Section, we {first} study chaotic dual-unitary circuits, {whose special property of ``dual-unitarity'' allow for a number of analytical results~\cite{Bertini2019,BertiniSFF,BertiniPRXEntanglementDU,claeys2021ergodic,piroli2020exact,bertini2025exactly} despite these systems being generically non-integrable and chaotic.}
Dual-unitarity can be colloquially described as the {property that the} ``space-time rotated'' {version of the} brickwork circuit ({i.e., when viewed along the sptial direction}, rather than the time direction) {also describes}  unitary {evolution}~\cite{Guhr1,Bertini2019,bertini2025exactly}. 
{In particular, the key results involving dual-unitary circuits relevant for us is that for a family of solvable initial states of the bath~\cite{piroli2020exact}, their IMs can be shown to take the perfect depolarizer form of Eq.~(\ref{eq:rho_PD}) (or perfect dephaser depending on the particular model considered), signifying that such a bath leads to perfectly Markovian dynamics~\cite{lerose2021Influence}.}
{Further}, Reference~\cite{FolignoBertiniPRX2023} analyzed TE for dual-unitary circuits with generic, non-solvable initial states, and proved that von Neumann TE exhibits volume-law scaling with~$T$, whereas all higher Rényi TE exhibit area-law scaling. 

Here, we {analytically} demonstrate the occurrence of a TE transition {upon} 
coarse-graining dual-unitary baths. For definiteness, we will consider a specific circuit family, the kicked Ising model. The unitary dynamics $U$ that defines the bath's influence matrix is given by the Floquet operator 
\begin{gather}\label{eq:KIM}
    {U}=  \prod\limits_{j\ge 0}{P}_{j+1/2} \prod\limits_{j\ge 1}{W}_j \\
    {P}_{j+1/2}=e^{iJ{\sigma}_j^z{\sigma}_{j+1}^z}, \;\; 
{W}_j=e^{ih{\sigma}^z_j}e^{ig {\sigma}^x_j}, 
\end{gather}
where $\sigma_j^{x,y,z}$ are Pauli matrices acting on site $j=0,1,2,\dots$ of a semi-infinite qubit chain. For $h=0$, the kicked Ising model dynamics is mappable to a free Majorana chain, while for $h\ne0$ it is expected to be chaotic~\cite{kim2014testing}. The dual-unitary property is realized for $|J|=|g|=\pi/4$ or $3\pi/4$ and arbitrary~$h$.  

The probe system is chosen to be the boundary qubit at $j=0$, while all other qubits at $j\ge1$ constitute the bath. As discussed in Section~\ref{sec:im_formalism}, due to the product-operator structure 
of qubit-qubit interactions in the kicked Ising model, its IM 
takes a diagonal form with respect to input-output degrees of freedom (see Eq.~\ref{eq:Ising_constraint}). Defining the reduced 
Floquet operator 
\begin{eqnarray}\label{eq:U_boundary}
U_{s}= e^{iJs\hat{\sigma}^z_1}   \prod\limits_{j\ge 1}{P}_{j+1/2} \prod\limits_{j\ge 1} {W}_j, 
\end{eqnarray}
conditioned on the state $s=\pm1$ of the probe qubit 
at $j=0$, 
the IM is defined by Eq.~\eqref{eq:IM_kicked_ising}.The  IM elements have a transparent physical interpretation as generalized Loschmidt echoes, given by two different time-dependent boundary magnetic fields in the forward-in-time and backward-in-time evolutions~\cite{lerose2021Influence}.


The initial states considered are simple product pure states
\begin{equation}\label{eq:tilted_in1}
\rho^0_{\mathcal{B}}=|\psi_0\rangle \langle \psi_0|=\otimes_{j=1}^\infty \rho_j^0. 
\end{equation}
For definiteness, we take all the bath qubits pointing along the direction $(\cos\phi, \sin \phi, 0)$, i.e.,
\begin{eqnarray} \label{eq:tilted_in2}
\rho^0_j=\frac{1}{2}\begin{pmatrix}
    1 & e^{i\phi} \\
    e^{-i\phi} & 1
\end{pmatrix}.
\end{eqnarray}
{At dual-unitary points of the kicked Ising model}, this initial state is solvable for $\phi=0,\pi$, {whereupon its IM is provided by a perfect dephaser solution from Eq.~\eqref{eq:IPD_KIM}.
For parameters $\phi \neq 0,\pi$, the initial state is non-solvable; this will be the focus of our subsequent analysis.}

\subsection{IM structure and entanglement}

In this Subsection we review the argument from Ref.~\cite{FolignoBertiniPRX2023} for the linear scaling of TE, and adapt it to the case of kicked Ising model.
{The discussion is technical, but it is important to isolate the mechanism responsible for the growth of TE and provides the key ingredients needed to determine how temporal coarse-graining modifies the structure of the influence matrix. We present only the steps required for this purpose and refer the reader to Ref.~\cite{FolignoBertiniPRX2023} for the full derivation.
}

\begin{figure}
    \centering
    \includegraphics[width=0.9\linewidth]{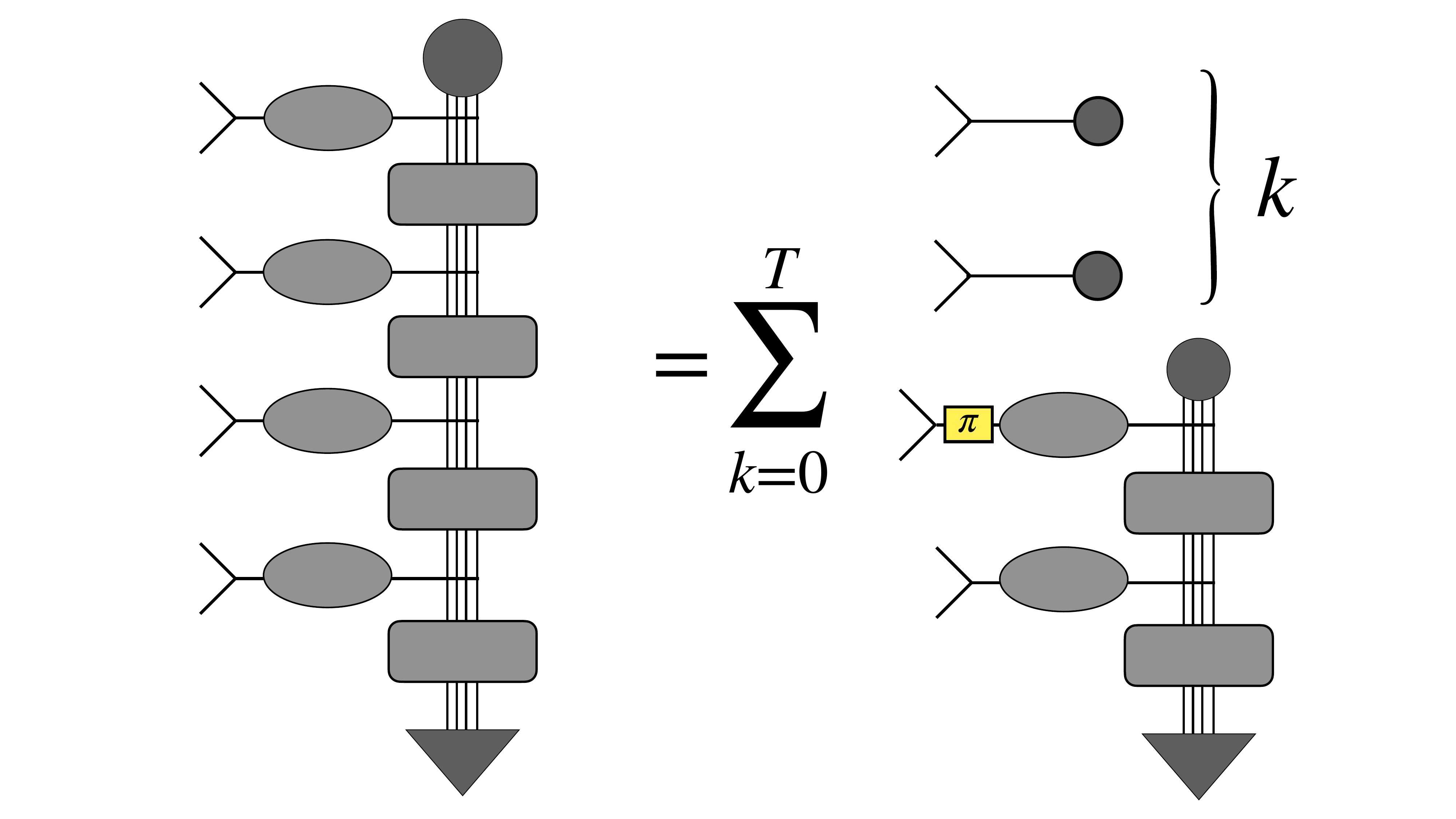}
    \caption{{Due to the recursive structure of the IM, as expressed by Eq.~\eqref{eq:recursive_IM}, an IM at any time $T$ necessarily contains IMs corresponding to all shorter times. By recursively inserting the resolution of the identity $\mathbbm{1}_{s,\bar{s}}^{s',\bar{s}'}=\delta_{s',\bar{s}'}\delta_{s,\bar{s}}+\pi_{s,\bar{s}}^{s',\bar{s}'},$ we obtain the decomposition shown here.
    }} 
    \label{fig:Bertini_decomp}
\end{figure}

Firstly, we decompose the IM in terms of IMs associated with shorter evolution time windows. 
To obtain this decomposition, we first note that the components of IM which are diagonal at the last time step simplify as
\begin{equation}\label{eq:recursive_IM}
\delta_{s_T,\bar s_T}\mathcal{I}_{s_1\dots s_{T-1},s_T}^{\bar{s}_1\dots \bar{s}_{T-1},\bar s_T}= \delta_{s_T,\bar s_T}\mathcal{I}_{s_1\dots s_{T-1}}^{\bar{s}_1\dots \bar{s}_{T-1}} 
\end{equation}
(no summation implied).
This property, which follows straightforwardly from Eq.~\eqref{eq:IM_kicked_ising}, implies that the last time step's diagonal component of the IM $\mathcal{I}_T$ for evolution time $T$ factorizes as tensor product of $\mathcal{I}_{T-1}$ and a single-qubit perfect dephaser. 
Using this observation and recursively decomposing the IM into its diagonal and off-diagonal components at the final time step, we obtain the following decomposition:
\begin{equation}\label{eq:IM_decomposition}
\mathcal{I}_T=\sum_{k=0}^{T} \tilde{\mathcal{I}}_{T-k} \otimes \mathcal{I}_{k,\rm deph}, \quad \tilde{\mathcal{I}}_{T-k}= \pi_{T-k} \mathcal{I}_{T-k} \, .
\end{equation}
In this equation we have defined the single-qubit projector $\pi$ on the off-diagonal components $s\neq\bar s$, i.e.,
\begin{equation}
\left( \pi_\tau \mathcal{I}_T \right)_{s_1\dots s_T}^{\bar{s}_1\dots \bar{s}_T}= \delta_{s_\tau,-\bar{s}_\tau} \left(\mathcal{I}_T \right)_{s_1\dots s_T}^{\bar{s}_1\dots \bar{s}_T} \, .
\end{equation}
The $k$-th term in the decomposition~\eqref{eq:IM_decomposition} is the tensor product of two factors: the perfect-dephaser IM, cf.~Eq.~\eqref{eq:IPD_KIM}, acting over the final $k$ time steps, and the bath IM for the shorter evolution time $T-k$, projected at its final time step onto the off-diagonal subspace $s_{T-k}\neq\bar{s}_{T-k}$, {as illustrated in Fig.~\ref{fig:Bertini_decomp}.}

We now analyze TE scaling of the IM viewed as a $2^{2T}$-dimensional vector $|\mathcal{I}_T\rangle$. We define the TE and the ``reduced density matrix'' according to Eq.~\eqref{eq:defTE}.
%
First, using the orthogonality of the vectors $|\widetilde{\mathcal{I}}_{T-k}\rangle$, 
we can write 
\begin{equation} \label{eq:Foligno_decomposition_rho}
\rho_{\mathcal{P}} = w\frac{|\mathcal{I}_p\rangle\langle\mathcal{I}_p|}{\langle\mathcal{I}_{p}|\mathcal{I}_{p}\rangle}+ \sum\limits_{k=p+1}^{T} w_k\tilde{\rho}_{k} ,
\end{equation}
where $\tilde{\rho}_{k} \propto\Tr_{\mathcal{F}}(|\widetilde{\mathcal{I}}_k\rangle\langle\widetilde{\mathcal{I}}_k|)$ are normalized density matrices with unit trace. The weights $w_k$ and $w$ are defined consistently with the normalization requirements of each term: 
Defining the squared norm of each vector 
\begin{equation}
N_k^2=
\langle\widetilde{\mathcal{I}}_{k}|\widetilde{\mathcal{I}}_{k}\rangle \langle\mathcal{I}_{\rm deph,T-k}|\mathcal{I}_{\rm deph,T-k}\rangle=\langle\widetilde{\mathcal{I}}_{k}|\widetilde{\mathcal{I}}_{k}\rangle 2^{T-k},
\end{equation}
the weights can be expressed as
\begin{gather}
w_k=\frac{N_k^2}{\sum\limits_{k=0}^T N_k^2}, \;\;\;
w = \sum\limits_{k=1}^p w_k.
\end{gather}

Evaluating the weights in the decomposition~\eqref{eq:Foligno_decomposition_rho} plays a central role in the computation of TE. It is natural to expect that the matrix elements of $\tilde{\mathcal{I}}_{k}$ 
 are, effectively, exponentially small (in $k$) random numbers.
{Indeed, to gain an intuitive understanding of these scalings, we use the Ising-specific structure of the IM and interpret its matrix elements as generalized Loschmidt echoes following Ref.~\cite{lerose2021Influence}. Namely, the matrix elements correspond to a forward-in-time evolution subject to the time-dependent boundary field $h_\tau = h + s_\tau J$, followed by a backward-in-time evolution subject to the field $\bar h_\tau = h + \bar s_\tau J$, with $J=\pi/4$ in the present case. This can be seen directly by specializing Eq.~\eqref{eq:IM_kicked_ising} to a pure initial state:

\begin{equation} 
\mathcal{I}_{s_1\dots s_T}^{\bar{s}_1\dots \bar{s}_T} = \langle\psi| U_{\bar{s}_1}^\dagger\dots U_{\bar{s}_T}^\dagger U_{s_T}\dots U_{s_1} |\psi\rangle . 
\end{equation}

 Combined with a rapidly entangling evolution in chaotic dual-unitary circuits, this suggests that the generalized Loschmidt echo is, in effect, an overlap of two random vectors $v_1=U_{s_T}\dots U_{s_1}|\psi\rangle$ and $v_2=U_{\bar{s}_T}\dots U_{\bar{s}_1}|\psi\rangle$ in the bath portion that is causally connected with the boundary.} This effective bath has Hilbert space dimension $2^{k}$, and hence the typical overlap is expected to be of order $2^{-k/2}$ in magnitude
{\begin{equation}\label{eq:IM_magnitude}
\|\mathcal{I}_{s_1\dots s_{T-1},s_T}^{\bar{s}_1\dots \bar{s}_{T-1},\bar s_T}\| \sim 2^{-k/2}.    
\end{equation}}
This expectation is consistent with the conclusions of Ref.~\cite{FolignoBertiniPRX2023} for generic dual-unitary circuits, and we verified it  numerically, cf. Fig.~\ref{fig:IM_norm}. 

We are now in a position to estimate the norm of $\tilde{\mathcal{I}}_{k}$ as
\begin{equation}\label{eq:tildeI_norm}
\langle\widetilde{\mathcal{I}}_k|\widetilde{\mathcal{I}}_k\rangle \approx C 2^k,
\end{equation}
where $C$ is a constant that depends on the initial state (in particular, for solvable initial states, see Ref.~\cite{piroli2020exact}, {corresponding to a perfect dephaser solution,} we have $C=0$). This yields 
\begin{gather}
N_k^2 \approx C 2^T\quad \text{for} \ k\ne T, \\
N_0^2=2^T,
\end{gather}
where $N_0^2=2^T$ is the contribution of the perfect dephaser part in the IM decomposition. We thus see that all the $T+1$ orthogonal vectors appearing in the IM decomposition have comparable norms. The overlap with the non-entangled perfect dephaser term ($k=T$), in particular, is responsible for the area-law higher Rényi entropies~\cite{FolignoBertiniPRX2023}. 

As the last ingredient, we note that the von Neumann entanglement entropy of a statistical mixture of states may be upper and lower bounded as follows~\cite{FolignoBertiniPRX2023},
\begin{equation}\label{eq:entropy_bound}
S_{\text{mix}} \le S_{\text{TE}} ( \rho_{\mathcal{P}})\le S_{\text{mix}}+S_{\text{cl}},
\end{equation}
where we introduced the mixture of entanglement entropies $S_{\text{mix}}$ and the classical mixing entropy  $S_{\text{cl}}$, defined as 
\begin{gather}\label{eq:mixture_bound}
S_{\text{mix}}=\sum\limits_{k=p+1}^{T} w_k S_{\text{TE}}(\tilde{\rho}_{k}) \, ,\\
S_{\text{cl}}=-\sum\limits_{k=p+1}^T w_k\log_2 w_k-w\log_2 w \, ,
\end{gather}
respectively.
In Eq.~\eqref{eq:entropy_bound}, the left-hand side inequality is a direct consequence of the concavity of the entanglement, and the right-hand side one follows from the property $\chi=S_{\text{TE}} ( \rho_{\mathcal{P}})-S_{\text{mix}}\le S_{\text{cl}}$ of Holevo's information (see for example Chapter 10.6 of \cite{preskill2016quantum}).

As a crucial step, Ref.~\cite{FolignoBertiniPRX2023} showed that the second Rényi entropy $S^{(2)}(\tilde{\rho}_{k})$ takes the maximal possible entropy value, which constrains the von Neumann entropy to take the same value:
\begin{equation}\label{eq:ent_rho_tilde}
S_{\text{TE}}(\tilde{\rho}_{k})= S^{(2)}(\tilde{\rho}_{k})=\text{min}\left(\frac{k-p}{2},p \right),
\end{equation}
{where $p$ denotes the number of `past' time steps before the cut, as defined above Eq.~\eqref{eq:past_future_steps}.}
This leads to the extensive entropy scaling 
\begin{equation}\label{eq:mix_entanglement}
S_{\text{mix}}=\begin{cases}      \frac{(T-p)(T-p+1)}{4(TC+1)}  & p> \frac{1}{3}T, \\
      \frac{p (2T-4p+1)}{2(TC+1)}  & p < \frac{1}{3} T.
\end{cases}
\end{equation}
Lastly, since the classical entropy is seen to scale subextensively (logarithmically) for long evolution times,
\begin{equation}
S_{\text{cl}}\approx\frac{pC+1}{TC+1}\log_2\frac{TC+1}{pc+1}+\frac{f}{1+TC}\log_2 (TC+1),
\end{equation}
then TE asymptotically equals the result~\eqref{eq:mix_entanglement}, i.e,   $S_{\text{TE}}\sim S_{\text{mix}}$ for large $T$.

\begin{figure}[t]
    \centering
    \includegraphics[width=0.9\linewidth]{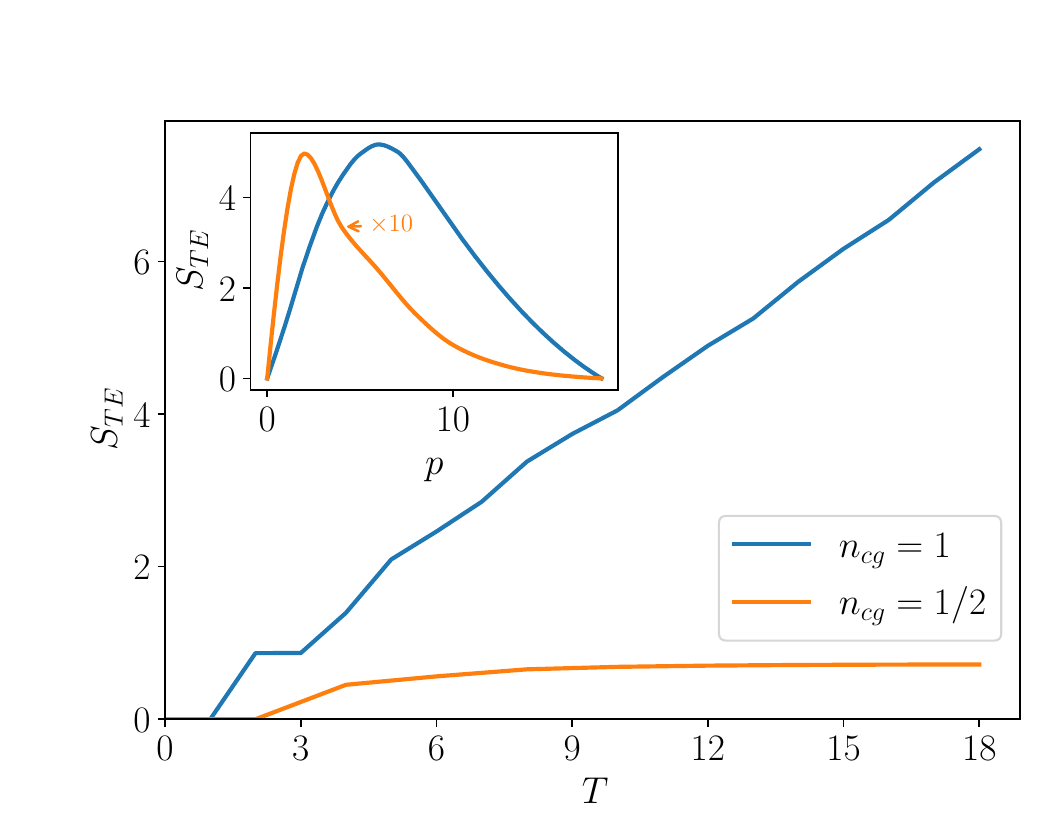}
    \caption{Von Neumann TE before and after coarse-graining with density $n_{\rm{cg}}=1/2$, for the dual-unitary kicked Ising model bath with $h=0.5$ and initial pure state described in Eqs.~(\ref{eq:tilted_in1},\ref{eq:tilted_in2}) with tilt angle $\phi=0.7$.}
    \label{fig:TEE_DU}
\end{figure}

In Fig.~\ref{fig:TEE_DU} we report a numerically exact computation of the von Neumann TE as a function of $p$ and $T$, obtained by constructing a truncation-free MPS representation of the IM using the light-cone growth algorithm (LCGA). {This algorithm iteratively constructs the IM adding one time-step per iteration. At each evolution time, the corresponding IM includes only the portion of the bath contained within the probe’s past causal light cone. Thus, although the effective bath is finite at any fixed time, its size grows with the maximal evolution time, allowing the dynamics to be simulated directly in the thermodynamic limit. We refer the reader to Refs.~\cite{lerose2023overcoming,FriasBanulsPRB2022}, where this algorithm was first introduced}. We observe volume-law scaling with $T$, as well as an asymmetric profile of TE  as a function of $0\le p \le T$ at fixed evolution time $T$, reported in the inset. Both properties are in agreement with the analytical estimates above.

\subsection{Coarse-graining and TE transition}

We now turn to our central question: the effect of temporal coarse-graining on the IM. 
For simplicity, here we choose the simplest coarse-graining unitary $V_\tau=\mathbbm{1}$; we expect, however, that our conclusions do not depend on the particular choice of $V_\tau$.
 Under coarse-graining, the decomposition in Eq.~\eqref{eq:Foligno_decomposition_rho} becomes
\begin{equation}
\rho_{\mathcal{P}}^{\rm{cg}}=w^{\rm{cg}}\frac{|\mathcal{I}_{p,cg}\rangle\langle \mathcal{I}_{p,\rm{cg}}|}{\langle\mathcal{I}_{p,cg}|\mathcal{I}_{p,cg}\rangle}+\sum\limits_{k=n_{\rm{cg}}(p+1)}^{n_{\rm{cg}}T}\omega^{\rm{cg}}_{k/n_{\rm{cg}}} \tilde{\rho}_{k/n_{\rm{cg}}},
\end{equation}
where we assumed that any non-integer index is rounded to its closer integer neighbor; in the long-time limit, this is expected to introduce a small relative error suppressed as $1/T$. Crucially, different terms in the above equation `flow' differently under coarse-graining. As we discuss below, this implies that under an arbitrarily mild coarse-graining, TE collapses to an area law scaling. This observation agrees with the heuristic guess in Eq.~\eqref{eq:n_CG}, which implies $n_{\rm{cg}}^\star = 1$ for systems with maximal butterfly velocity $v_B=1$.

\begin{figure}
    \centering
    \includegraphics[width=0.9\linewidth]{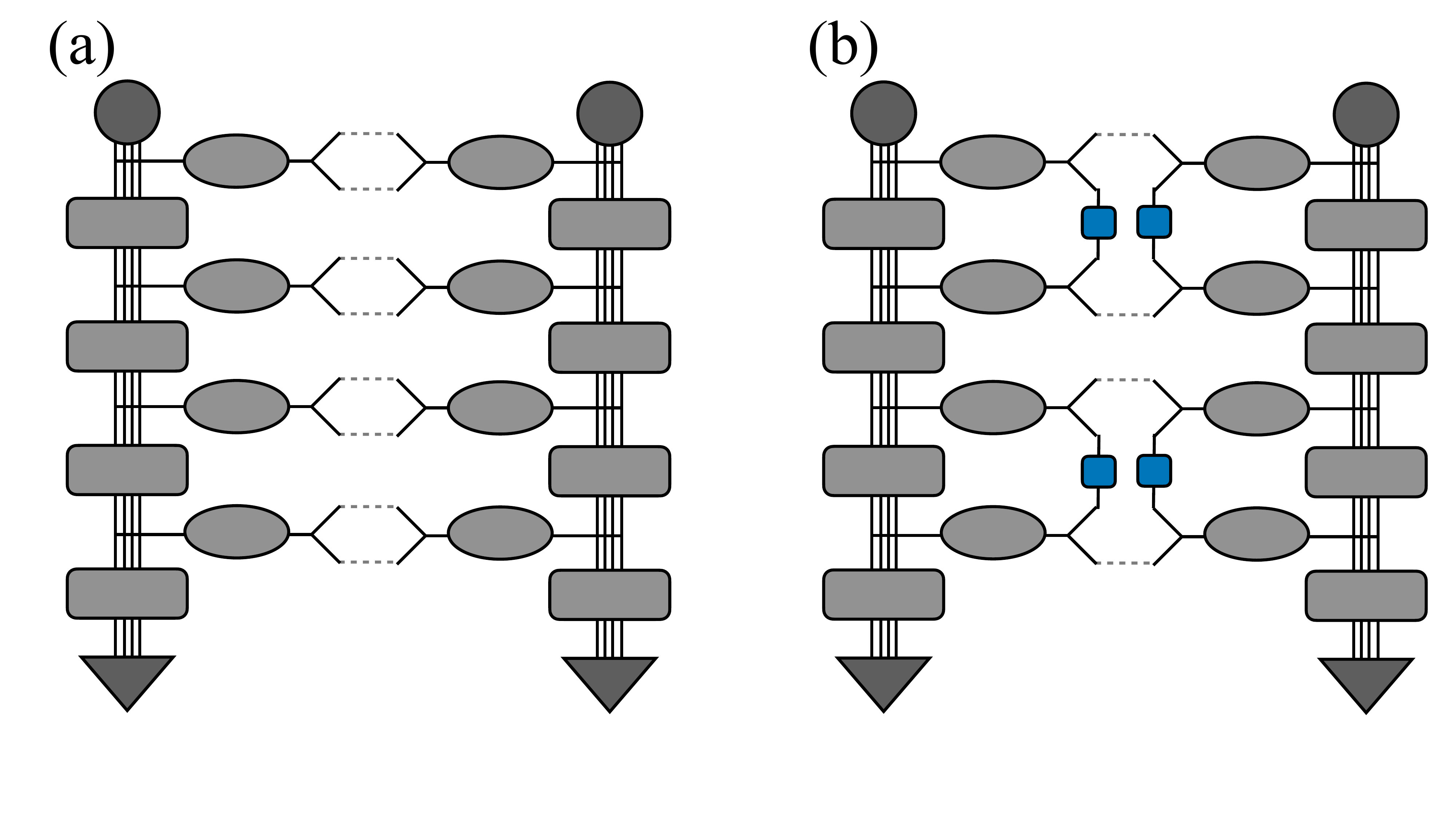}
    \caption{{Comparison between the (a) norm of IM from Eq.~\eqref{eq:tildeI_norm} and the (b) norm of coarse-grained IM from Eq.~\eqref{eq:norm_cg}. The insertion of each vertical contraction (coarse-graining step) reduces the norm by the factor of $2^{-2}$.}}
    \label{fig:IM_norm_cg}
\end{figure}

Let us first estimate the norms of the coarse-grained IM components. Our previous estimate of the magnitude of each matrix element of the IM still applies -- in fact, matrix elements of the coarse-grained IM are a subset of matrix elements of the original IM. However, the total number of matrix elements has decreased from $2^{2T}$ to $2^{2n_{\rm{cg}}T}$. Thus, for the norms, we obtain [cf. Eq.~(\ref{eq:tildeI_norm}), and also a diagrammatic comparison from Fig.~\ref{fig:IM_norm_cg}]:
\begin{gather}\label{eq:norm_cg_PD}
\langle\mathcal{I}^{\rm{cg}}_{T-k,\rm deph}|\mathcal{I}^{\rm{cg}}_{T-k,\rm deph}\rangle =  2^{n_{\rm{cg}}(T-k)} \, ,
\\
\langle\widetilde{\mathcal{I}}_{k}^{cg}|\widetilde{\mathcal{I}}_{k}^{cg}\rangle\approx C_{\rm{cg}} 2^{(2n_{\rm{cg}}-1)k} \, .\label{eq:norm_cg}
\end{gather}

\begin{figure}[t]
    \centering
    \includegraphics[width=0.9\linewidth]{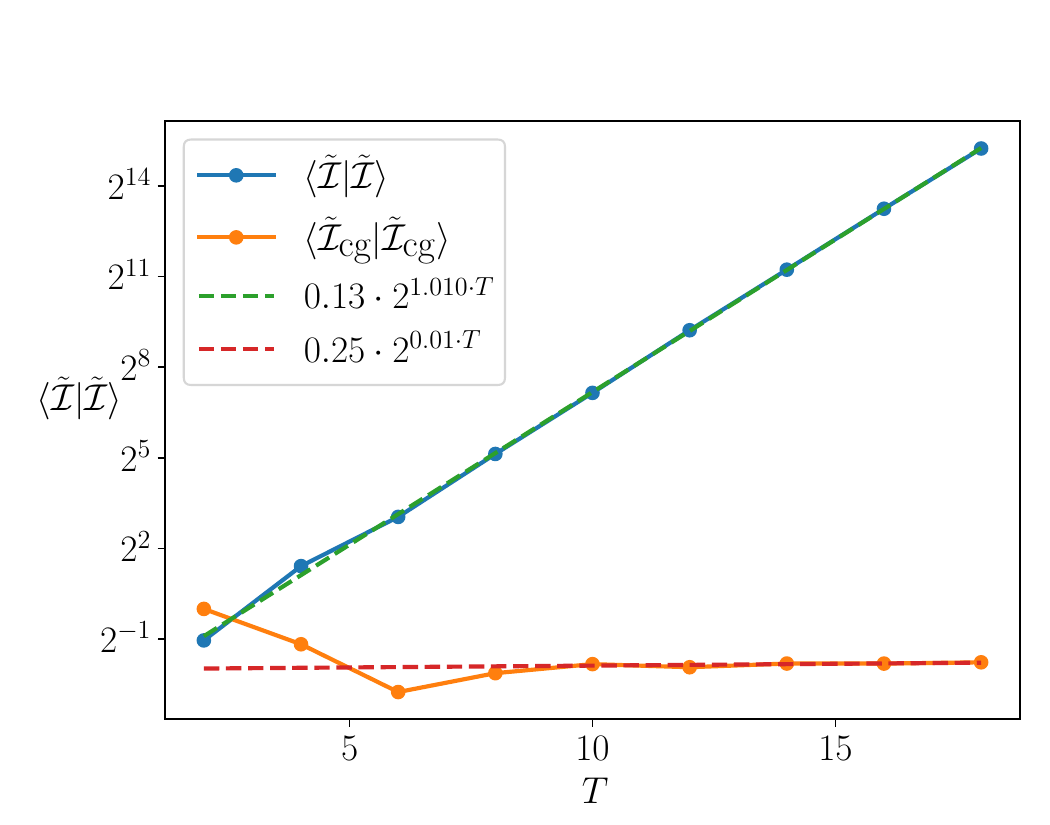}
    \caption{Numerical calculation of the norm of the IM at dual-unitary kicked Ising model with $h=0.5$ and initial tilt angle $\phi=0.7$, before and after coarse-graining with density $n_{\rm{cg}}=1/2$.  The constants $C\simeq 0.13$ and $C_{\rm{cg}}=0.25$ are not universal and depend both on $\phi$ and $h$. 
   }

    \label{fig:IM_norm}
\end{figure}

While Eq.~\eqref{eq:norm_cg_PD} is exact, Eq.~\eqref{eq:norm_cg} is based on the heuristic Loschmidt-echo {estimation of matrix elements from Eq.~(\ref{eq:IM_magnitude})}. To make sure that the norms $\langle\widetilde{\mathcal{I}}_{k}^{cg}|\widetilde{\mathcal{I}}_{k}^{cg}\rangle$ are not affected by possible neglected correlations between IM elements, we additionally verified Eq.~\eqref{eq:norm_cg} numerically, as reported in Fig.~\ref{fig:IM_norm}.
The coarse-grained IM decomposition weights thus become
\begin{gather}\label{eq:w_k_cg}
w^{\rm{cg}}_k=\tilde{C}_{T,cg}2^{-(1-n_{\rm{cg}})k} \, ,\\
w^{\rm{cg}}=1-\sum\limits_{k=n_{\rm{cg}}(p+1)}^{n_{\rm{cg}}T} w^{\rm{cg}}_{k/n_{\rm{cg}}} \, ,\\ \label{eq:w_cg}
\tilde{C}_{T,cg}=\left(C_{cg}\frac{1-2^{-(1-n_{\rm{cg}})T}}{2^{\frac{1-n_{\rm{cg}}}{n_{\rm{cg}}}}-1}+1\right)^{-1} \, .
\end{gather}

Crucially, from Eqs.~\eqref{eq:norm_cg_PD} and~\eqref{eq:norm_cg}, we see that the norm of the perfect dephaser components in the IM vector decomposition are suppressed exponentially \textit{slower} than the orthogonal ``random'' components.
Thus, coarse-graining remodulates the decomposition weights in favor of the perfect dephaser components. Such a remodulation is accompanied by a decrease of TE.

In order to prove the saturation of TE at a finite coarse-graining density, it is sufficient to use the right-hand side inequality from Eq.~\eqref{eq:mixture_bound}. We will separately bound the mixture of entanglement entropies and the classical entropy of the mixture. The former is given by 
\begin{equation}\label{eq:Smix_cg}
S_{\text{mix}}=\sum\limits_{k=n_{\rm{cg}}(p+1)}^{n_{\rm{cg}}T} w^{\rm{cg}}_{k/n_{\rm{cg}}} S_{\text{TE}}(\tilde{\rho}^{\rm{cg}}_{k/n_{\rm{cg}}}).
\end{equation}
Due to the exponential suppression of the weights in Eq.~\eqref{eq:w_k_cg}, it is sufficient to bound each term by the maximal possible entanglement entropy, that is, the number of qubits $2n_{\rm{cg}}p$,
\begin{equation}
S_{\text{TE}}(\tilde{\rho}^{\rm{cg}}_k)<2n_{\rm{cg}} p \, .  
\end{equation}
The sum in Eq.~(\ref{eq:Smix_cg}) then can be bounded as
\begin{equation}
S_{\text{mix}}< p 2^{-p(1-n_{\rm{cg}})}C_{\text{mix}},
\end{equation}
where $C_{\text{mix}}=\frac{2\log(2)n_{\rm{cg}} }{1-2^{-(1-n_{\rm{cg}})/n_{\rm{cg}}}}\tilde{C}_{T,\rm{cg}}$.
Thus, the quantity $S_{\text{mix}}$ is finite for arbitrarily large $p$ and $T$, as soon as $n_{\rm cg}<1$.
In addition, the classical entropy from Eq.~\eqref{eq:entropy_bound} is given by the sum
\begin{equation}
S_{\text{cl}}^{\rm{cg}}=-\sum\limits_{k=n_{\rm{cg}}(p+1)}^{n_{\rm{cg}}T} w^{\rm{cg}}_{k/n_{\rm{cg}}}\log_2 w^{\rm{cg}}_{k/n_{\rm{cg}}}-w^{\rm{cg}}\log w^{\rm{cg}},
\end{equation}
which is also convergent for large $T$, again thanks to the exponential decay of the coefficients $w^{\rm{cg}}_k$. 

Putting everything together, we have shown that $S_{\text{TE}}(\rho_{\mathcal{P}}^{\rm{cg}})$ stays finite for arbitrarily large $T$ as soon as $n_{\rm cg}<1$. This corresponds to the claimed transition to area-law scaling in the presence of an arbitrarily mild coarse-graining.  
This TE transition is illustrated in Fig.~\ref{fig:TEE_DU} for $n_{\rm cg}=1/2$; we further verified this occurrence for other choices of $n_{\rm cg}$ (not shown).



Let us now discuss the consequences of our result for approximating the IM.
Since the kicked Ising model IM is in diagonal form (see Eq.~\ref{eq:Ising_constraint}) and we assumed a diagonal unitary evolution  $V_\tau=\mathbb{1}$ for the probe spin, nontrivial correlation functions must involve off-diagonal observables such as $O=\sigma^x$ or $\sigma^y$.
Two-point autocorrelation functions, for example, can be written as
\begin{equation}\label{eq:correlator_bound}
\langle O(t_2)O(t_1)\rangle = \tfrac12\,\delta_{t_1,t_2}\,\Tr\!\big(O^{2}\big) + R(t_1,t_2),
\end{equation}
where the first term is determined by the perfect depolarizer component of the IM, whereas the remainder, $R(t_1,t_2)$, {is a strongly fluctuating function} arising from the highly entangled component $\widetilde{\mathcal{I}}_{t_2}$.
Noting that two-point functions are obtained by coarse-graining all sites except the two insertions, and using Eq.~\eqref{eq:norm_cg} with $n_{\rm cg}=0$, {we estimate the magnitude of this term as $|R(t_1,t_2)| \lesssim 2^{-t_2/2}$ for $t_2>t_1$, uniformly in $t_1$}. 
Now, let us turn to the MPS approximation that allows us to compute the correlation function up to a fixed error $\epsilon$. We may truncate $R(t_1,t_2)$ by setting it to zero for $t_2 > 2\log_2(1/\epsilon) = t_2^{\rm max}$, and the corresponding IM can then be naturally represented as an MPS with bond dimension $2^{t_2^{\rm max}}$, yielding a polynomial relation between the absolute error and the required MPS bond dimension. Owing to the high entanglement of $\widetilde{\mathcal{I}}_{t_2}$ [cf. Eq.~\eqref{eq:ent_rho_tilde}], we do not expect further compression to be compatible with the allowed error.


\section{Generic Floquet circuits}

\label{sec:generic}

In this Section, we study TE in a more generic class of non-integrable Floquet circuits: the kicked Ising model away from dual-unitary points, which is {considered a paradigmatic model of quantum chaos \cite{kim2014testing, banuls2009Matrix}.} It is known, however, to provide an example of non-trivial thermalization dynamics of local observables. Specifically, its dynamics in an open chain (or in a semi-infinite chain, as considered here) exhibits four distinct regimes, which stem from the existence of prethermal $0$- and $\pi$-edge modes~\cite{FendleyPRXPreth, Mitra19, XiaoMiScience2022}. Such modes are approximate integrals of motion localized at the boundary, and give rise to parametrically slower, compared to a random or dual-unitary circuit, relaxation dynamics of a probe system coupled at the edge. 

Below we study TE in three regimes: (I) a `plain vanilla' regime without edge modes, (II) a regime with an edge $\pi$-mode, and (III) a regime with an edge $0$-mode. [We do not consider the regime where $0$- and $\pi$-modes coexist, since it shows a similar phenomenology to (II).] We compute TE by numerically constructing the IM via the LCGA. Consistent with previous work and our analysis above, we generally find volume-law scaling of TE in all three regimes. As expected, upon coarse-graining, TE gets significantly reduced. For all parameter values we consider, TE 
transitions to a {subvolume-}law scaling below a finite critical coarse-graining threshold, consistent with the scenario outlined above.

{The coarse-graining transition further suggests that few-point correlation functions should be captured by a compact MPS approximation to the influence matrix, consistent with the subvolume growth of TE. Motivated by this observation, we study the dynamics of local observables and quantify the errors introduced by replacing the full influence matrix—whose exact bond dimension grows exponentially with the evolution time—by a lower-bond-dimension approximation obtained through standard singular-value truncation.}

\subsection{TE across dynamical regimes}
\begin{figure}[t]
\centering
\includegraphics[width=0.9\columnwidth]{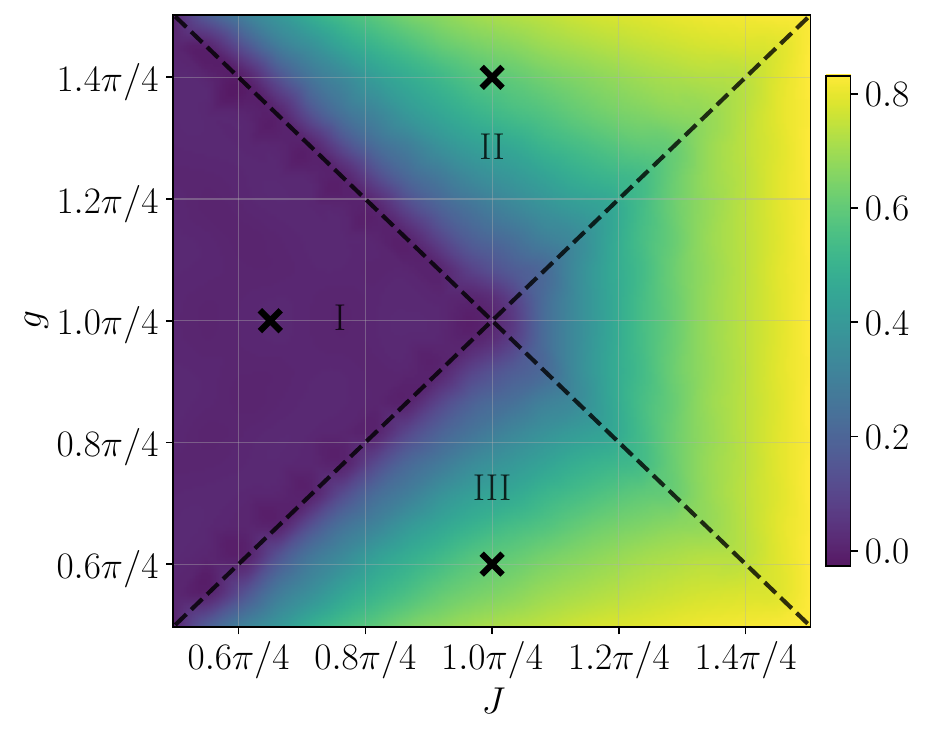}
\caption{Phase diagram of the integrable kicked Ising model ($h=0$). Brighter color indicates a higher saturated value of the autocorrelator $\lim_{T\to\infty}\left\lvert\langle \sigma^z_0(T)\sigma^z_0(0)\rangle\right\rvert$, reflecting a more localized edge mode. The three crosses correspond to representative values of $J$ and $g$ chosen for numerical simulations.}
\label{fig:Phase_diagram}
\end{figure}

\begin{figure*}[t]
    \centering
\includegraphics[width=.3\linewidth]{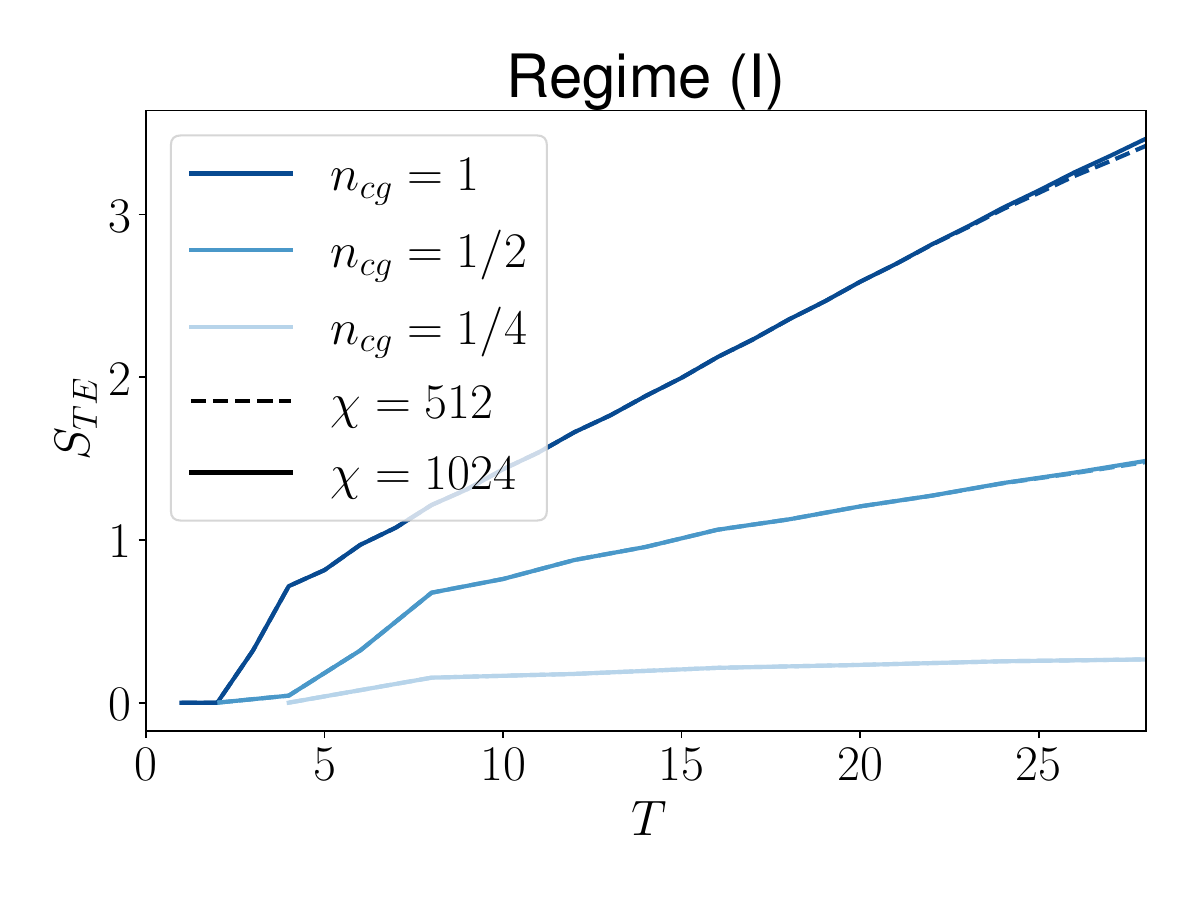}
\includegraphics[width=.3\linewidth]{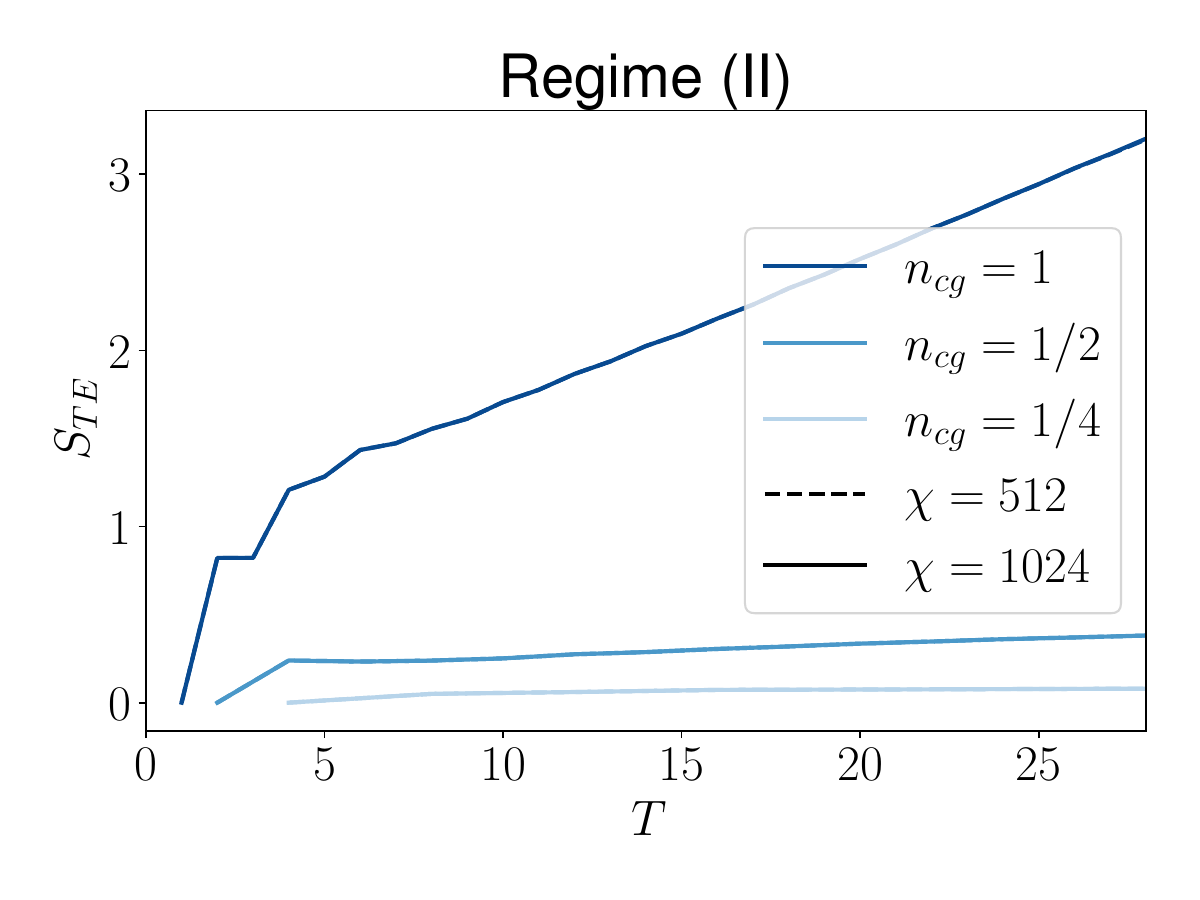}
\includegraphics[width=.3\linewidth]{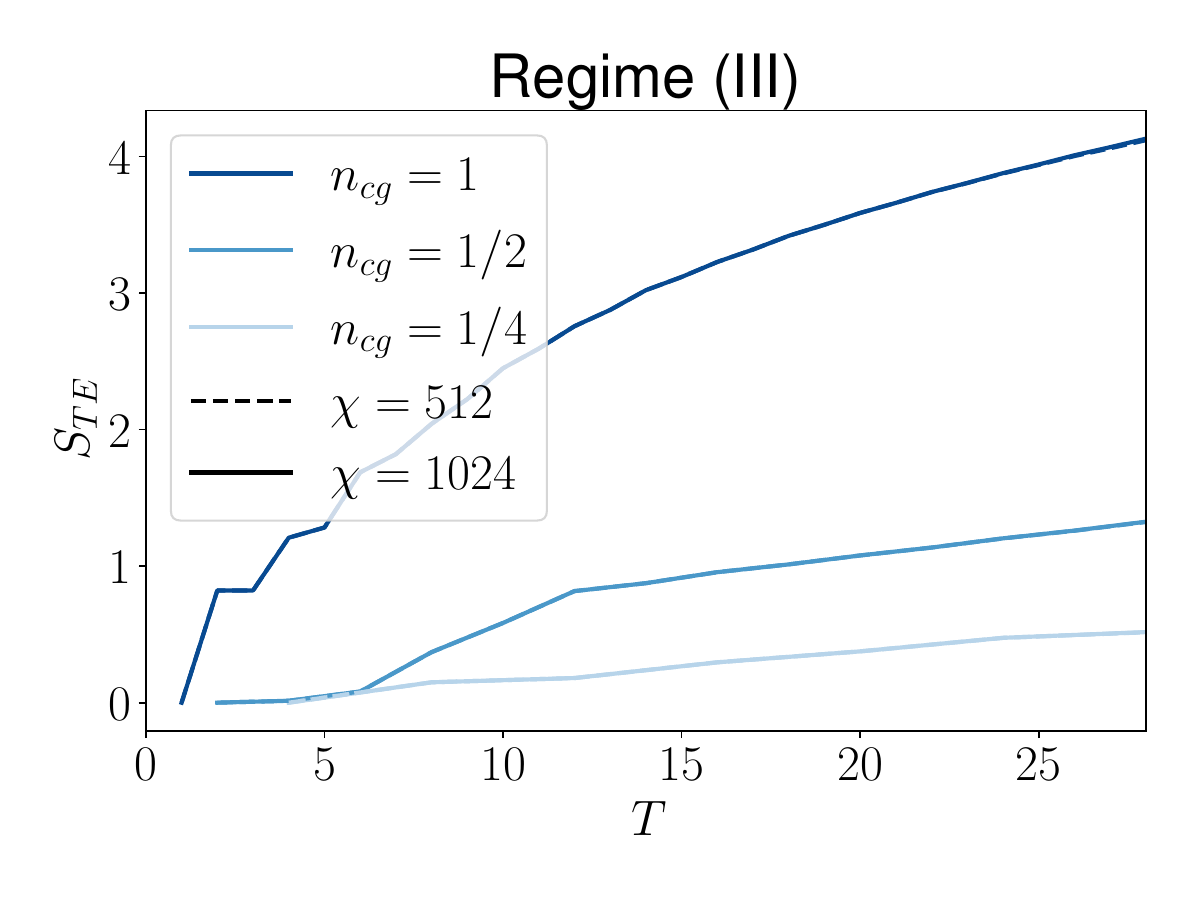}
\includegraphics[width=.3\linewidth]{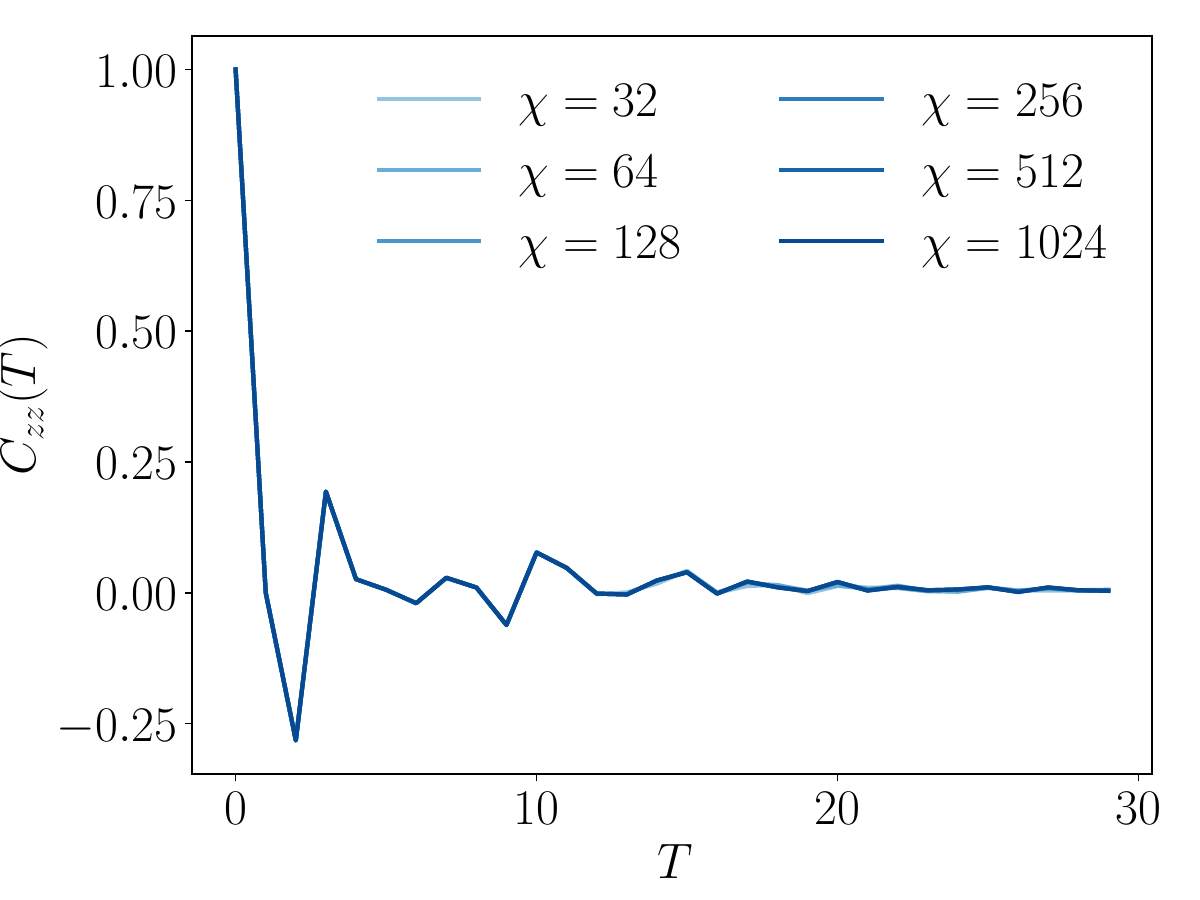}
\includegraphics[width=.3\linewidth]{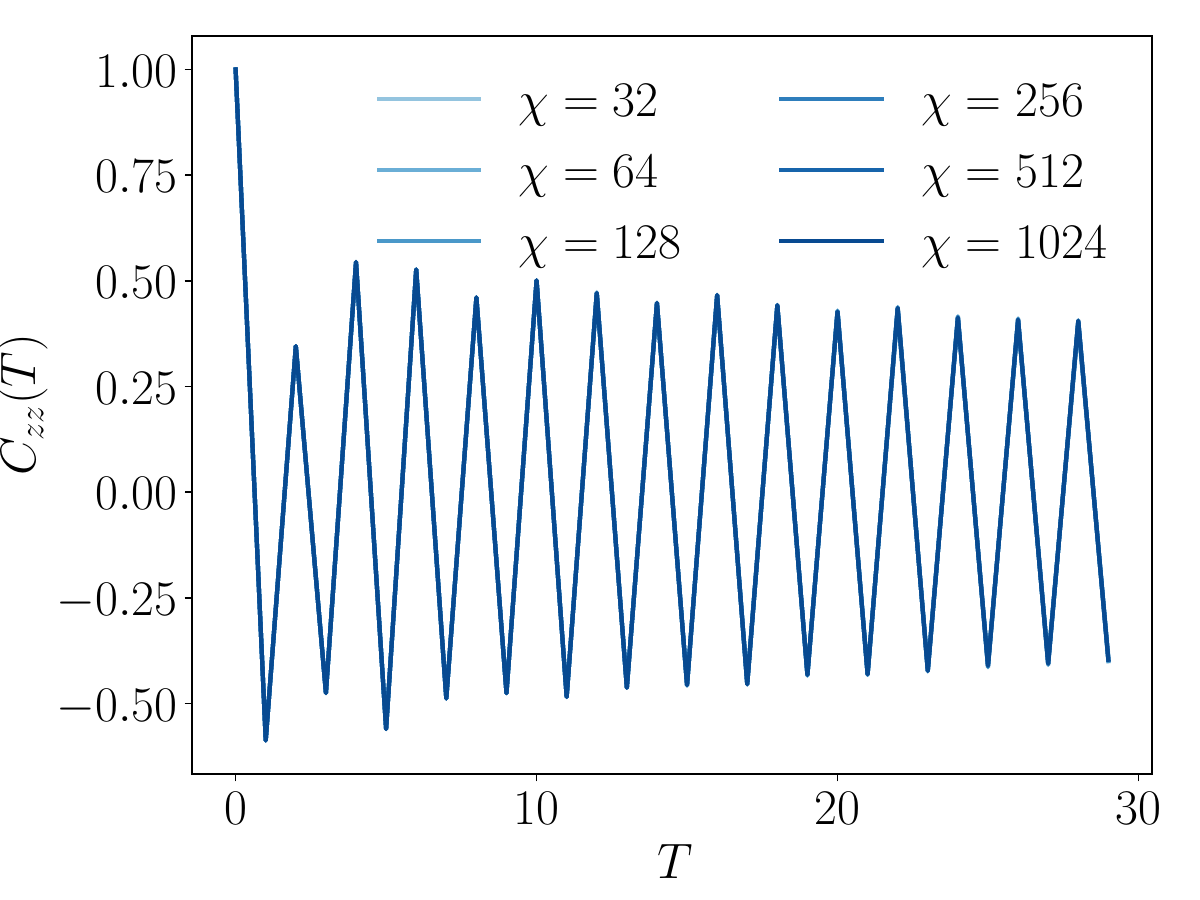}
\includegraphics[width=.3\linewidth]{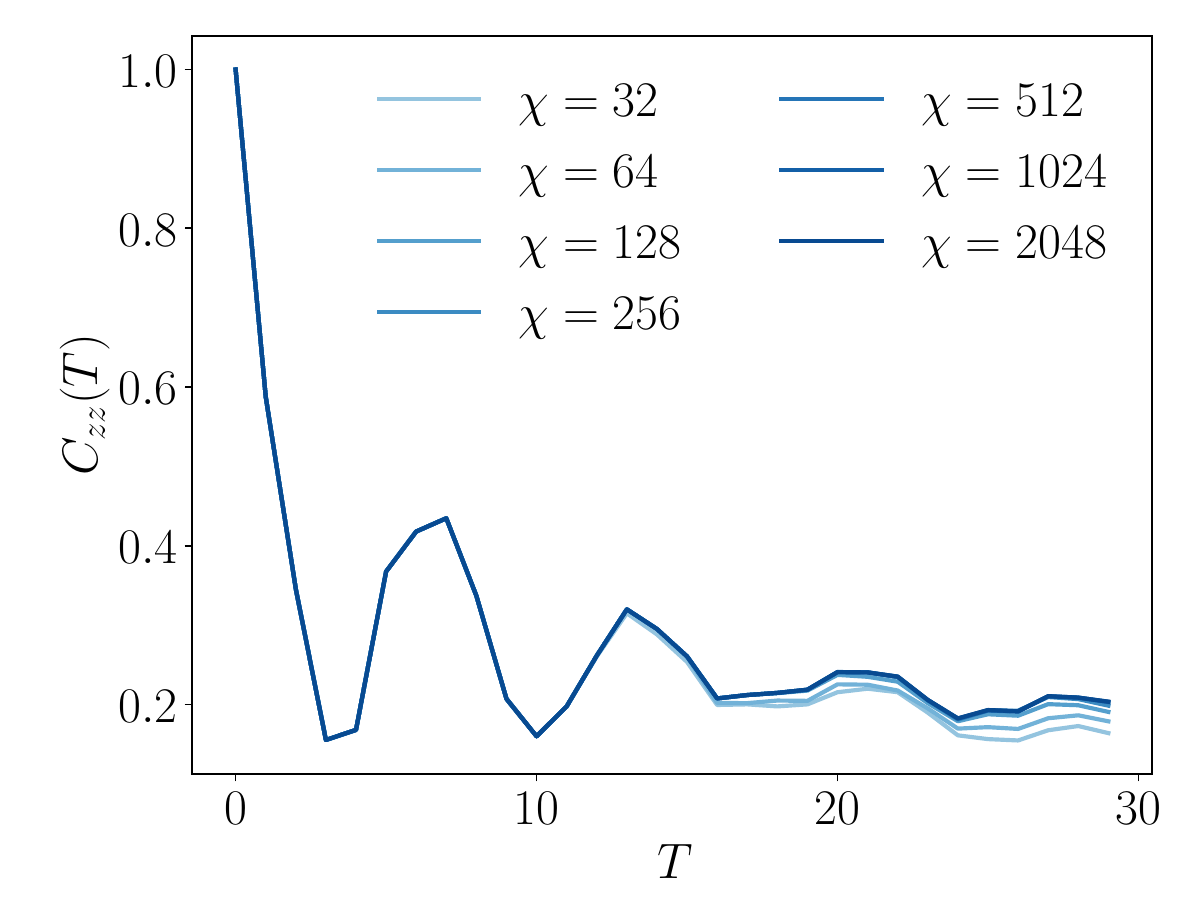}
\caption{\emph{Top panels:} Temporal entanglement entropy, computed using the LCGA, in different dynamical regimes of the non-integrable KIM: (I) no edge modes, (II) edge $\pi$-mode, and (III) edge $0$-mode.  \emph{Bottom panels:} The local temporal correlator $C_{zz}(T)$ in the three regimes shows qualitatively different behaviors. While it quickly decays to zero in the ``plain vanilla'' thermalization regime (I), the decay is slow in the other two regimes due to the overlap of the $\sigma^z$ operator with the edge mode operator.
In the main text we explain the observed quantitatively different behavior of coarse-grained TE in (II) and (III), and compare the convergence of observables with increasing $\chi$.
}\label{fig:TEE_KIM}
\end{figure*}

We consider a fully mixed initial state of the kicked Ising model bath. In the integrable limit $h=0$ the kicked Ising model can be mapped onto a kicked Kitaev (free Majorana) chain~\cite{Kitaev_2001}, whose edge modes are exact integrals of motion. The phase diagram of the non-interacting kicked Ising model~\cite{DuttaPRB13,lerose2021Scaling} is reported in Fig.~\ref{fig:Phase_diagram}. On breaking integrability, $h\neq 0$, the edge modes acquire a finite lifetime, which varies parametrically from infinity on approaching the top and bottom sides of the phase diagram to zero at the central point $g=J=\pi/4$ (dual-unitary point), qualifying the edge modes as prethermal~\cite{FendleyPRXPreth}. Notably, for sizable fields $h$, the parameter region where prethermal edge modes are long-lived may differ somewhat compared to the non-interacting phase boundaries. 
We distinguish three distinct regimes, which are deformations of the regimes (I), (II) and (III) from Fig.~\ref{fig:Phase_diagram}, respectively. 
We choose the following representative parameter values in the three regimes: 
\begin{gather}\label{eq:parameters_plain_vanilla}
{\rm (I)}\, (g=\pi/4, J=0.65\pi/4, h=0.5), \quad v_B=0.61\pm 0.05\, , \\ 
{\rm (II)} \, (g=1.4\pi/4, J=\pi/4, h=0.5), \quad v_B=0.62\pm 0.01 \, , \\
{\rm (III)} \, (g=0.6\pi/4, J=\pi/4, h=0.5), \quad v_B=0.62 \pm 0.02\, .  
\end{gather}
These three choices of $J$ and $g$ are marked by crosses in Fig.~\ref{fig:Phase_diagram}. 
For each of them, we numerically construct the IM via the LCGA up to times where TE is converged using a maximum bond dimension $\chi=1024$. The three points have been chosen to have roughly equal values of the butterfly velocity $v_B$, which we computed independently by fitting the spatiotemporal spreading of the out-of-time-ordered correlator~\cite{Bohrdt_2017,luitz2017information,nahum2018operator} 
\begin{equation}\label{eq:C_AB}
\mathcal{C}_{AB}(r,t) = \tfrac{1}{2}\,\mathrm{tr}\!\left( [A_0(t),B_r(0)]^{\dagger}[A_0(t),B_r(0)] \right)
\end{equation}
averaged over Pauli operators $A,B \in \{\sigma^x,\sigma^y,\sigma^z\}$ using exact diagonalization for a chain of length $L=12$ as explained in Appendix~\ref{app:Butterfly_velocity}. 

The behavior of TE and the effect of coarse-graining is reported in the top panels of Fig.~\ref{fig:TEE_KIM}, and it follows the intuition developed above. In regime (I) observables decay to zero over a short timescale, similarly to random and dual-unitary circuits.
Here, however, coarse-graining with density $n_{\rm cg}=1/2$ reduces TE to a less steep but still volume-law scaling. A further round of coarse-graining, $n_{\rm cg}=1/4$, finally yields a TE scaling consistent with a {subvolume-}law. This indicates that $1/4<n_{\rm cg}^\star<1/2$ for this parameter choice. The smaller threshold for the TE transition is expected given the reduced butterfly velocity $v_B\approx 0.61 \pm 0.05$ of this model compared to the dual-unitary value $v_B=1$. The relation between the critical coarse-graining and $v_B$ was examined in the toy model of a growing random unitary bath, which yielded Eq.~(\ref{eq:n_CG}); see also Appendix \ref{app:growing_bath}. In a more structured bath model, such as a local unitary circuit, the simple dependence between $v_B$ and the critical coarse-graining in Eq.~(\ref{eq:n_CG}) may give room to a more complicated relation; nevertheless, the basic picture of a competition between bath and temporal degrees of freedom remains valid and predicts a transition at a different $n_{\rm cg}^\star$ \textit{at least} as small as $v_B$, in agreement with our numerical results. 
As discussed in Appendix~\ref{app:TE_plain_vanilla}, tuning the model parameters closer to a dual-unitary point,  $J=0.8{\pi}/4$ with $g$ and $ h$ fixed, where the butterfly velocity $v_B=0.95\pm 0.02$ is larger, we observe a clear collapse of TE to {subvolume-}law scaling after the first round of coarse-graining, which implies  $1/2<n_{\rm cg}^\star<1$ for this parameter choice.

\begin{figure*}[t]
  \centering
  \subfloat[]{\includegraphics[width=0.45\linewidth]{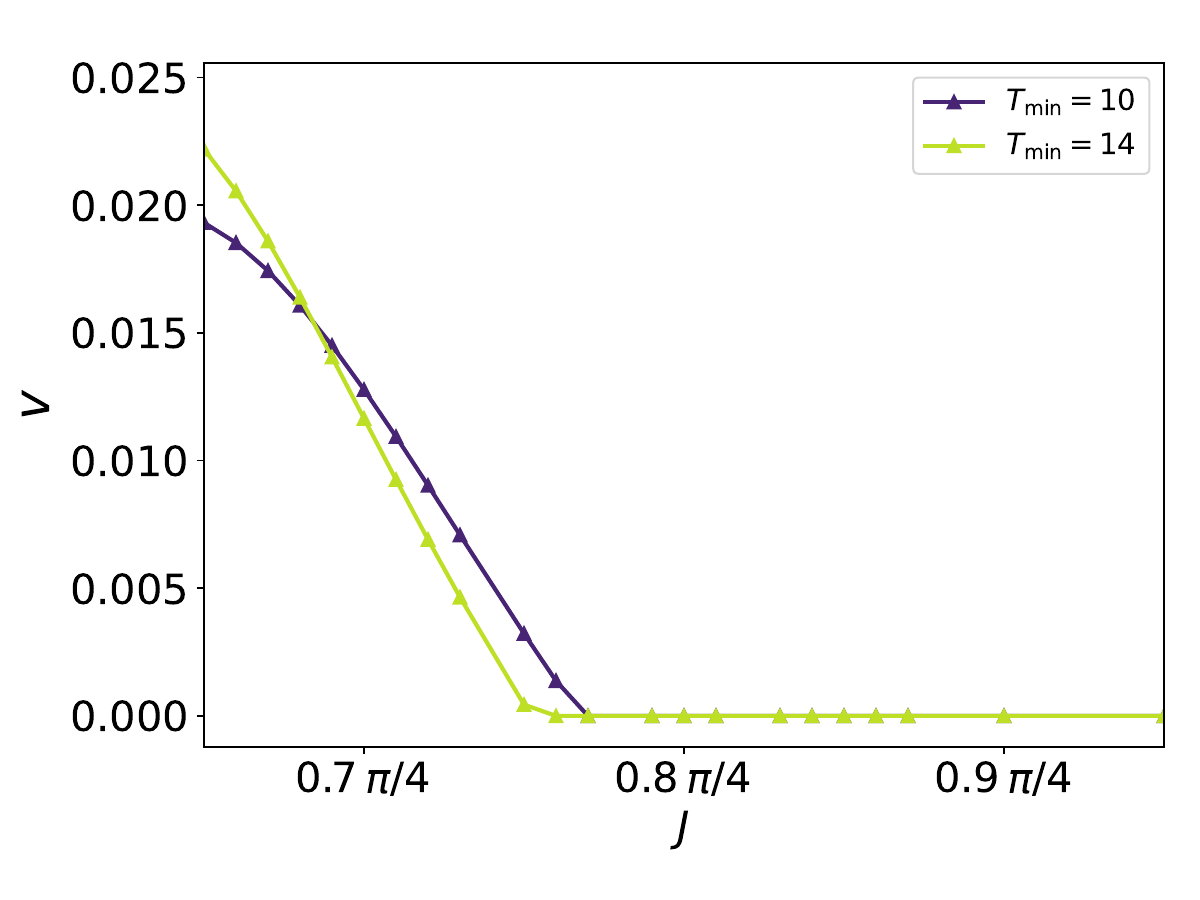}}%
  \hfill
  \subfloat[]{\includegraphics[width=0.45\linewidth]{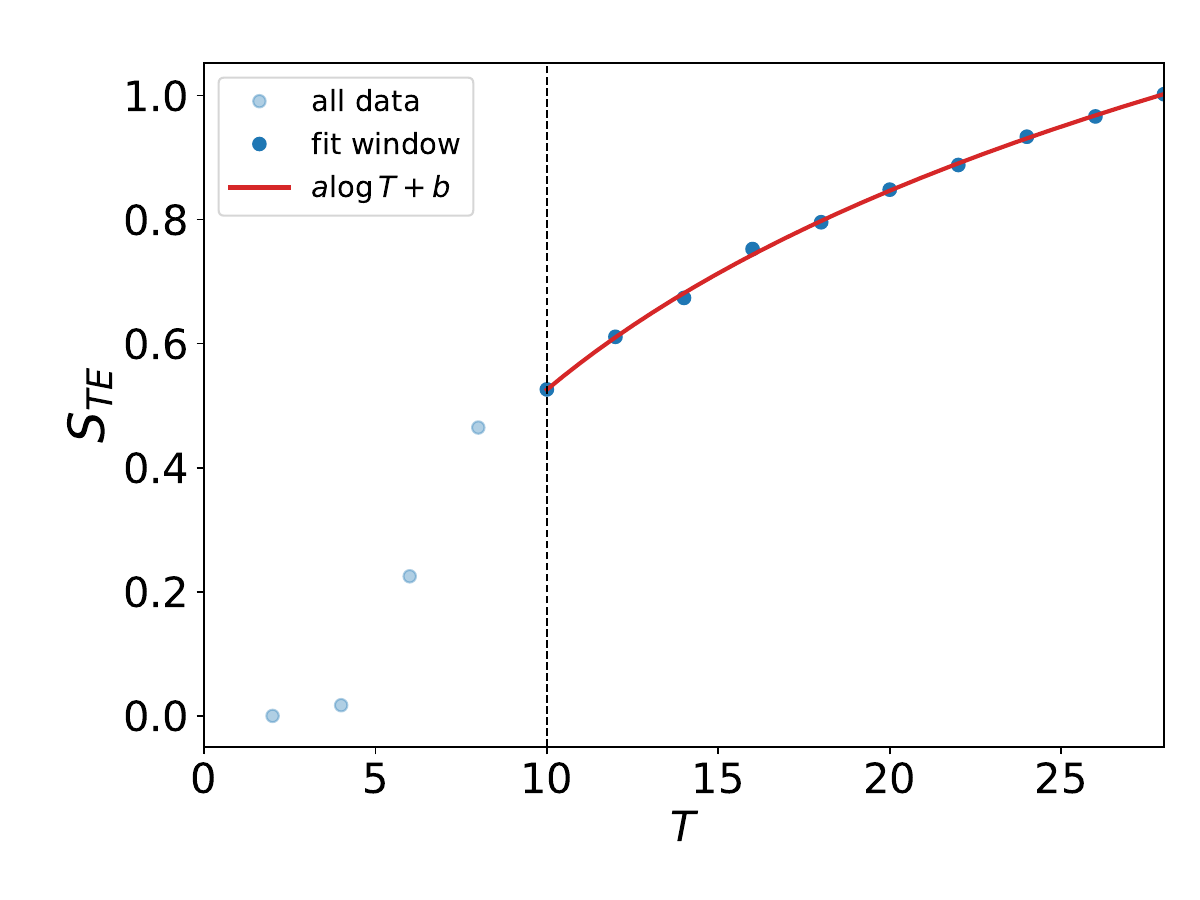}}
  
\caption{(a) The linear fit coefficient as a function of $J$ abruptly vanishes at $J^\star_{1/2} \simeq 0.77 \frac{\pi}{4}$. The slope extraction was made for $T$ in range $[T_{\textrm{min}}=\{10,14\},T_{\textrm{max}}=28]$, (b) log fit for a critical $J^\star_{1/2}$ with $a \simeq 0.37, b \simeq -0.43$}.
\label{fig:transition_point}
\end{figure*}

{
A more direct way to establish 
the existence of a transition is to vary $n_{\textrm{cg}}$ continuously while keeping the model parameters $g$, $h$, and $J$ fixed. For each value of $n_{\textrm{cg}}$, one can extract the linear coefficient $v$ by fitting the TE to
\begin{equation}
\label{eq:fit}
S_{\mathrm{TE}}=vT+a\log T+b.
\end{equation}
This ansatz is natural from the membrane perspective \cite{zhou2020entanglement}: the linear term is associated with the membrane tension, while fluctuations of the membrane generate a subleading logarithmic correction. Its application to TE is, however, more subtle. As shown in Ref.~\cite{FolignoBertiniPRX2023} volume-law scaling can be established analytically at the dual-unitary point, whereas away from this point the membrane interpretation is less direct. Nevertheless, our numerical results indicate that the temporal entanglement is well described by Eq.~\eqref{eq:fit}.

Within this fitting procedure, the critical coarse-graining density $n_{\textrm{cg}}^\star$ can then be identified as the value beyond which the fitted slope $v$ vanishes. In practice, our ability to vary $n_{\textrm{cg}}$ continously at a fixed evolution time $T$ is limited due to the constraint that $n_{\textrm{cg}}T$-is an integer number of time-steps. It is more reliable to fix $n_{\textrm{cg}}=1/2$ and study $v$ as a function of $J$. This allows us to determine the critical coupling $J^\star_{1/2}$ at which a coarse-graining density of $n_{\textrm{cg}}=1/2$ becomes sufficient to drive the IM from volume-law scaling into a subvolume-law, logarithmic regime. The corresponding results are shown on Fig.~\ref{fig:transition_point} and show a clear transition at $J^\star_{1/2}\simeq 0.77,\frac{\pi}{4}$ .}

Regimes (II) and (III) exhibit qualitatively similar phenomenology, but with striking quantitative differences, as shown in Fig.~\ref{fig:TEE_KIM}. In particular, in regime (II), even coarse-graining at $n_{\textrm{cg}}=1/2$ is consistent with a subvolume-law scaling of the TE, whereas regime (III) exhibits a pronounced resilience of large TE. As we briefly discuss below, these differences can be qualitatively attributed to prethermal phenomena.

Next, we use the IM to compute the dynamics of a probe spin coupled to a half-chain kicked Ising bath. {The dynamics is fully specified only once the boundary evolution of the probe spin is fixed. We choose the probe-spin kick to match the bulk kicks, setting $g_0=g$, so that $V_{\tau}=e^{ig_0\sigma_x}$. This choice differs from the trivial boundary evolution, $V_{\tau}=\mathbbm{1}$, used in defining the coarse-graining procedure. Although the same boundary protocol could in principle be used in both settings, we keep them distinct in order to probe more generic boundary dynamics. Nevertheless, as shown below, the resulting correlation functions exhibit behavior consistent with the temporal-entanglement evolution obtained under coarse graining.
} Namely, the autocorrelator 
\begin{equation}
C_{zz}(T) =  \langle  \sigma^z_0(T)\sigma^z_0(0) \rangle 
\end{equation} 
 behaves differently in the three regimes, see the bottom panels of Fig.~\ref{fig:TEE_KIM}. 
We note that in regime (I), $C_{zz}(T)$ exhibits irregular fluctuations around zero with exponentially decaying magnitude for $T>0$, similar to the case of random and dual-unitary circuits. In contrast, $C_{zz}(T)$ is long-lived in regimes (II), (III), as can be expected from the sensitivity of the edge spin operator $\sigma^z_0$  to the prethermal edge mode~\cite{FendleyPRXPreth}; we note that such slow decay for the edge $\pi$-mode was observed experimentally on a superconducting quantum processor~\cite{XiaoMiScience2022}. {The $C_{xx}$ and $C_{yy}$ correlators behaves similarly; we report the numerical results for $C_{xx}$ in Appendix~\ref{app:C_xx}.}

The similarities in edge dynamics in regimes (II) and (III) (up to the presence or absence of the staggered sign-changing behavior) are not representative, however, of the markedly different bulk dynamics in the two regimes. This contrast, discussed below, leads us to a natural conjecture on the origin of the different behavior of TE under coarse-graining in the two regimes despite the similar $v_B$.

In regime (III) 
the Ising coupling $J$ is larger than the transverse field $g$. Together with a symmetry-breaking longitudinal field this gives rise to domain-wall confinement~\cite{Kormos:2017aa, RobinsonNonthermalStatesShort,LeroseSuraceQuasilocalization} -- a phenomenon that has been studied in detail in the Hamiltonian version of the model, but has a very similar phenomenology in the kicked Ising model~\cite{collura2022discrete}. In contrast with a symmetry-preserving integrability-breaking interaction, which conserves a prethermal edge mode but otherwise coexists with fast bulk dynamics, a symmetry-breaking confining field gives rise to slow dynamics uniformly across the system -- a phenomenon that can be linked to an emergent strong Hilbert-space fragmentation in the limit of small transverse field~\cite{LeroseSuraceQuasilocalization,yang2020hilbert}.
It was shown that, throughout this regime, dynamics from certain highly excited states gives rise to long-lived oscillations of local observables, suppressed entanglement growth, and transient many-body localization of domain walls~\cite{LeroseSuraceQuasilocalization}. {Recalling the discussion in Section~\ref{Sec:TE_structureless_bath}, we note that, for the toy model, the critical coarse-graining density is proportional to the butterfly velocity. In a fragmented or quasi-fragmented Hilbert space, however, the dynamics may involve a hierarchy of approximately isolated subspaces, each characterized by its own effective growth rate, $D_k\sim d^{2v_k t}$. Rather than producing a single sharply defined coarse-graining threshold, this hierarchy may lead to a broadened or multi-stage crossover, as different subspaces become progressively disentangled when $n_{\mathrm{cg}}$ becomes comparable to their respective effective rates $v_k$. Slowly thermalizing subspaces with small $v_k$, such as those associated with dynamics confined to a fixed domain-wall sector, may therefore keep being entangled at comparatively low coarse-graining densities.
} 
{This results in a longer thermalization scale, in particular linear fit with logarithmic correction from Eq.~\eqref{eq:fit} suggests the domination of logarithmic contribution, signaling that the real thermodynamic behavior of TE might be observable at a much longer times only.} This explains why even $n_{\rm cg}=1/4$ does not eliminate the growth of
TE, and why the convergence of observables with bond-dimension is slower in this regime. {One intriguing aspect of our study is the unexpected robustness of prethermal physics in the kicked Ising model, even in a regime of strong integrability breaking. This observation motivates the investigation of less constrained models, such as spin-1 Ising models or brickwork quantum circuits, which we hope to pursue in future work.}

In contrast, in regime (II) the kick can be viewed as a perturbation of the $\pi$-kick (realized for $g=\pi/2$), which perfectly flips all spins. Studies of prethermal discrete time crystals in the kicked Ising model~\cite{VonKeyserlingkPRB16_AbsoluteStability,ElsePRX17_PrethermalDTC,khemani2019brief,collura2022discrete,mi2022time} have clarified that symmetry-breaking interactions such as the longitudinal field get ``echoed out'' at all orders in perturbation theory, effectively restoring the symmetry. Therefore, domain-wall confinement is fully suppressed dynamically, while integrability breaking persists in the form of effective symmetry-preserving interactions, giving rise to a long-lived prethermal edge mode coexisting with fast bulk dynamics. This argument explains why the resilience of high TE in regime (III) does not occur in regime (II), while the prethermal edge mode continues to affect edge temporal correlations over long timescales.

\subsection{Local observables and IM compression}

Finally, we turn to the important question regarding the possibility of using a compressed IM to approximate few-point local temporal correlators for general choices of probe-system dynamics $V_\tau$. For concreteness, we again choose probe-spin kicks of equal strength to those in the bath, i.e., $g_0=g$, and we will focus on the practical issue of determining the IM bond dimension necessary to approximate local autocorrelators of the kicked Ising model up to time $T$ with either a given absolute or relative error. We quantify the latter via
\begin{equation}
    \delta_{\chi}(T)=\min \left( \left| \frac{ C_{zz}^{\chi}(T) -C_{zz}^{\chi_{\rm max}}(T)}{C_{zz}^{\chi_{\rm max}}(T)} \right|, 1 \right), 
\end{equation}
which is truncated to a maximum value 1.

We first observe that the IM of the kicked Ising model bath with fully mixed initial state can be represented exactly by an MPS with bond dimension 
\begin{equation}\label{eq:chi_exact}
\chi_{\rm exact}=2^{T/2}.
\end{equation}
We further expect that for $v_B<1$, the exponential growth rate decreases accordingly, $\chi_{\rm exact}\approx 2^{v_BT/2}$.

\begin{figure*}[t]
    \centering
\includegraphics[width=.3\linewidth]{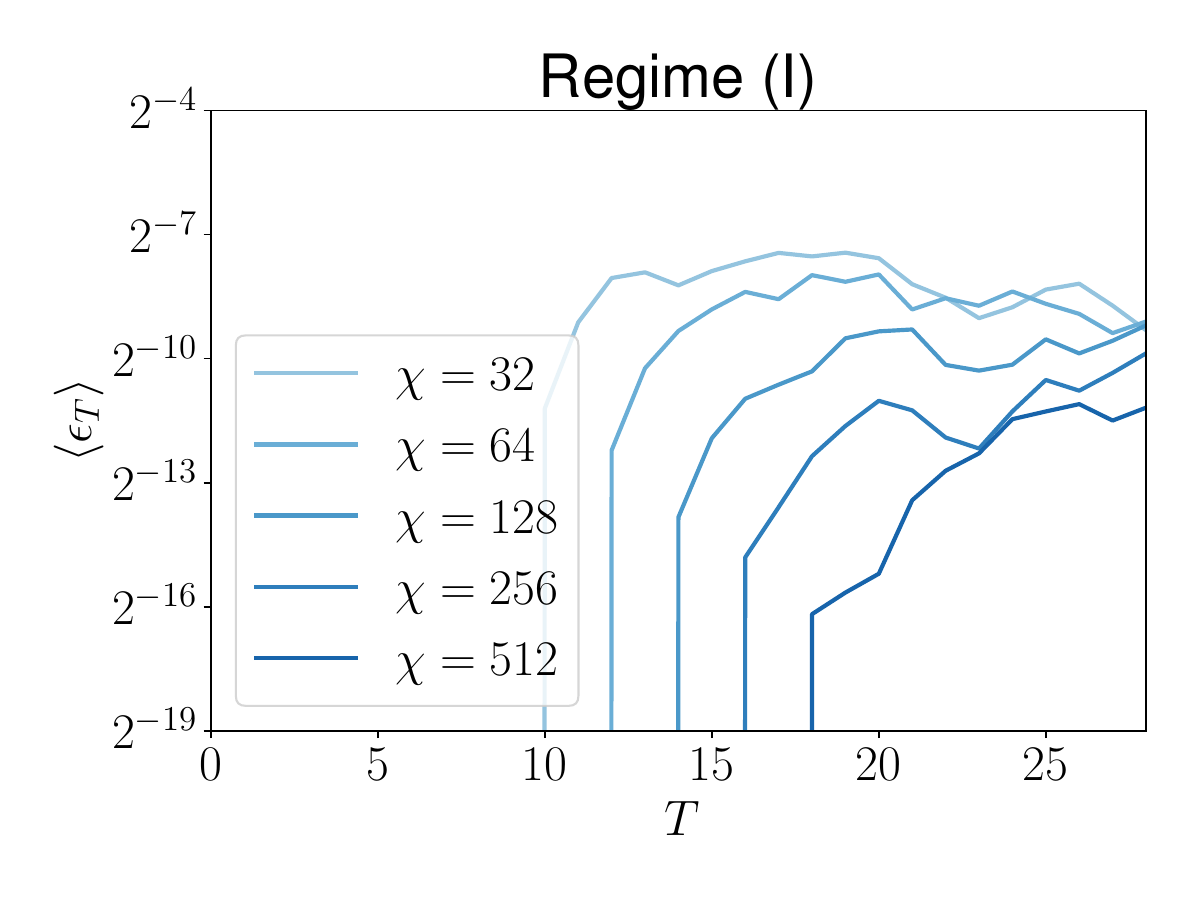}
\includegraphics[width=.3\linewidth]{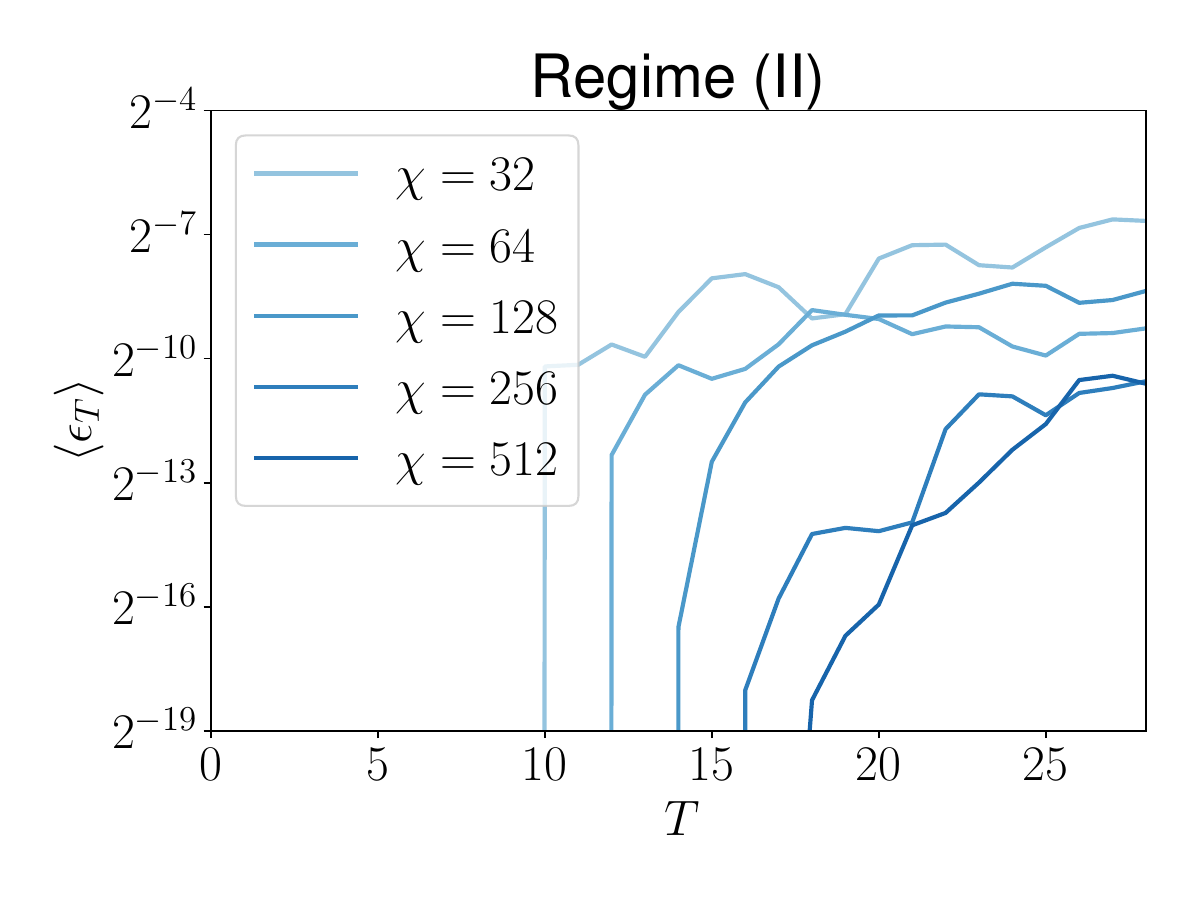}
\includegraphics[width=.3\linewidth]{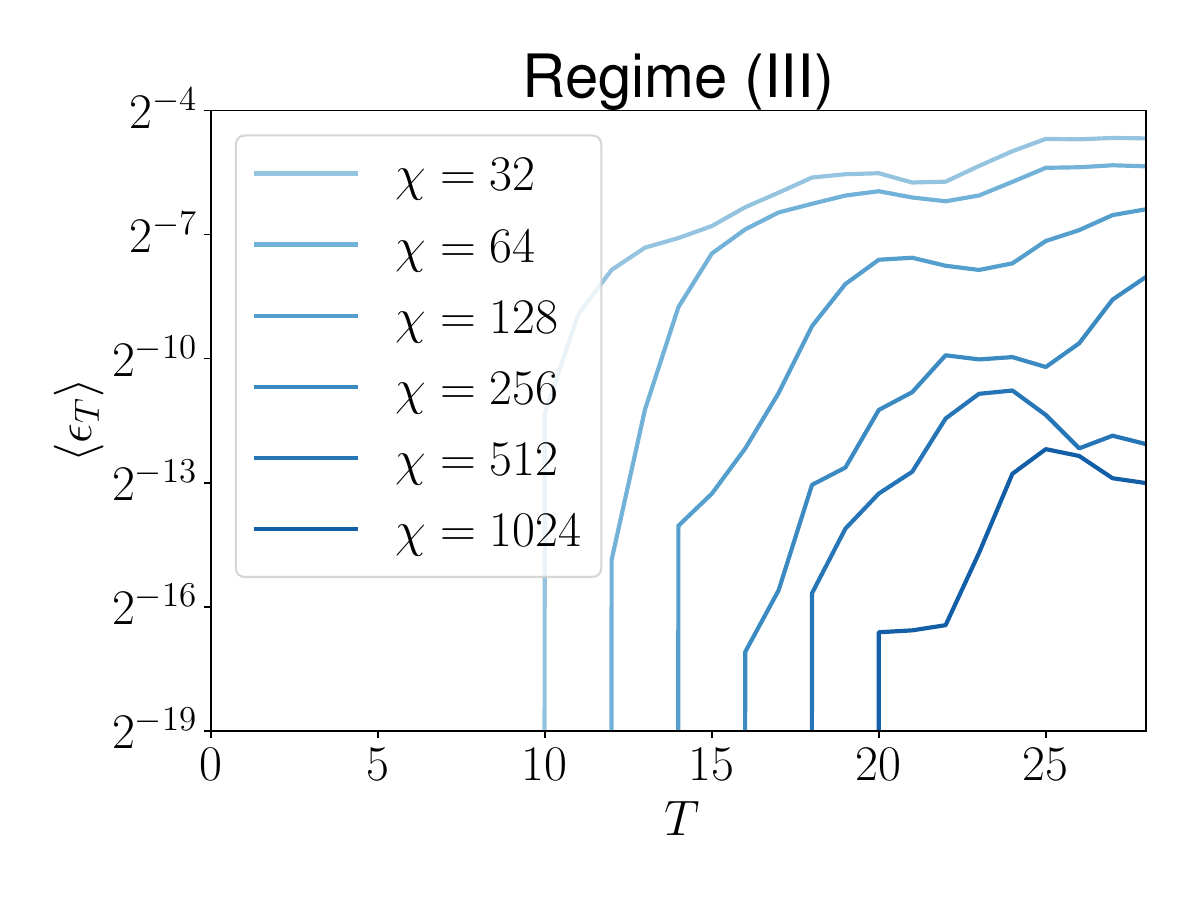}
\includegraphics[width=.3\linewidth]{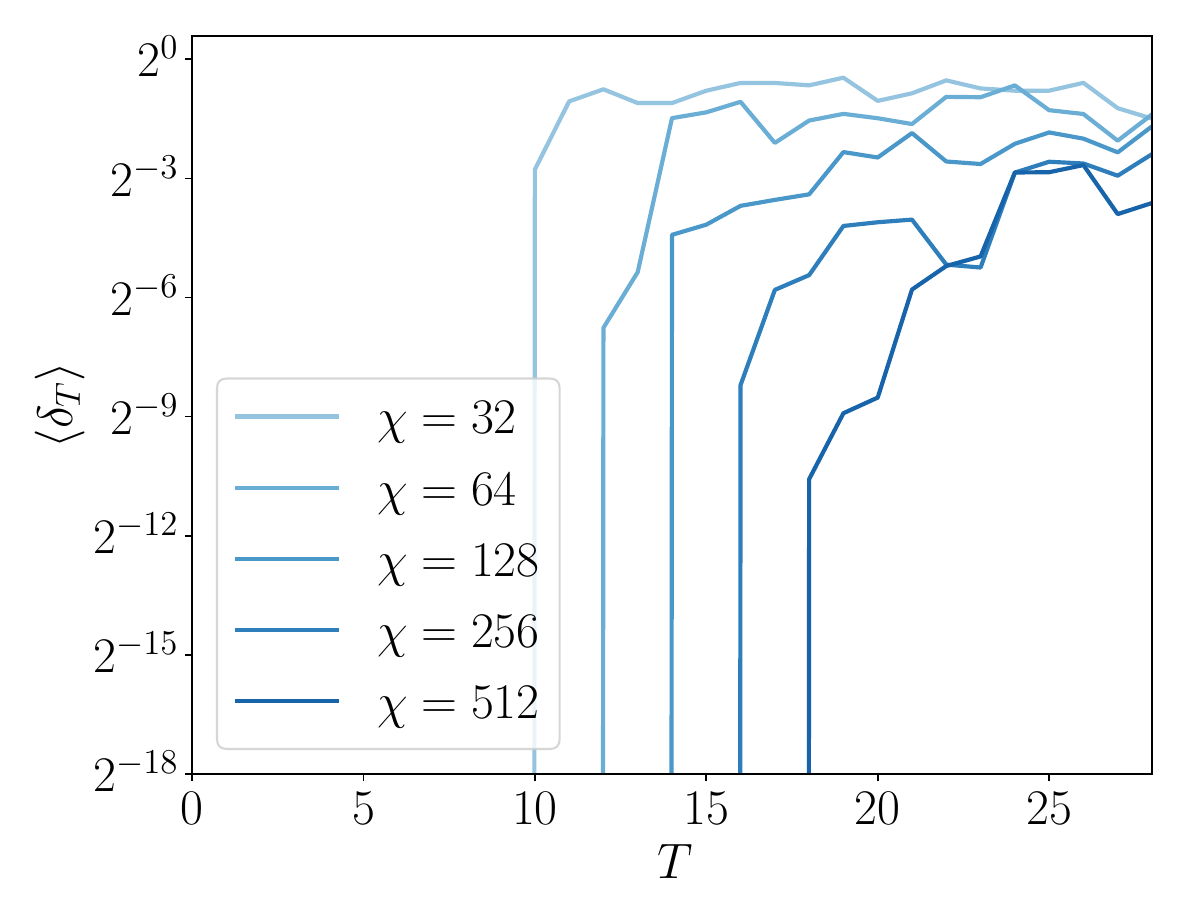}
\includegraphics[width=.3\linewidth]{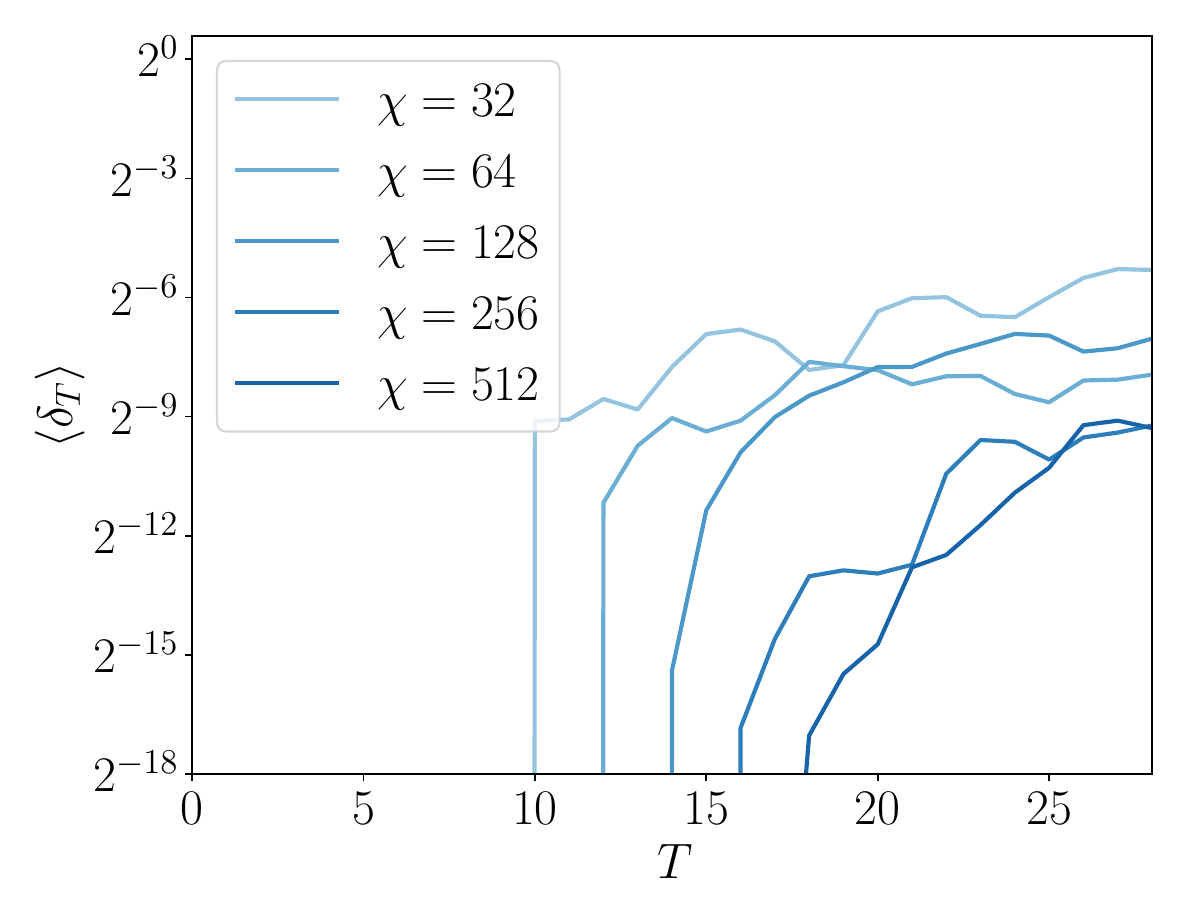}
\includegraphics[width=.3\linewidth]{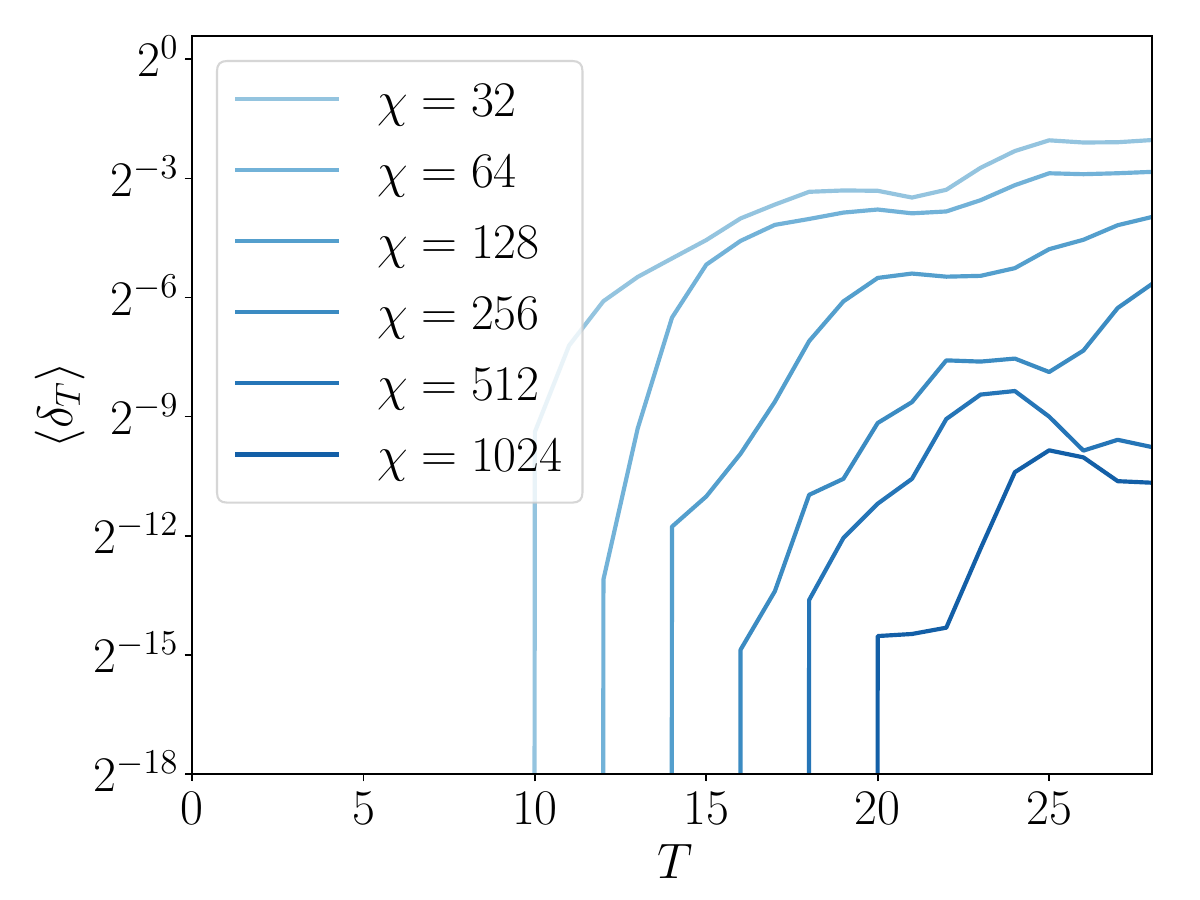}
\caption{Comparison of absolute (\textit{top panels}) and relative (\textit{bottom panels}) errors for different bond dimensions, with $\chi_{\max}=1024$ in three regimes (I), (II), (III) from left to right. 
In both figures, to make the graphs smoother and easier to interpret, we compute the moving average of the instantaneous error  with the next two values.}\label{fig:ZZ_error}
\end{figure*}

Our analysis so far indicates that the volume-law TE scaling reflects highly complex temporal correlations corresponding to a high rate of interrogations of the probe system, and it has very little bearing on few-point temporal correlations. Coarse-graining provides a way to project out the IM components responsible for the complex correlations, giving rise to a new, {subvolume-}law IM. It is natural to expect that the ``nonlocal'' IM components are associated with the small Schmidt values of the IM vector. Then, we expect that truncation to the largest $\chi$ Schmidt values within LCGA performs an effective coarse-graining procedure, yielding a compressed IM $\mathcal{I}_{\chi}$ 
that retains accuracy on few-point temporal correlators while satisfying a {subvolume-}law scaling.

In Fig.~\ref{fig:ZZ_error} we report empirical tests of this expectation. 
In regime (I) local observables decay in around one Floquet period, subsequently displaying irregular fluctuations around zero, with exponentially decaying magnitude. Thus, if we set an absolute error threshold~$\epsilon$, we only need to approximate dynamics up to time $t(\epsilon)\propto |\log\epsilon|$. This corresponds to a polynomial scaling of the required bond dimension $\chi$ with~$\epsilon$. In general, the irregular fluctuations around zero of observables arise from complex operator growth process~\cite{YoshimuraGarrattPRB2025OperatorsFloquet}, and would require an exact description of the IM, so we do not expect further compression to be possible. Accordingly, we expect the {\it relative} error to rise rapidly to a value of order 1 as $\chi$ is reduced below $\chi_{\rm exact}$. We observed this behavior most clearly in the `plain vanilla' regime close to the dual-unitary point (see Appendix~\ref{app:TE_plain_vanilla}). Slightly farther away from the dual-unitary point, for parameter as in Eq.~(\ref{eq:parameters_plain_vanilla}), the absolute errors remain extremely small, and relative errors rise rapidly with $T$ at fixed $\chi$, as illustrated in Fig.~\ref{fig:ZZ_error}.

In contrast, in regimes (II), (III), an approximate localized integral of motion exists, leading to a slow decay of the autocorrelator. In this case, we expect that the slow regular decay $C_{zz}(T) \sim c \, 2^{-\gamma T}$ will be accompanied by irregular and rapidly decaying fluctuations similar to the regime (I); these fluctuations stem from the highly-entangled components of the IM, and would only lead to a small relative error for local correlators. We indeed found that in both regimes the IM's bond dimension can be significantly reduced while keeping high accuracy on the slow exponential decay. The relative and absolute errors for the autocorrelator computed for different values of $\chi$ are reported in Fig.~\ref{fig:ZZ_error}. We conclude that, in line with our expectations, for chaotic baths na\"ive Schmidt value truncation projects out the IM components responsible for complex nonlocal temporal correlations and volume-law TE scaling, while keeping high accuracy on the edge (probe system) dynamics. For example, in regime (II), we find a bond dimension of just $\chi=256$ is sufficient to capture dynamics of local observables with accuracy $\delta_T<2^{-10}$ up to times $T\approx 30$.

\section{Discussion}\label{sec:discussion}



In summary, we have studied TE properties of thermalizing many-body baths. We provided an intuitive explanation of the previously observed volume-law scaling of TE in 1d chaotic brickwork circuits~\cite{FolignoBertiniPRX2023} and its relation to non-Markovianity and local temporal correlation functions. We found that the volume-law scaling does reflect genuine non-Markovianity of dynamics, and, by employing quantum-information techniques, we established a link with distillable entanglement between the past and the future. Yet, the volume-law TE manifests itself only in highly complex temporal correlation functions. Such correlations are effectively projected out through a coarse-graining procedure, leading to a sharp collapse of TE to a {subvolume-}law scaling, which we dubbed TE transition, at a \textit{finite} critical density of coarse-graining. We have studied this transition both analytically and numerically in a toy random unitary bath model, as well as in 1d dual-unitary and generic Floquet circuits.  

Furthermore, we considered a Floquet model that displays non-trivial thermalization dynamics, 
leading to a parametrically slow decay of local autocorrelators of a probe spin at the edge. In this model, we found that TE undergoes a similar transition from volume- to {subvolume-}law under coarse-graining. Turning to local observables, we demonstrated that the slow thermalization dynamics could be captured accurately with a compressed representation of the IM via a low-bond dimension MPS. This is consistent with the intuition that, when the system is quickly thermalizing apart from quasi-conserved edge operators,  truncating the highly non-local temporal correlations responsible for volume-law TE does not significantly affect the accuracy  of the description of few-point correlators. An unexpected high accuracy of the method of transverse contractions (``folding algorithm'') for long-time dynamics of local observables with moderate bond dimensions -- far beyond the times allowed by convergence of TE -- was reported for certain instances of Hamiltonian (continuous-time)  dynamics in the first explorations~\cite{banuls2009Matrix,muller-hermes2012Tensor,hastings2015Connecting}, and more recently for discrete-time Floquet dynamics~\cite{lerose2023overcoming}. We hypothesize that the leading Schmidt vectors carry information on the {subvolume-}law coarse-grained IM, whereas the small Schmidt values are related to highly complex multi-time correlators, providing the underlying mechanism behind the empirically observed accuracy. 

Our results suggest that, although the exact IM is highly entangled, for the goal of computing $N$-point temporal correlators, there exists a greatly compressed MPS representation of IM that captures dynamics of slowly decaying operators. In the future, it will be interesting to explore different strategies for constructing such {effective} MPS {representations}, either via a generalization of the coarse-graining procedure, or via introducing weak dissipation, similar to the dissipation-assisted operator evolution (DAOE) approach~\cite{daoe}.

Finally, we note that coarse-graining effectively increases the range of circuit gates by a factor of $1/n_{\mathrm{cg}}$. A direct prediction of our analysis is that IMs of $k$-local circuits with $k>2$ obey {subvolume-}law scaling, provided the elementary gates are sufficiently entangling. We can apply the same argument to higher-dimensional baths: The bath size $\sim 2^{T^{d}}$ greatly exceeds the probe’s accessible temporal Hilbert space $\sim 2^T$ whenever the dimensionality of the bath is $d>1$.  Thus, in chaotic higher-dimensional systems, we predict {subvolume-}law TE scaling. This suggests that their IM  can be approximated by an MPS with a moderate bond dimension. Developing approximate algorithms for constructing MPS representation of higher-dimensional interacting quantum baths is an exciting direction for future research. A recently introduced influence-functional belief-propagation method for 2d dynamics provides a first promising step in that direction~\cite{Park2025IFBeliefPropagation}.

\acknowledgments

We thank Matan Lotem, Tianci Zhou, Jacopo De Nardis, Bruno Bertini, and Sarang Gopalakrishnan for insightful discussions. 
This work was supported by a Brown Investigator Award (I.~V. and D.~A.~A.) and by the European Research Council (ERC) under the European Union's Horizon 2020 research and innovation program (grant agreement No.~864597) (I.~V., A.~L., D.~A.~A.). M.~S.~is supported by the Swiss National Science Foundation (SNSF) through a Postdoc.Mobility fellowship (Grant P500PT\_225372).
W.~W.~H.~is supported by the National Research Foundation (NRF), Singapore through the NRF Fellowship NRF-NRFF15-2023-0008 and through the National Quantum Office, hosted in A*STAR, under its Centre for Quantum Technologies Funding Initiative (S24Q2d0009).
A.~L.~was supported by a Leverhulme-Peierls fellowship at the University of Oxford and by the Flemish Research Foundation (FWO) through an Odysseus grant (project G0ASY25N).
This research was supported in part by grant NSF PHY-2309135 to the Kavli Institute for Theoretical Physics (KITP). 

\section{Data availability}
The data used in this manuscript are available at Ref. \cite{data}. Additional data generation code is available on request.
\bibliography{TemporalTransition}


\appendix
\section{Calculation of temporal Rényi entropy}
In this Appendix, we provide details on the calculation of the temporal Rényi entropy for the structureless random unitary bath, leading to the result in Eq.~\eqref{eq:Renyi_2_pure}.

\subsection{Definition of Temporal Entanglement}
We start by revising the definition of the IM provided by Eq.~\eqref{eq:IM_definition}. 
In order to define TE, we vectorize the IM and treat it as a pure state. Formally this can be done by introducing a vector $|\mathcal{I}\rangle$:
\begin{eqnarray}\label{eq:IM_vec}
\langle s_T,\bar{s}_T\dots s_1,\bar{s}_1|\mathcal{I}\rangle = \mathcal{I}_{s_1\dots s_T}^{\bar{s}_1\dots \bar{s}_T}.
\end{eqnarray}
This vector is not normalized: rather, its squared norm $N^2$ equals
\begin{eqnarray}
N^2=\langle\mathcal{I}|\mathcal{I}\rangle.
\end{eqnarray}

Let us also introduce the corresponding unnormalized density matrix $\tilde\rho = |\mathcal{I}\rangle\langle\mathcal{I}|$, which satisfies
\begin{equation}
\text{tr} \tilde\rho = N^2,
\end{equation}
and we denote its partial trace as
\begin{eqnarray}
\tilde\rho_\mathcal{F}=\text{tr}_\mathcal{P}\left( |\mathcal{I}\rangle\langle \mathcal{I}| \right).
\end{eqnarray}

With these definitions,  temporal Rényi entropy takes the standard expression
\begin{eqnarray}
S^{(\alpha)}=\frac{1}{1-\alpha}\log_d \frac{\text{tr} \tilde\rho_\mathcal{F}^\alpha}{N^{2\alpha}}.
\end{eqnarray}
For brevity, we will henceforth drop $d$ in the base of logarithms.

\subsection{Weingarten calculation of Rényi TE for the structureless random unitary bath}\label{app:Renyi}
\paragraph{Static bath}
In this section we concentrate on the computation of the temporal Rényi entropy for the structureless random unitary bath, in the limit of large bath $D_\mathcal{B}\gg 1$. In principle, we are interested in Rényi entropy averaged over the Haar-random unitaries ${U}_t$:
\begin{equation}
\mathbb{E}[S^{(\alpha)}]=\frac{1}{1-\alpha}\mathbb{E}\left[\log \frac{\text{tr} \tilde\rho_\mathcal{F}^\alpha}{N^{2\alpha}}\right].
\end{equation}
Instead, we calculate a related but easier quantity, i.e., the annealed entropy:
\begin{eqnarray}
S^{(\alpha)}_{\text{ann}}=\frac{1}{1-\alpha}\log \mathbb{E}\left[\frac{\text{tr} \tilde\rho_\mathcal{F}^\alpha}{N^{2\alpha}}\right].
\end{eqnarray}
The latter provides a good approximation to the former if fluctuations are small, as we discuss below. The calculation of averages can be done with the help of Weingarten calculus \cite{collins2022weingarten}, with the central formula
\begin{equation}
\left\langle\left(\mathcal{U}\otimes \mathcal{U^\dagger}\right)^{\otimes k}\right\rangle = \sum\limits_{\tau,\sigma}\textrm{Wg}_q(\tau^{-1}\sigma) \ |\tau\rangle_{\rm t}|\tau\rangle_{\mathcal{B}} \times {}_{\mathcal{B}}\langle \sigma|{}_{\rm t}\langle \sigma|,
\end{equation}
where the sum is taken over all possible permutations $\tau, \sigma$ of the set of $k$ tensor copies. Here $q=d\times D_\mathcal{B}$ is the Hilbert space dimension of the composite system of the probe and bath at a single time step, namely the dimensionality of $\mathcal{U}$; the states $|\tau\rangle_{\rm t},|\tau\rangle_ \mathcal{B}$ correspond to permutation of replica spaces in probe and bath spaces, they can be explicitly defined via matrix elements:
\begin{equation}
    {}_a\langle s_k, \bar{s}_k, \cdots, s_1, \bar s_1|\tau\rangle_a = \prod_{i=1}^k\delta_{s_{\tau(i)},\bar{s}_{i}},
\end{equation}
where $a=\rm t$ or $\mathcal{B}$; and $s_i$ ($\bar s_i$) is the index of computational basis in the $i$-th forward (backward) branch, respectively. These permutation states are neither orthogonal nor normalized: the scalar product of two such states is given by
\begin{equation}
{}_{a}\langle\tau|\sigma\rangle_{a}=C_{\sigma,\tau}=q_{a}^{\#\text{cycles}(\tau^{-1}\sigma)},
\end{equation}
where $q_{\rm t,\mathcal{B}}=d,D_\mathcal{B}$ is dimension of the probe system or bath.

Below, we will concentrate on the case $\alpha=2$, for which the annealed entanglement is nothing but the logarithm of the average purity. Analysis for general $\alpha$ is similar. Moreover, one can estimate the fluctuations of $N^2$, concluding that they are suppressed as $\frac{1}{D_{\mathcal{B}}}$, allowing us to further approximate the entropy as:
\begin{eqnarray}\label{eq:S_annealed}
S^{(2)}_{\text{ann}}\simeq -\log \frac{\mathbb{E}[\text{tr} \tilde\rho_\mathcal{F}^2]}{\mathbb{E}[N^{4}]}.
\end{eqnarray}
We will establish this approximation later; for now, let us concentrate on  numerator and denominator separately. Each of them involves four copies of the IM and can be calculated using Weingarten calculus. The two terms are represented pictorially as:
\begin{figure}[h]
    \centering
    %
    %
    %
    
    \subfloat[\label{fig:F_P_splitting_a}]{\includegraphics[width=0.1\textwidth]{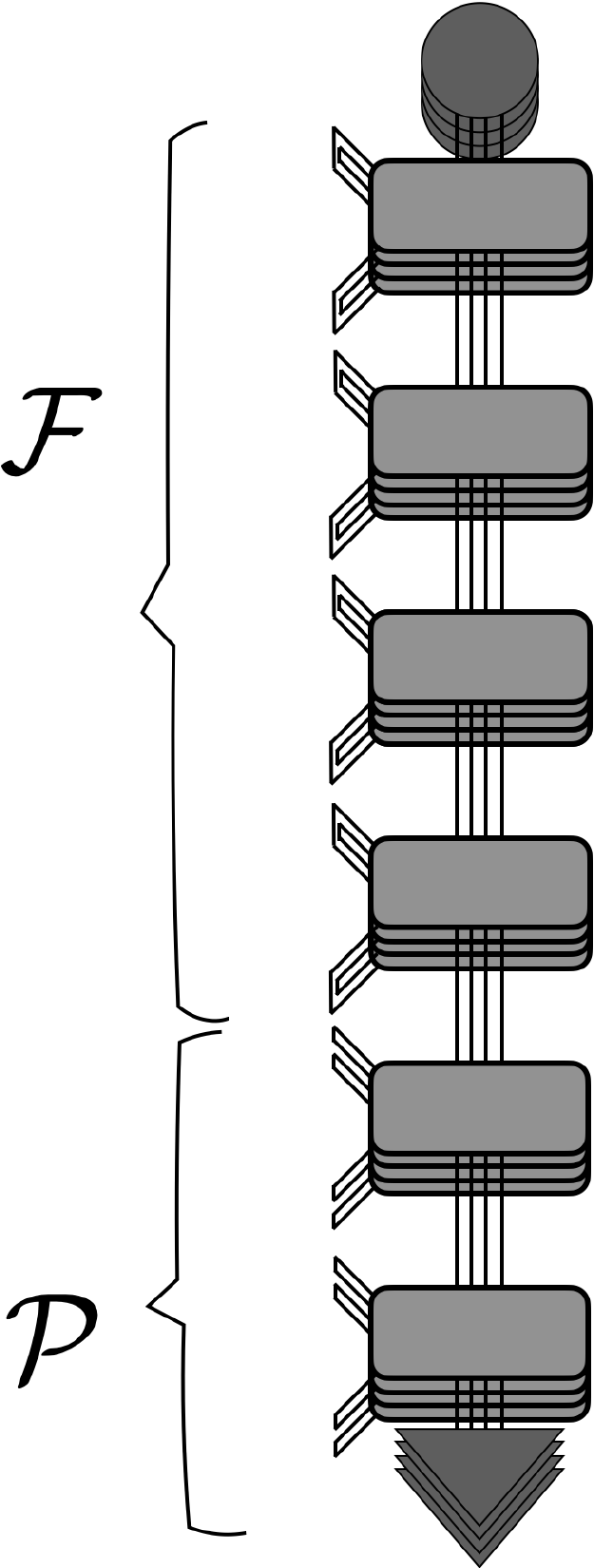}} \qquad \qquad \qquad  
    \subfloat[\label{fig:F_P_splitting_b}]{\includegraphics[width=0.045\textwidth]{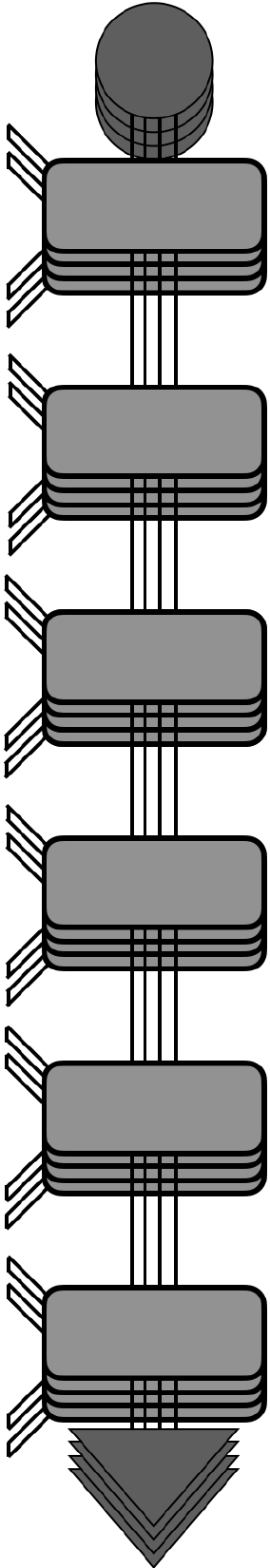}}
    
    \caption{(a) The pictorial definition of the numerator $\text{tr} \tilde\rho_\mathcal{F}^2$. (b) Pictorial definition of the normalization term $N^4$.}
    \label{fig:F_P_splitting}
\end{figure}

Based on these figures, we can compute the averaged quantities using the corresponding transfer matrices:
\begin{equation}\label{eq:TEE_purity}
    \mathbb{E} \left[ \mathrm{tr}(\tilde\rho_\mathcal{F}^2) \right] = {}_\mathcal{B}\langle{\varnothing}|\mathbb{T}_{\pi_\mathcal{F}}^{f}\mathbb{T}_{\pi_\mathcal{P}}^{p}|{\rho_0}\rangle_\mathcal{B},
\end{equation}

\begin{equation}\label{eq:TEE_norm}
    \mathbb{E}\left[ N^4 \right] = {}_\mathcal{B}\langle{\varnothing}|\mathbb{T}_{\pi_\mathcal{P}}^{T}|{\rho_0}\rangle_\mathcal{B},
\end{equation}
where we introduced two permutations $\pi_\mathcal{P}=(1~2)(3~4)$, $\pi_\mathcal{F}=(1~4)(2~3)$. The state ${}_\mathcal{B}\langle\varnothing|$ corresponds to the identity permutation, representing the final tracing over the bath degrees of freedom, while $|\rho_0\rangle_\mathcal{B}$ is the initial state of the bath. Two relevant choices are a pure state, $|\rho_0\rangle_\mathcal{B} = (|0\rangle_\mathcal{B}\otimes|0^*\rangle_\mathcal{B})^{\otimes 4}$ --- in this case ${}_\mathcal{B}\langle\sigma|\rho_0\rangle_\mathcal{B} = 1,\forall\sigma\in S_4$ --- or a maximally mixed state, $|\rho_0\rangle_\mathcal{B} = (\frac{1}{D_\mathcal{B}})^4|\varnothing\rangle_\mathcal{B}$, --- in which case ${}_\mathcal{B}\langle\sigma|\rho_0\rangle_\mathcal{B} = \frac{1}{D_\mathcal{B}^{|\sigma|}}$.  In the latter equation $|\sigma|$ denotes the minimal number of transpositions to obtain $\sigma$ from the identity permutation, which can alternatively be written in terms of the number of cycles in $\sigma$ as
\begin{equation}
|\sigma| = 4-\#\text{cycles}(\sigma).
\end{equation}

For the long products Eq. (\ref{eq:TEE_purity}, \ref{eq:TEE_norm}) under consideration, the transfer matrices are given explicitly as
\begin{equation}\label{eq:transfer_matrix_X}
\mathbb{T}_\textrm{X}=\sum\limits_{\tau,\sigma}(\mathcal{T}_{\textrm{X}})_{\tau,\sigma}|\tau\rangle_{\mathcal{B}} {}_\mathcal{B}\langle \sigma|,
\end{equation}
where we defined component-wise
\begin{equation}\label{eq:T_X}
(\mathcal{T}_{\rm{X}})_{\tau,\sigma}=\sum\limits_{\tau,\sigma}\textrm{Wg}_q(\tau^{-1}\sigma) \ {}_{\rm t}\langle \pi_{\rm{X}}|\tau\rangle_{\rm t}{}_{\rm t}\langle\sigma|\pi_{\rm{X}}\rangle_{\rm t},
\end{equation}
and $\textrm{X}\in\{\mathcal{P},\mathcal{F}\}$ denotes the type of boundary condition imposed. Let us fix the basis of permutation states, and switch to a matrix representation of transfer matrices provided by
\begin{equation}\label{eq:transfer_matrix_X_matrep}
    (C\mathcal{T}_\textrm{X})_{\tau,\sigma} =\sum_\lambda {}_\mathcal{B}\langle\tau|\lambda\rangle_\mathcal{B}(\mathcal{T}_\textrm{X})_{\lambda,\sigma},
\end{equation}
which takes into account the inner products between permutation states. Both Eqs.~(\ref{eq:TEE_purity}--\ref{eq:TEE_norm}) can be computed in terms of eigenvalues and eigenvectors of $(C\mathcal{T}_{\pi_\mathcal{P}})$ and $(C\mathcal{T}_{\pi_\mathcal{F}})$. Below, we examine these eigenvalues. In the limit of large $D_\mathcal{B}$, the leading order contribution is provided by diagonal terms of the Weingarten matrix $\textrm{Wg}_q(\sigma^{-1}\sigma)=\textrm{Wg}_q(\textrm{id})$, for which the suppression $1/q^{4+|\lambda^{-1}\sigma|}$ is minimal. By again selecting the leading contribution in ${}_\mathcal{B}\langle\tau|\lambda\rangle_\mathcal{B}$, we obtain a condition $\sigma=\lambda=\tau$. Thus, the eigenvalues (to leading order) of the matrix $(C\mathcal{T}_\textrm{X})$ are labeled by single permutations $\sigma$:
\begin{equation}\label{eq:eigenvalues_TEE}
E_{\sigma}=d^{4-2|\pi_X\sigma^{-1}|}+o(\frac{1}{D_\mathcal{B}}).
\end{equation}
Examining the permutation group $S(4)$, we summarize the eigenvalues into a following table:
\begin{table}[H]
  \centering
  \caption{Eigenvalue spectrum and degeneracies}
  \label{tab:spectrum}
  \begin{tabular}{cc}
    \toprule
    \textbf{Eigenvalue } & \textbf{Degeneracy} \\
    \midrule
    $d^{4}$   & 1  \\
    $d^{2}$   & 6  \\
    $1$       & 11 \\
    $d^{-2}$  & 6  \\
    \bottomrule
  \end{tabular}
\end{table}
 It can be demonstrated that the leading eigenvalue $d^4$ remains unchanged to all orders in $D_\mathcal{B}$ --- in fact, this is a consequence of the unitarity condition of the Haar-random ensemble matrices $U_\tau$. The remaining eigenvalues receive corrections.

According to Table \ref{tab:spectrum}, it is sufficient to compute the leading order contributions to the terms proportional to $d^{kT}$, for $k=1,\dots ,4$. Omitting the details, we provide the results of an explicit expansion in $D_\mathcal{B}^{-1}$ using the Wolfram Mathematica package:
\begin{multline}
\mathbb{E} \left[ \mathrm{tr}(\tilde\rho_\mathcal{F}^2) \right]= 1 + \frac{d^{4T}}{D_\mathcal{B}^4}+2\frac{d^{4f+2p}}{D_\mathcal{B}^3}+\\ + \frac{d^{4f}+\frac{3}{4}d^{4p}f}{D_\mathcal{B}^2}+2\frac{d^{2p}+d^{2f-2p}}{D_\mathcal{B}},
\end{multline}
and
\begin{equation} \label{eq:Norm_TEE}
    \mathbb{E} \left[ \mathrm{tr}(N^4) \right] = (1+\frac{d^{2T}}{D_\mathcal{B}})^2.
\end{equation}
Armed with these results, let us go back to entanglement, given by Eq.~\eqref{eq:S_annealed}. In the regime $r<1$ or $T<\log_d D_\mathcal{B}$, the numerator is dominated by $1$, and the entanglement is symmetric with respect to $f$ and $p$. In the regime $r>1$, the leading term in the numerator is $\frac{d^{4T}}{D_\mathcal{B}^4}$ which is again symmetric. We conclude that the maximal entanglement is reached around $f=p$ in all regimes, with possible corrections at $r\simeq 1$. Thus, we have reproduced Eqs.~(\ref{eq:Renyi_2_pure}--\ref{eq:Renyi_epsilon}) of the main text.

Additionally, by examining Eq.~\eqref{eq:Norm_TEE} and the 2-copy Weingarten calculation from Eq.~\eqref{eq:purity_pure}, we conclude the following equality:
\begin{equation}
\mathbb{E} \left[ \mathrm{tr}(N^4) \right]=\mathbb{E} \left[ \mathrm{tr}(N^2) \right]^2 +o(D_\mathcal{B}^{-1}). 
\end{equation}
This suppression of fluctuations of $\text{tr}(N^2)$ justifies the approximation in Eq.~\eqref{eq:S_annealed}.
\paragraph{Dynamically growing bath.}\label{app:growing_bath}
In the case of a dynamically growing random unitary bath, we increase the bath dimension   at every time step by feeding in the bath evolution a maximally mixed state of dimension $d^{2v_B}$. The modification of the transfer matrices from Eqs.~(\ref{eq:transfer_matrix_X}--\ref{eq:T_X}) is given by the following formula:
\begin{equation}\label{eq:transfer_matrix_X_dyn}
\mathbb{T}_\textrm{X}= \sum_{\tau, \sigma}(\mathcal{T}_{\textrm{X}})_{\tau,\sigma}\times |\tau\rangle_{\mathcal{B}}{}_{\mathcal{B}}\langle \sigma|,
\end{equation}
with
\begin{equation}\label{eq:T_X_dyn}
(\mathcal{T}_{\textrm{X}})_{\tau,\sigma}=d^{-4v_B}\sum\limits_{\tau,\sigma}\mathrm{Wg}_q(\tau^{-1}\sigma) {}_{\rm t}\langle \pi_{\rm{X}}|\tau\rangle_{\rm t}{}_{\rm t}\langle\sigma|\pi_{\rm{X}}\rangle_{\rm t} {}_{\rm e}\langle\sigma|\varnothing\rangle_{\rm e},
\end{equation}
where we introduced an additional space ${\rm e}$ corresponding to the newly introduced spin to the bath at each time step.
where the transfer matrices now explicitly depend on $n$, and $D_n = d^{2v_B n}$. The average Rényi entropy can be computed along the same lines as before with the modified transfer matrices, the numerical results are presented in Fig.~\ref{fig:growing_bath}(c).

To estimate the effect of CG, we note the following graphical equation:
\begin{figure}[H]
    \centering
    \includegraphics[angle=-90, width=0.6\linewidth]{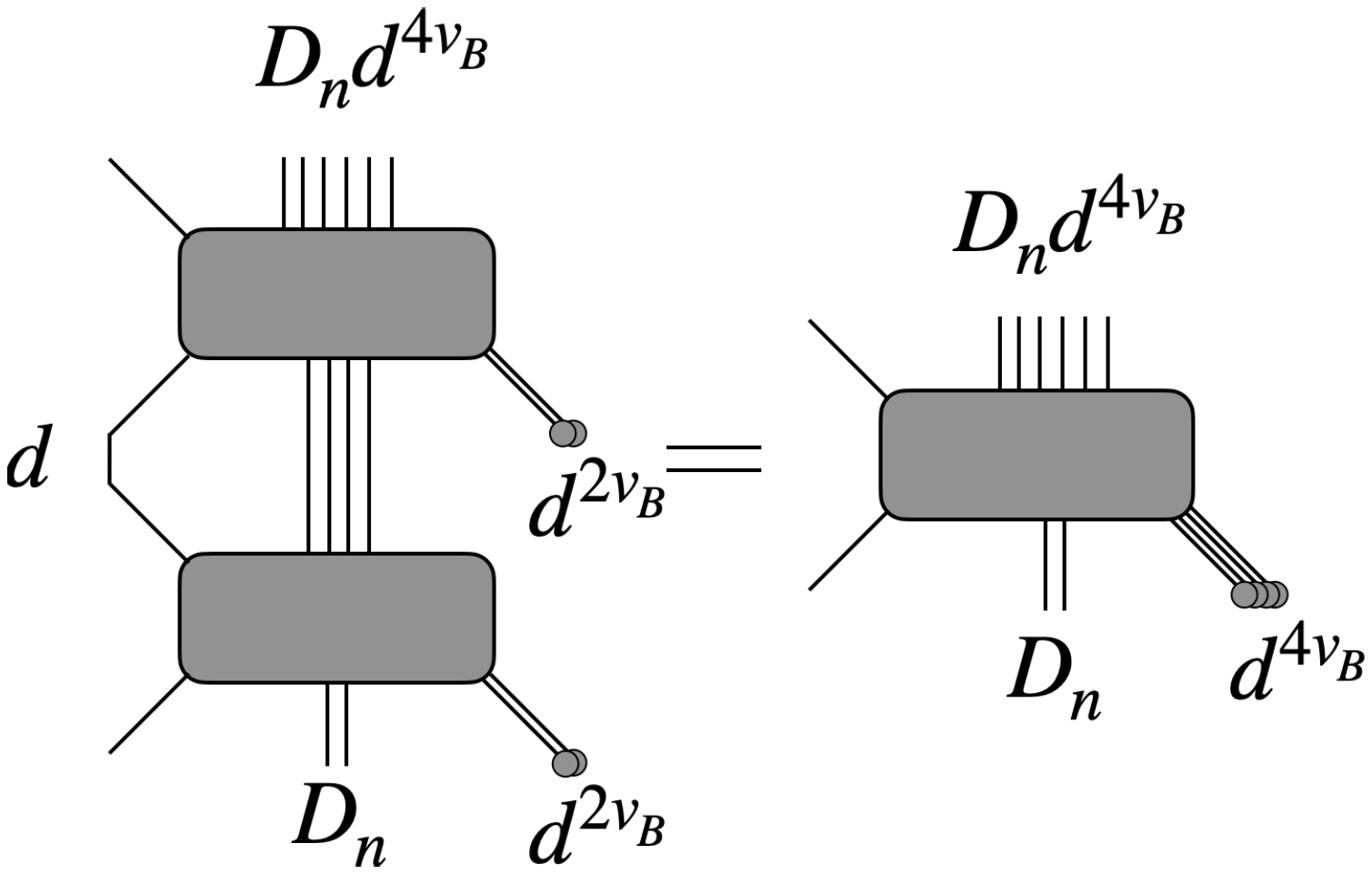}
    \caption{Illustrating that the coarse-graining  two consecutive gates is equivalent to a single gate with twice larger extra spin fed into the bath.}
    \label{fig:Product_of_two_gates}
\end{figure}
Mathematically, this equation can be formalized by introducing $\mathbb{T}(q,d_{\rm e})$, which acts on the temporal state as well as on the bath state:
\begin{equation}
\mathbb{T}(q,d_{\textrm{e}})=d^{-2v_B}\sum\limits_{\sigma,\tau}\mathrm{Wg}_{q}(\tau^{-1}\sigma)|\tau\rangle_{\textrm{t}}|\tau\rangle_{\mathcal{B}}\ {}_{\mathcal{B}}\langle \sigma|\ {}_{\textrm{t}}\langle \sigma|\ {}_{\text{e}}\langle\sigma|\varnothing\rangle_{\rm e},
\end{equation}
Here, $d_\text{e}$ denotes the dimension of extra spin, and $q=dd_{\text{e}}D_{\mathcal{B}}$ denotes the total dimension of the space. The pictorial equation from the figure reads as:
\begin{equation}
\mathbb{T}(D_{n+1}d,d^{2v_B})\mathbb{T}(D_{n+2}d,d^{2v_B})=\mathbb{T}(D_{n+2}d,d^{4v_B}),
\end{equation}
where we used $D_n=d^{2v_B n}$. This observation demonstrates that the coarse-grainig reduces the number of tim-steps, effectively upscaling the the $v_B\to\frac{v_B}{n_{\rm cg}}$.
\section{Estimation for the norms of IMs}\label{app:replicas}

In this section we provide details on the Weingarten calculations leading to the estimates for the density matrices $\rho_{\mathcal{I}}$ and $\rho_\mathrm{in,out}$, used in Eqs.~(\ref{eq:Pure_purity_bound}) and in Appendix \ref{app:information_recovery}, respectively. To obtain these estimates, we first use Eq.~\eqref{eq:trace_Frobenius_ineq} to bound the trace norm by the Frobenius (Hilbert–Schmidt) norm, paying a dimensional penalty for switching norms. The Frobenius norm then allows for a direct Weingarten calculation, which is related closely with the purities.

We start by fixing the notations. As discussed in Section~\ref{sec:Non-Markovianity}, we group the $2T$ temporal degrees of freedom in a `reference' set $Q$ and a `system' $S$ each containing $T$ qudits and initialize them in pairs of maximally entangled states $|\Phi^{\text{max}}_{SQ}\rangle$. Analogously, we split the bath degrees of freedom into `in' and `out' parts, also preparing them in a maximally entangled state $|\Phi^{\text{max}}_{\mathcal{B}_{\textrm{in}}\mathcal{B}_{\textrm{out}}} \rangle$. Then, we act on a `system' and bath `out' degrees of freedom by a sequence of unitary operators, preparing the precursor state that belongs to the tensor product of three states: $\textrm{in, t, out}$ (see the illustration in Fig.~\ref{fig:Tripartite_system}):
\begin{equation}\label{eq:precursor_state}
|\Psi_{\textrm{in},\rm t,\textrm{out}}\rangle={U}_T\dots {U}_1|\Phi^{\text{max}}_{\mathcal{B}_{\textrm{in}}\mathcal{B}_{\textrm{out}}} \rangle |\Phi^{\text{max}}_{SQ}\rangle.
\end{equation}
\begin{figure}[H]\label{fig:Tripartite_system}
    \centering
\includegraphics[width=0.45\columnwidth]{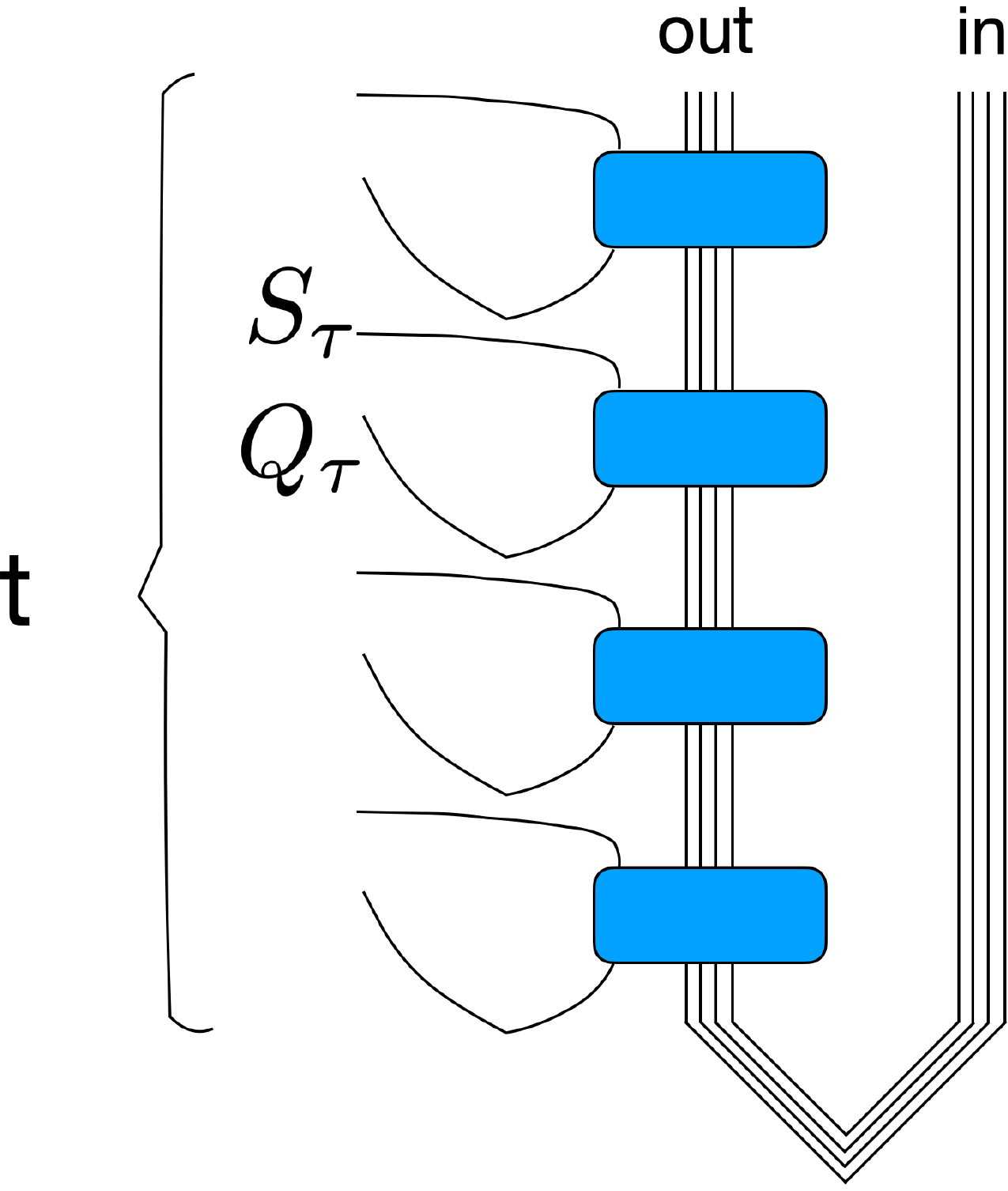}
    \caption{Partition of the precursor state from Eq.~\eqref{eq:precursor_state} into impurity degrees of freedom and bath degrees of freedom; the latter are also split into in and out degrees of freedom. The blue gates correspond to unitaries $U_\tau$ (forward branch only).}
    \label{fig:Tripartite_system}
\end{figure}
Tracing out the bath degrees of freedom defines the density matrix $\rho_\textrm{t}$, which is nothing but the IM in a density matrix form (i.e., its Choi dual $\rho_{\mathcal{I}}$). Alternatively, one can trace out the temporal degrees of freedom, defining the density matrix on bath `in' and `out' degrees of freedom $\rho_{\rm in,out}$. 
Below, we will provide explicit calculations of the purities of these corresponding density matrices.

\subsection{Purity calculations}\label{app:purity}
\paragraph{Structureless bath.}
Proceeding in analogy with Eqs.~(\ref{eq:TEE_purity}--\ref{eq:TEE_norm}), we may define:
\begin{align}
\mathbb{E}[\text{tr}\rho_{\mathcal{I}}^2] &= d^{-2T}{}_\mathcal{B}\langle \varnothing|\mathbb{T}_{\text{S}}^{T} |\rho_{0}\rangle_\mathcal{B},\\
\mathbb{E}[ \text{tr}\left({\rho}_{\text{in,out}}^2\right)] &= d^{-2T}{}_\mathcal{B}\langle \varnothing|\mathbb{T}_\text{I}^{T}|\rho_{0}\rangle_\mathcal{B},
\end{align}
where the transfer matrices are defined in analogy with Eqs.~(\ref{eq:transfer_matrix_X}--\ref{eq:T_X}), with a difference that in this section we consider only two replicas (not four). Hence, there are only two permutations: the identity permutation $\pi_{\rm I}$ and the transposition $\pi_{\rm S}=(1, 2)$. We consider maximally mixed (MM) or pure initial states. The MM state is defined by  $\rho_{0}=\frac{1}{D_\mathcal{B}^2}|\varnothing\rangle$, and the pure initial state $\rho_{0}$ is characterized by a unit overlap with all permutation states: $\langle \sigma|\rho_{0}\rangle=1$.
The Weingarten matrix reads
\begin{eqnarray}
    \mathrm{Wg}_q = \frac{1}{q^2-1}\begin{pmatrix} 1 & -q^{-1}\\
    -q^{-1} & 1
    \end{pmatrix},
\end{eqnarray}
and an explicit calculation provides the following expression for the transfer matrix components in the basis of permutation states:
\begin{eqnarray}
\mathcal{T}_{\text{S}}=\frac{d^2}{D_\mathcal{B}^2d^2-1}\begin{pmatrix}
 1 & -D_\mathcal{B}^{-1} \\
 -D_\mathcal{B}^{-1} & d^2
\end{pmatrix}.
\end{eqnarray}
The purity from Eqs.~(\ref{eq:purity_pure}-\ref{eq:purity_max_mixed}) can be written as
\begin{eqnarray}
\mathbb{E}[\text{tr}\rho_{\mathcal{I}}^2]= d^{-2T}\langle \varnothing|(C\mathcal{T}_\text{S})^T C|\sigma_{\text{in}}\rangle,
\end{eqnarray}
where we introduced the matrix of scalar products $C$:
\begin{eqnarray}
    C = \begin{pmatrix}
        D_\mathcal{B}^2 & D_\mathcal{B} \\
        D_\mathcal{B} & D_\mathcal{B}^2
    \end{pmatrix}.
\end{eqnarray}
The matrix product $C\mathcal{T}_{\text{S}}$ gives
\begin{equation}\label{eq:CT_S}
C\mathcal{T}_\text{S} = d^2\begin{pmatrix}
    \frac{D_\mathcal{B}^2-1}{D_\mathcal{B}^2d^2-1} & \frac{(d^2-1)D_\mathcal{B}}{D_\mathcal{B}^2d^2-1} \\
    0 & 1
\end{pmatrix} \sim \begin{pmatrix}
    1-\frac{1-d^{-2}}{D_\mathcal{B}^2} & \frac{d^2-1}{D_\mathcal{B}} \\
    0 & d^2
\end{pmatrix}.
\end{equation}
The leading contribution to the $T$-th power is then given by:
\begin{eqnarray}
d^{-2T}(C\mathcal{T}_\textrm{S})^T \sim  \begin{pmatrix}
    d^{-2T} & \frac{1}{D_\mathcal{B}} \\
    0 & 1
    \end{pmatrix}.
\end{eqnarray}
Taking into account that $C|\sigma_{\text{in}}\rangle$ equals to:
\begin{eqnarray}
\begin{cases}
(1,1)^\top \,, \quad \text{for pure initial state}\\
(1 , D_\mathcal{B}^{-1})^\top \,,\quad \text{for MM initial state},
\end{cases}
\end{eqnarray}
one gets for the purity:
\begin{align}\label{eq:purity_pure}
\mathbb{E}[\text{tr}\rho^2_\mathcal{I}]=&d^{-2T}+\frac{1}{D_\mathcal{B}}\,,\quad \text{for pure initial state}\,,\\
\label{eq:purity_max_mixed}
\mathbb{E}[\text{tr}\rho^2_\mathcal{I}]=&d^{-2T}+\frac{1}{D_\mathcal{B}^2}\,, \quad
\text{for MM initial state}.
\end{align}

This immediately reproduces the following estimate for the Frobenius distances:
\begin{equation}\label{eq:norm_pure_MM}
\mathbb{E}[\|\rho_{\mathcal{I}}-\rho_{\textrm{PD}}\|_2]\le \frac{1}{D_\mathcal{B}^{\xi}},
\end{equation}
where $\xi=1,2$ for the case of pure and MM initial state respectively. Combining this estimate with Eq.~\eqref{eq:trace_Frobenius_ineq} we reproduce Eq.~\eqref{eq:Pure_purity_bound}.

The calculation of the purity in the bath channel is analogous, with the difference that one has to consider $\mathcal{T}_{\text{I}}$:
\begin{equation}\label{eq:CT_I}
C\mathcal{T}_\text{I}=d^2\begin{pmatrix}
    1 & 0\\
    \frac{(d^2-1)D_{\mathcal{B}}}{d^2D_\mathcal{B}^2-1} & \frac{D_\mathcal{B}^2-1}{d^2D_\mathcal{B}^2-1}
\end{pmatrix}\sim
d^2\begin{pmatrix}
    1 & 0 \\
    \frac{d^2-1}{d^2D_\mathcal{B}} & d^{-2}
\end{pmatrix}.
\end{equation}
The purity of $\rho_{\text{in},\text{out}}$ from Eq.~\eqref{eq:rho_ino_out_purity} then equals
\begin{multline}\label{eq:purity_in_out}
\mathbb{E}[\text{tr}\rho_{\text{in},\text{out}}^2]=
d^{-2T}\begin{pmatrix}
    0 & 1
\end{pmatrix} (C\mathcal{T}_\text{I})^T
\begin{pmatrix}
D_\mathcal{B}^{-1} \\ 1
\end{pmatrix} =\\= d^{-2T} +D_\mathcal{B}^{-2} +o(D_\mathcal{B}^{-2}) \,.
\end{multline}
\section{Information recovery protocol}\label{app:information_recovery}
In this Appendix we describe the information transition protocol for the scenario described in Sec \ref{sec:Non-Markovianity}. This protocol bounds the distilled entanglement and quantum non-Markovianity of structureless and Hayden-Preskill bath.

We perform the information recovery in two steps. At first step we encode the information from the state in the past $\mathcal{P}$ to an intermediate $\textrm{out}^\prime$ state in the bath. At the second step we feed the intermediate $\textrm{out}^\prime$ state back to the bath, denoting it as $\textrm{in}^{\prime}=\textrm{out}^\prime$ and decode the information back to the future system registers $\mathcal{F}$. We call these two steps an \textit{encoding} and \textit{decoding}, respectively. In the next paragraphs, we first introduce the so-called \textit{decoupling lemma} -- our main tool to study the encoding/decoding properties, next we explain the decoding and encoding steps separately, and finally combine all together, providing the estimate in Eq.~\eqref{eq:N_shared}. In the end of this Appendix, we discuss the concentration of the probabilistic measure, which allows us to estimate the probability of a large deviation of the \textit{encoding/decoding} fidelity from the average value. We will observe that this deviation is suppressed in the scaling limit when both the bath and the observation time are large.

\paragraph{Decoupling lemma.}
The derivation in this appendix relies on an important lemma 1.1 from
\cite{hayden2008decoupling}. Assume that the initial bath density matrix has exactly the rank $D_{\text{in}^\prime}$. Consider the precursor tripartite pure state between the temporal space `t' that can be chosen as $\mathcal{P}$ or $\mathcal{F}$ and the bath `in' and `out' spaces introduced in Eq.~\eqref{eq:precursor_state}.

The following lemma ensures the existence of a decoding map:
\begin{lemma}[Decoupling lemma]\label{lemma:Decoupling}
Let $\rho_{\text{in},\text{out}}=\text{tr}_{t} \left(|\Psi_{\text{in},t,\text{out}}\rangle\langle \Psi_{\text{in},t,\text{out}}| \right)$. 
Then there exists a decoding operator $\mathcal{D}^{t\to t^\prime}$ acting on the temporal spin degrees of freedom, such that:
\begin{equation}
\it{F}\left(\Phi^{\text{max}}_{\text{in},t^\prime},\mathbbm{1}_{\text{in}}\otimes(\mathcal{D}^{t\to t^\prime})\rho_{\text{in},t} \right)\ge 1-\epsilon,
\end{equation}
provided that:
\begin{equation}\label{eq:epsilon}
\|\rho_{\text{in,out}}-\pi_{\text{in}}\otimes\pi_{\text{out}}\|_1<\epsilon,
\end{equation}
where $\pi_{\text{in}/\text{out}}$ is the density matrix of a maximally mixed (MM) $\text{in}/\text{out}$ state.
\end{lemma}
In the simplified Hayden-Preskill one step process, the encoding and decoding processes are symmetric, and the existence of both maps can be proven using the decoupling lemma. Indeed, specializing the purity calculation from Eq.~\eqref{eq:purity_max_mixed} to the case of $T=1, d=D_{\rm t}=d^{f}$, we may estimate average trace distance from the left hand side of Eq.~\eqref{eq:epsilon} bounding it by a Frobenius norm:
\begin{multline}
\|\rho_{\text{in,out}^\prime}-\pi_{\text{in}}\otimes\pi_{\text{out}^\prime}\|^2_1 \\ \le D_\mathcal{B}^2 \|\rho_{\text{in,out}^\prime}-\pi_{\text{in}}\otimes\pi_{\text{out}^\prime}\|_2^2=\frac{D_\mathcal{B}^2}{D_t^2},
\end{multline}
\begin{multline}
\|\rho_{\text{in}^\prime,\text{out}}-\pi_{\text{in}^\prime}\otimes\pi_{\text{out}}\|^2_1 \\ \le D_\mathcal{B}^2\|\rho_{\text{in}^\prime,\text{out}}-\pi_{\text{in}^\prime}\otimes\pi_{\text{out}}\|_2^2=\frac{D_\mathcal{B}^2}{D_t^2}.
\end{multline}
These two inequalities sets a bound on $\epsilon$ from Eq.~\eqref{eq:epsilon} in average and prove the existence of a perfect decoding of the initial state in the regime $D_\mathcal{B}\gg D_{t}$ for the majority of choices for the unitaries. More precisely, one can use the concentration of measure technique, to estimate the probability of a significant deviation of the fidelity from the average value. We will make this estimation at the end of this Appendix [see Eq.~\ref{eq:F_deviation}], assuming for now that the average trace distance is a good approximation for the trace distance of any typical process. 
\paragraph{Decoding map.}\label{app:decoding_map}
The same lemma can be used to prove the existence of a decoding map for the random unitary bath. The direct Weingarten calculation from Eq.~\eqref{eq:purity_in_out} provides on average:
\begin{multline}\label{eq:rho_ino_out_purity}
\|\rho_{\text{in}^\prime\text{,out}}-\pi_{\text{in}^\prime}\otimes\pi_{\text{out}}\|^2_1 \\
<D_\mathcal{B}D_{\text{in}^\prime}\left(\text{tr}\rho_{\text{in}^\prime\text{,out}}^2-\frac{1}{D_\mathcal{B}D_{\text{in}^\prime}}\right) = \frac{D_\mathcal{B} D_{\text{in}^\prime}}{d^{2f}},
\end{multline}
where we assumed that the internal $\text{in}^\prime$ space might have smaller dimension than $D_\mathcal{B}$. The relevance of this assumption will become clear later, when we consider the encoding algorithm.

This estimate guarantees that as long as $f>\frac{1}{2}\log_d(D_\mathcal{B} D_{\text{in}^\prime})$, it is possible to distill $\log_d(D_{\text{in}^\prime})$ entangled qudits between the bath initial state and the observer $\mathcal{F}$, with the fidelity $\it{F}>1-\frac{\sqrt{D_\mathcal{B} D_{\text{in}^\prime}}}{d^{2f}}$. Let us note that Hayden-Preskill setup is a particular example of a random unitary bath, with $f=1$ and large dimension of spin $D_{t}=d^{f}$. Both setups lead to the same scaling of critical time beyond which the recovery becomes possible.


\paragraph{Encoding of an intermediate bath state.}\label{app:encoding_map}
We first start by considering pure initial bath state. Let us first assume a very short time in the past $p\ll\frac{1}{2}b$. As we have a pure initial bath state, the whole evolution defines the pure two-partite precursor state $|\Psi_{t,\text{out}^\prime}\rangle$. We will prove now the possibility to distill $2p$ qudits between impurity and the out bath state. To do so, we use again the result from Eq.~\eqref{eq:Pure_purity_bound}, to show that the impurity density matrix $\rho_{\mathcal{P}}$ is close to a maximally mixed state:
\begin{eqnarray}\label{eq:estimate_rho_imp}
\|\rho_{\mathcal{P}}-\rho_{\rm depol}\|_1<\frac{d^{2p}}{D_\mathcal{B}}.
\end{eqnarray}
This trace distance gives the lower bound on the fidelity:
\begin{eqnarray}
\it{F}(\rho_{\mathcal{P}},\rho_{\rm depol})>1-\|\rho_{\mathcal{P}}-\rho_{\rm depol}\|_1.
\end{eqnarray}
According to Ulhmann's theorem, there exists a purification of a MM density matrix $\rho_{\text{PD}}$ by a maximally entangled state $|\Phi^{\text{max}}_{\mathcal{P},\text{out}^\prime}\rangle$  such that: 
\begin{equation}\label{eq:P_out_fidelity}
|\langle \Psi_{\mathcal{P},\text{out}^\prime}|\Phi^{\text{max}}_{\mathcal{P},\text{out}^\prime}\rangle|^2 >1-\frac{d^{2p}}{D_\mathcal{B}}.
\end{equation}
This proves the possibility to distill $2p$ qudits between past and $\text{out}^\prime$ bath state. No encoding operator $\mathcal{D}_{\text{enc}}$ is needed in this case. In case of $p\gg \frac{1}{2}b$, the estimate \eqref{eq:estimate_rho_imp} is not applicable. One however can simply reduce the number of legs in $\mathcal{P}$ by coarse-graining the $2f-b+2s$ qudits. Which guarantees the distillation of $b-2s$ Bell pairs with the fidelity $\it{F}=1-2^{-2s}$.



The case of MM initial state provides a much weaker bound, stating that it is possible to distill $b$ Bell pairs between the temporal spin and bath out state, as long as $p\gg b$ with the fidelity $\it{F}=1-\frac{D_\mathcal{B}^2}{d^{2p}}$ which is an implication of decoupling lemma \eqref{lemma:Decoupling}.
\paragraph{Distillation of information between the future and the past.}
Summarizing the encoding and decoding steps, for the pure bath state, we may first decode $N_{\mathcal{PB}}=\text{min}(2p,b)$ qudits between the past and the intermediate bath. Moreover, according to Eq.~\eqref{eq:P_out_fidelity}, in the regime where $2p<b$ the $\text{out}^\prime$ approximately belongs to a subspace of dimension $d^{2p}$ making it possible to consider $D_{\text{in}^\prime}=d^{2p}<D_\mathcal{B}$ in Eq.~\eqref{eq:rho_ino_out_purity}. Taking it into account, Eq.~\eqref{eq:rho_ino_out_purity} guarantees that having $f=\frac{1}{2}b +p$ time steps on $\mathcal{F}$ is enough to distill Bell pairs between $\mathcal{F}$ and $\text{in}^\prime$. Thus, we provide a protocol of distilling $N_{\mathcal{PF}}$ Bell pairs with $N_{\mathcal{PF}}=T-\frac{1}{2}b$ time steps, reproducing Eq.~\eqref{eq:N_shared}.

\begin{figure}[t]
    \centering
    \includegraphics[width=\linewidth]{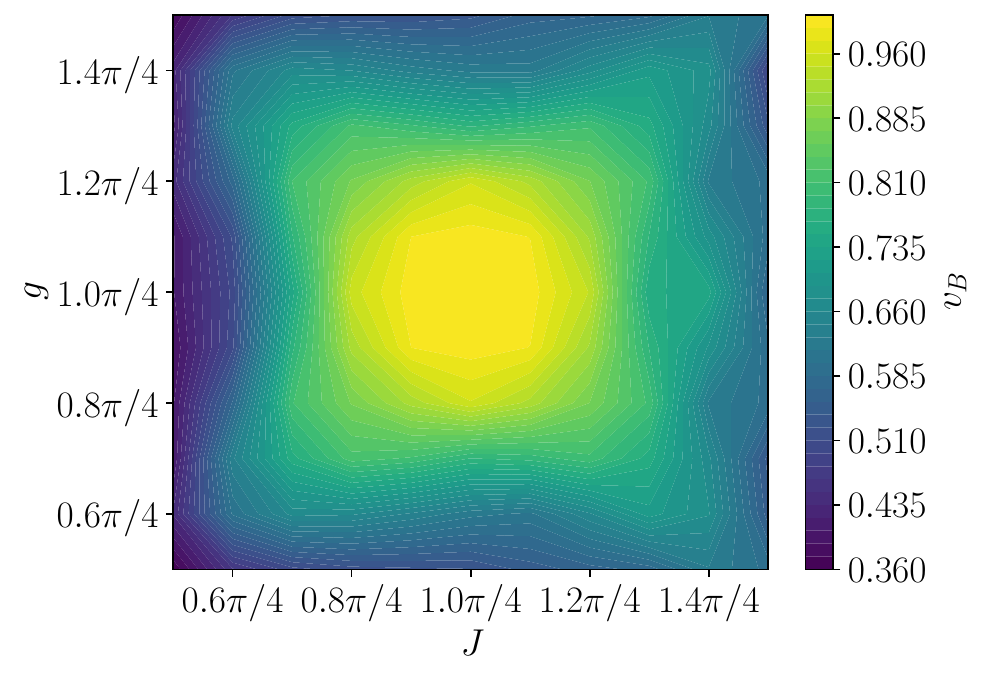}
    \caption{Butterfly velocity $v_B$ extracted from the out-of-time-order correlator front for system size $L=10$.} 
    \label{fig:v_B}
\end{figure}

\begin{figure*}[t]
  \centering
  \subfloat[]{\includegraphics[width=0.33\linewidth]{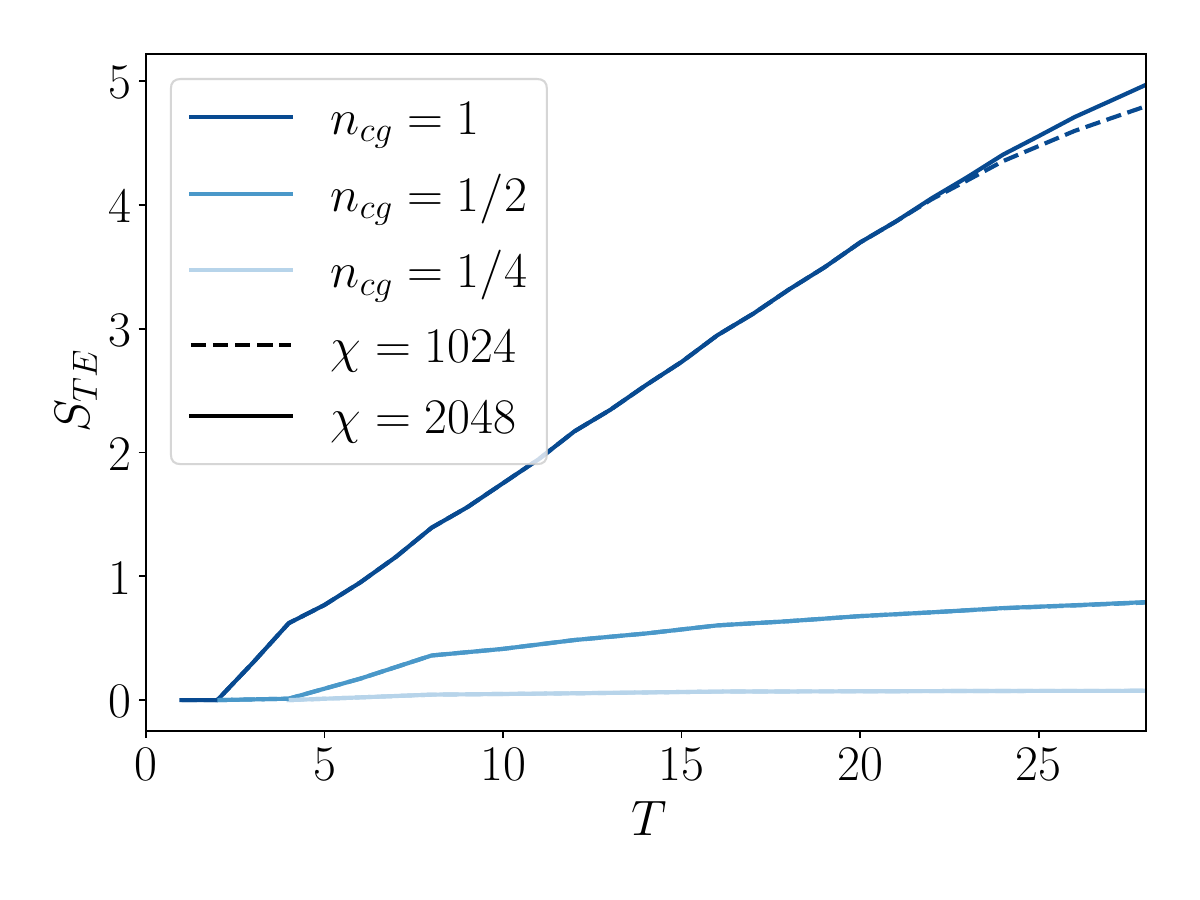}}%
  \hfill
  \subfloat[]{\includegraphics[width=0.33\linewidth]{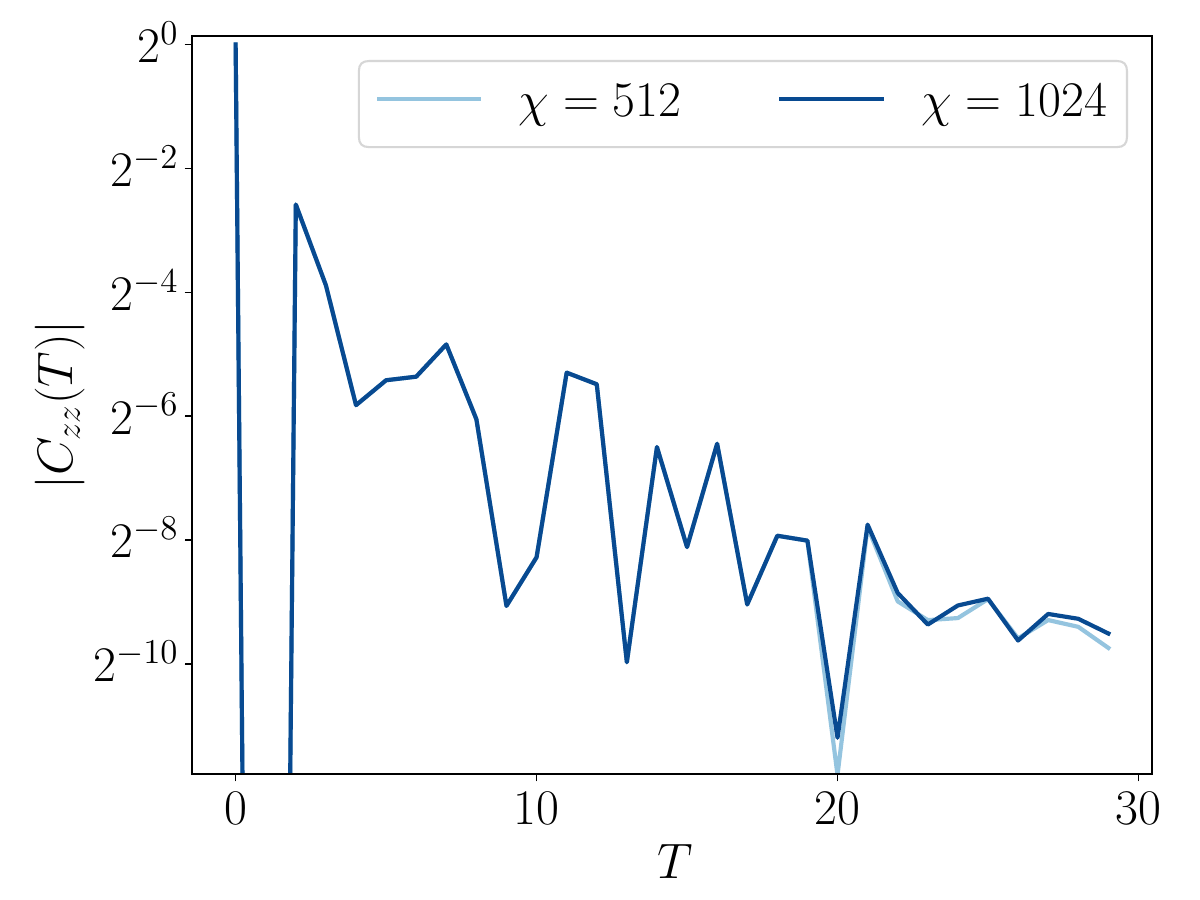}}
  \hfill
  \subfloat[]{%
    \includegraphics[width=0.33\linewidth]{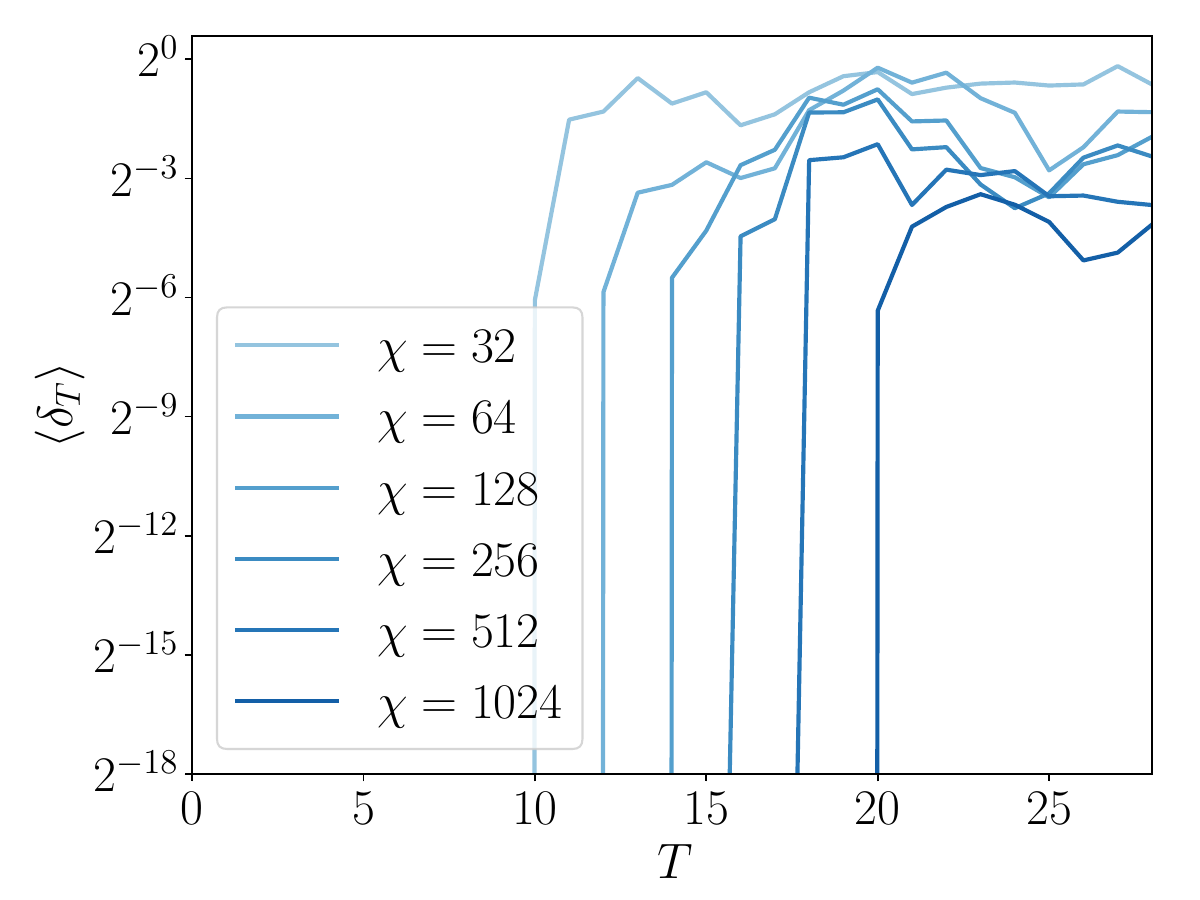}%
  }%

  \caption{`Plain vanilla' regime near the dual-unitary point, parameters $g=\tfrac{\pi}{4}$, $J=0.8\,\tfrac{\pi}{4}$, $h=0.5$.
  (a) (TE), (b) Absolute correlator $|C_{zz}(T)|$,
  (c) Moving average of relative error $\langle\delta_T\rangle$.}
  \label{fig:plain_vanilla}
\end{figure*}

\paragraph{Concentration of measure.} In this paragraph, we estimate the probability of the trace distances to deviate from the average values estimated above. The main estimation tool is the Levy's lemma \cite{Ledoux2001Concentration}:
\begin{lemma}[Lévy concentration inequality]
Let $(X,\textrm{dist},\mu)$ be a metric probability space and let $f\colon X \to \mathbb{R}$ be $L$-Lipschitz, then for all $\varepsilon>0$,
\[
\mu\big(|f-\mathbb{E}_\mu f| \ge \varepsilon\big) \le 2\,\alpha_X\!\left(\,\frac{\varepsilon}{L}\right).
\]
\end{lemma}
Here $\alpha_X\!\left(\,\frac{\varepsilon}{L}\right)$ - is a concentration function defined below:
\begin{definition}[Concentration function]
Let $(X,\textrm{dist},\mu)$ be a metric probability space. For $\varepsilon \ge 0$ and $A \subseteq X$, write
\[
A_\varepsilon := \{x \in X : \textrm{dist}(x,A) \le \varepsilon\}.
\]
The (Lévy) concentration function of $X$ is
\[
\alpha_X(\varepsilon) := \sup\{\,1-\mu(A_\varepsilon) : \mu(A) \ge 1/2\,\}.
\]
\end{definition}

In our case, the manifold $X$ is the $T$-fold Cartesian product of unitary groups of size $dD_{\mathcal{B}}\times dD_{\mathcal{B}}$, each equipped with the Haar probability measure, and endowed with the metric induced by the operator norm:
\begin{equation}
\textrm{dist}(\boldsymbol{U},\boldsymbol{V})=\|\boldsymbol{U}-\boldsymbol{V}\|,
\end{equation}
where $\|A\|$ is the largest eigenvalue of $\sqrt{A^{\dagger}A}$.
As proven in the Appendix of Ref.~\cite{figueroa2019almost} the concentration function can be bounded as:
\begin{equation}
\alpha_X\!\left(\,\frac{\varepsilon}{L}\right)\le e^{-\frac{\epsilon^2(T+1) dD_{\mathcal{B}}}{4L^2}}.
\end{equation}
Therefore, we only need to estimate the Lipschitz constant $L$ of the function
\begin{equation}
f(\boldsymbol{U})=\|\rho_{\text{in,out}}-\pi_{\text{in}}\otimes\pi_{\text{out}}\|_1.
\end{equation}
By definition, Lipschitz constant is defined through a bound connecting the difference between the function at different points and the distance between these points:
\begin{equation}
|f(\boldsymbol{U})-f(\boldsymbol{V})|\le L \|\boldsymbol{U}-\boldsymbol{V}\|.
\end{equation}
To estimate the LHS of this equation, let us recall that $\rho_{\textrm{in,out}}$ is obtained by tracing out the temporal degrees of freedom; therefore, it may be represented as
\begin{equation}
\rho_{\textrm{in,out}}=M_T\cdot\dots M_1\cdot |\Phi^{\textrm{max}}_{\mathcal{B}_{\textrm{in}},\mathcal{B}_{\rm out}}\rangle\langle \Phi^{\textrm{max}}_{\mathcal{B}_{\textrm{in}},\mathcal{B}_{\rm out}}|,
\end{equation}
where $M_\tau$ is a quantum channel defined from the unitary $U_{\tau}$:
\begin{eqnarray}\label{eq:M_def}
M_\tau\cdot \rho= \frac{1}{d}\sum\limits_{i=1}^d\Tr_{\rm s}{U_{\tau}\left( |i\rangle_{\rm s} {}_{\rm s}\langle i| \otimes \rho_{\mathcal{B}}\right)U^{\dagger}_{\tau} }.
\end{eqnarray}

In this equation, the trace is taken over the temporal spin at timestep $\tau$. To proceed with the bound, let us suppose that we have two sequences of quantum channels $\boldsymbol{M}$ and $\boldsymbol{W}$ corresponding to the unitaries $\boldsymbol{U}$, $\boldsymbol{V}$ respectively. Using the triangle inequality, we may estimate:
\begin{multline}
\|M_T\cdot \dots M_1 - W_T\cdot \dots W_1\|_1\le \\ \le  \|(M_T-W_T)M_{T-1}\cdot \dots M_1\|_1+ \\ + \|W_T(M_{T-1}\cdot \dots M_1-M_{T-1}\cdot \dots M_1)\|_1\le \\  \le \|M_T-W_T\|_1 +\|M_{T-1}\cdot \dots M_1 - W_{T-1}\cdot \dots W_1\|_1\le \\ 
\le \sum\limits_{\tau=1}^T \|M_\tau-W_\tau\|_1.
\end{multline}
The distance between single $M_\tau$ and $W_\tau$ can be further bounded by a distance between corresponding unitaries. By definition from Eq.~\eqref{eq:M_def}, we have:
\begin{eqnarray}
M_\tau=\frac{1}{d} \sum\limits_{i,j} U^i_{\tau,j} \cdot (U^j_{\tau,i})^\dagger,
\end{eqnarray}
where $U^i_{\tau,j} \cdot (U^j_{\tau,i})^\dagger$ is a superoperator acting on the space of bath density matrices. Using the triangle inequality again we have:
\begin{equation}
\|M_\tau-W_\tau\|_1\le \sum\limits \|U^i_{\tau,j} \cdot (U^j_{\tau,i})^\dagger-V^i_{\tau,j} \cdot (V^j_{\tau,i})^\dagger\|_1\,,
\end{equation}
now, using the lemma 12 from Ref.~\cite{aharonov1998quantum} we have:
\begin{equation}
\|U^i_{\tau,j} \cdot (U^j_{\tau,i})^\dagger-V^i_{\tau,j} \cdot (V^j_{\tau,i})^\dagger\|_1\le 2\|U^i_{\tau,j} -V^i_{\tau,j} \|\le 2\|U-V \|.
\end{equation}
This finally guarantees:
\begin{equation}
\|M_\tau -W_\tau\|\le 2d\| U_\tau - V_\tau\|\le 2d\| \boldsymbol{U} - \boldsymbol{V}\|.
\end{equation}
Putting everything together, we find $L\le 2Td$, and the probability to violate the average estimates from the first part of the Appendix is suppressed as:
\begin{equation}\label{eq:F_deviation}
    P(|F-F_{\rm av}|>\varepsilon)<e^{-\frac{\varepsilon^2 D_{\mathcal{B}}}{16}}.
\end{equation}

\section{Butterfly velocity}\label{app:Butterfly_velocity}
The front of the out-of-time-order correlator in Eq.~\eqref{eq:C_AB} is defined operationally as the locus where the signal drops to a fraction $p$ of its maximum. We estimate the butterfly velocity $v_B$ as the speed of this front. For each parameter set, we compute $v_B$ at thresholds $p$ uniformly sampled in $p\in[0.3,0.6]$; the reported value is the mean over $p$, and the error bar is the variance across these estimates. As expected, $v_B \to 1$ near the dual-unitary point, and it decreases as the $g$ or $J$ approach $0$ or $\pi/2$. The resulting landscape is summarized in Fig.~\ref{fig:v_B}.

\section{TE in the `plain vanilla' regime}\label{app:TE_plain_vanilla}


In this Appendix, we report numerical results for an additional `plain vanilla' point, $g=\tfrac{\pi}{4},\; j=0.8\,\tfrac{\pi}{4},\; h=0.5$, for which the butterfly velocity is $v_B=0.95\pm0.02$ as provided by Fig.~\ref{fig:plain_vanilla}. At such a high $v_B$, even a coarse-graining of $n=1/2$ is sufficient to drive the IM into an {subvolume-}law regime. In contrast to regimes with nontrivial boundary modes, the coarse-grained IM is dominated by perfect dephaser components with suppressed correlations. This picture is consistent with a rapid decay to zero of $|C_{zz}|$ accompanied by large irregular relative fluctuations. The latter are not expected to be efficiently captured by MPS approximations of the IM.

\section{$C_{xx}$  correlators} \label{app:C_xx}
\begin{figure*}[t]
    \centering
\includegraphics[width=.3\linewidth]{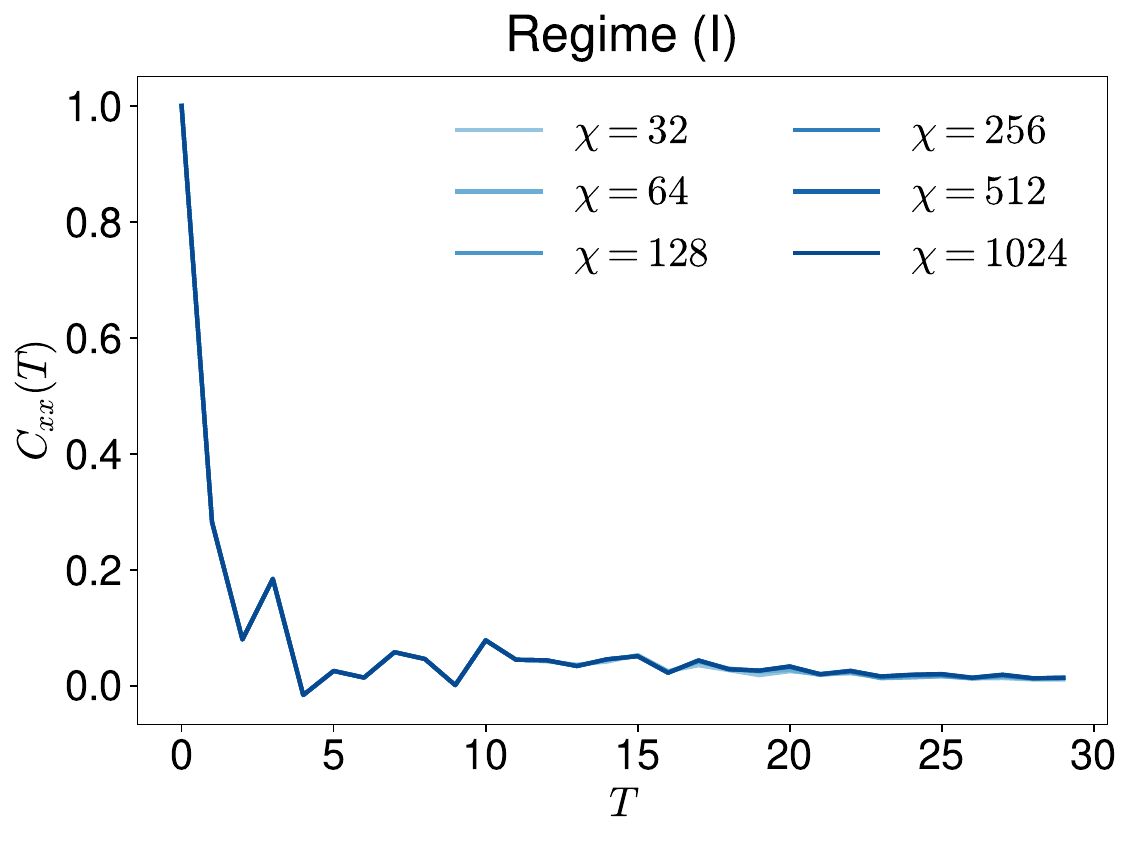}
\includegraphics[width=.3\linewidth]{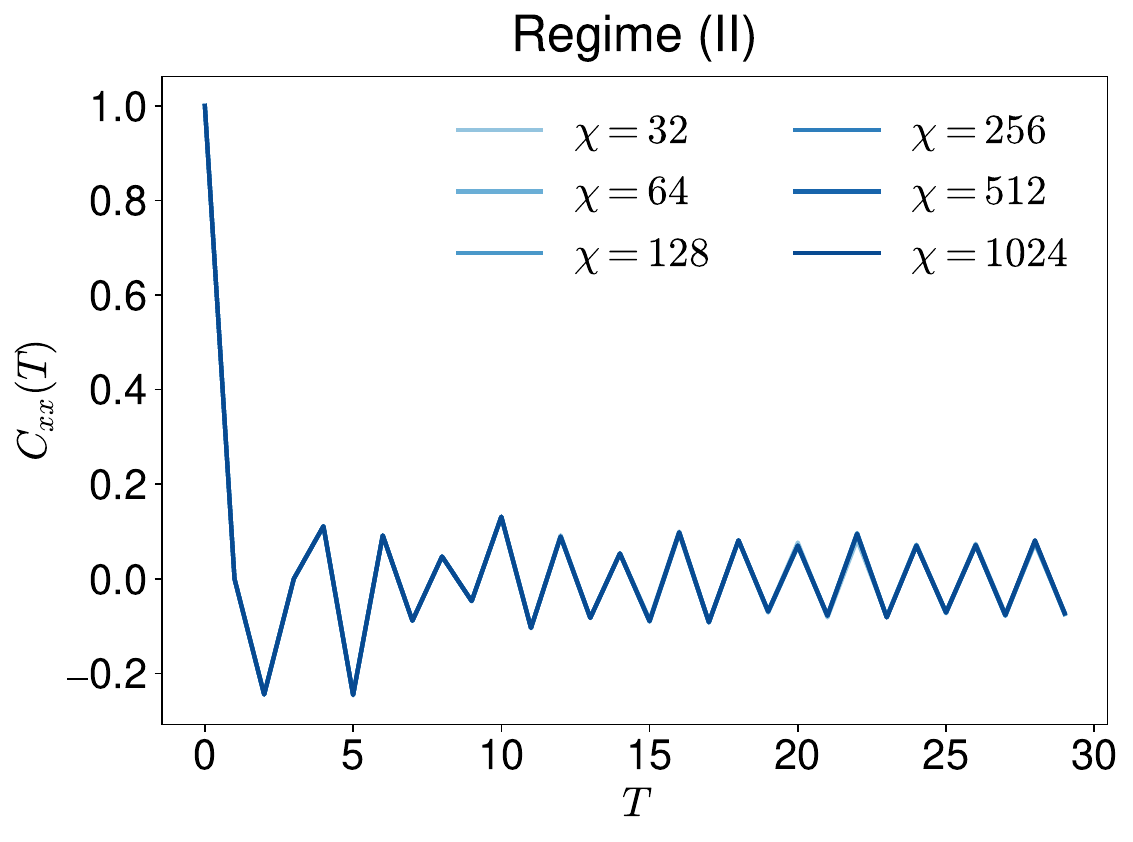}
\includegraphics[width=.3\linewidth]{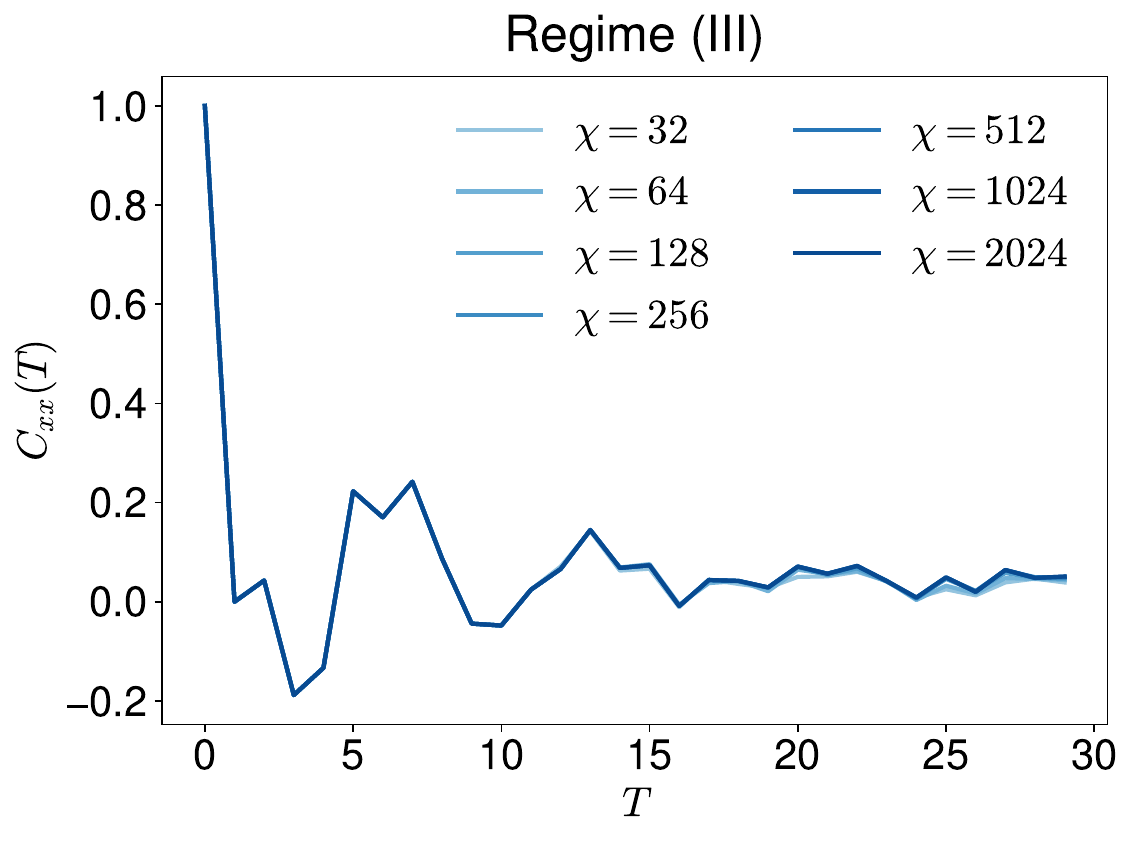}
\includegraphics[width=.3\linewidth]{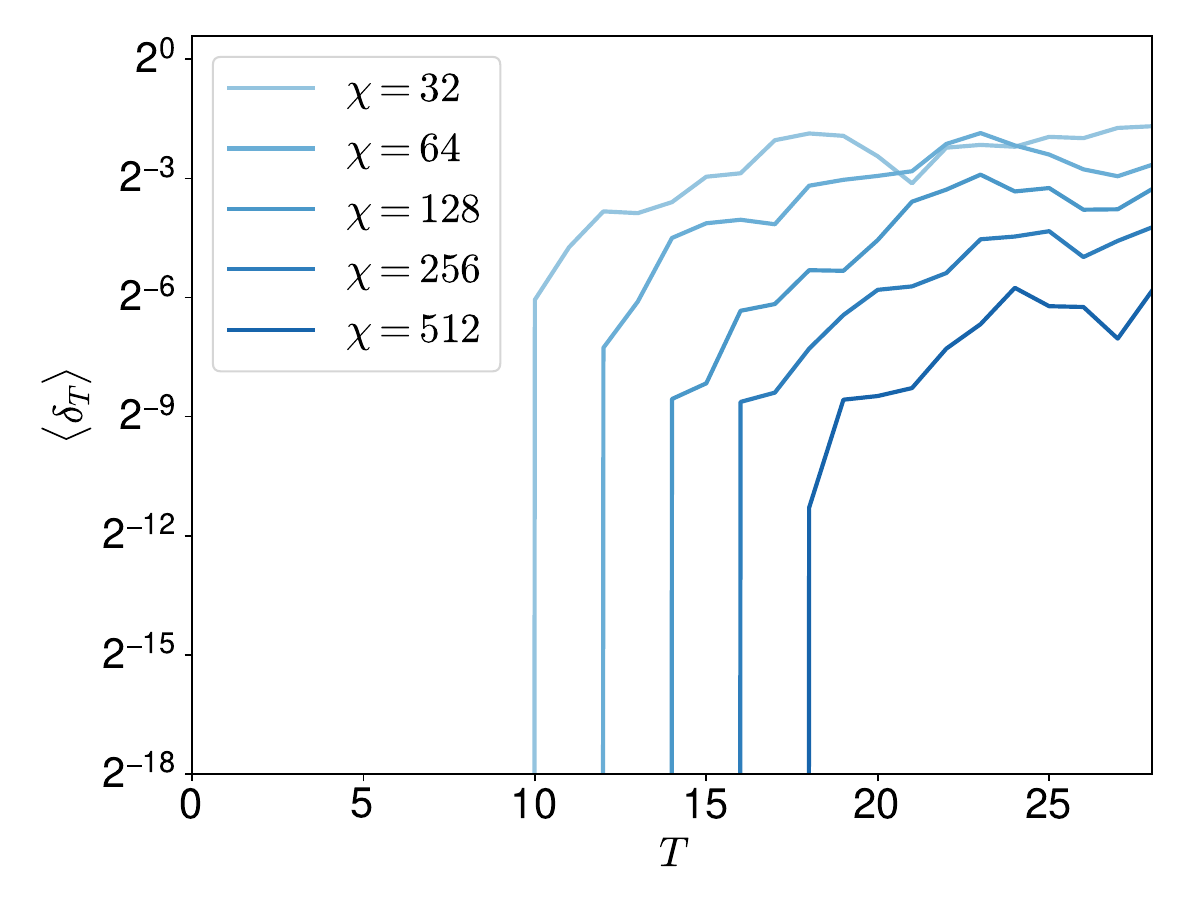}
\includegraphics[width=.3\linewidth]{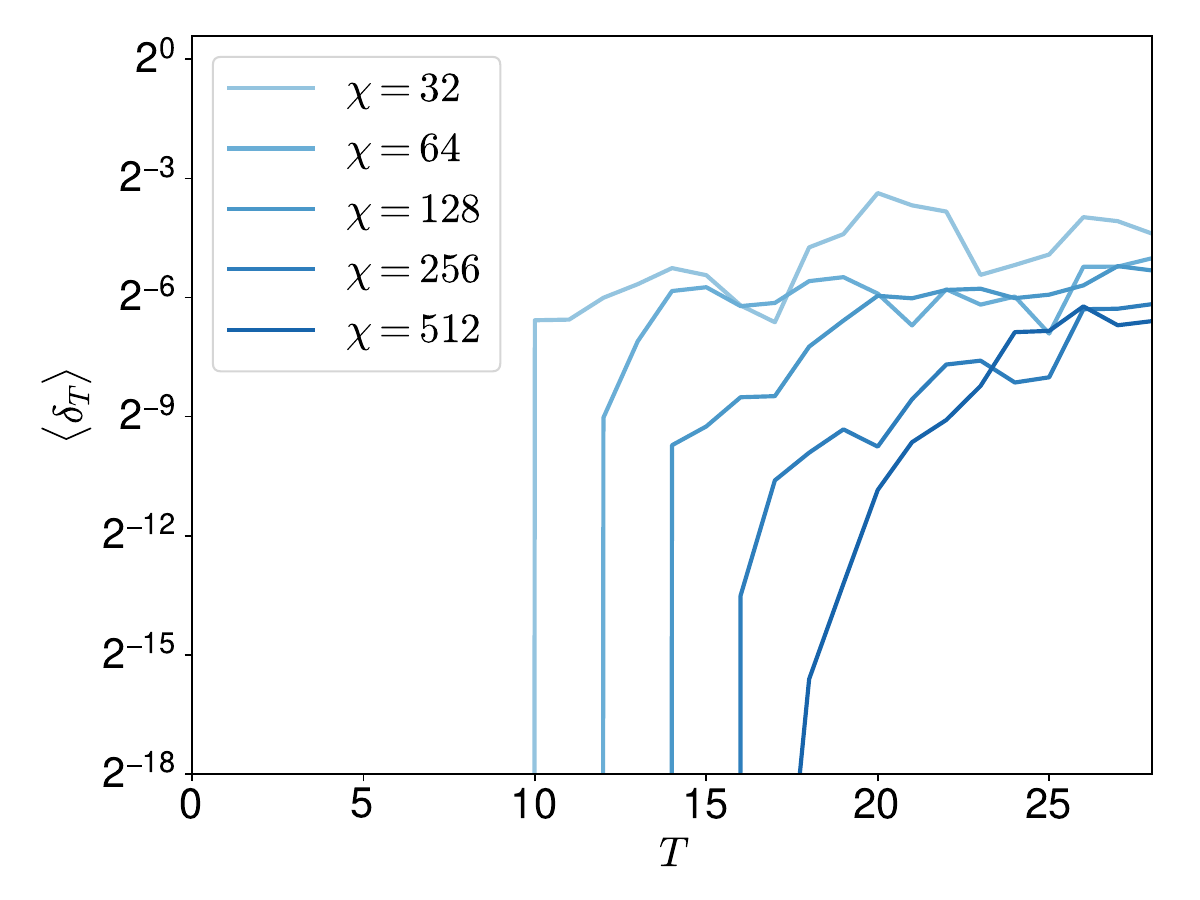}
\includegraphics[width=.3\linewidth]{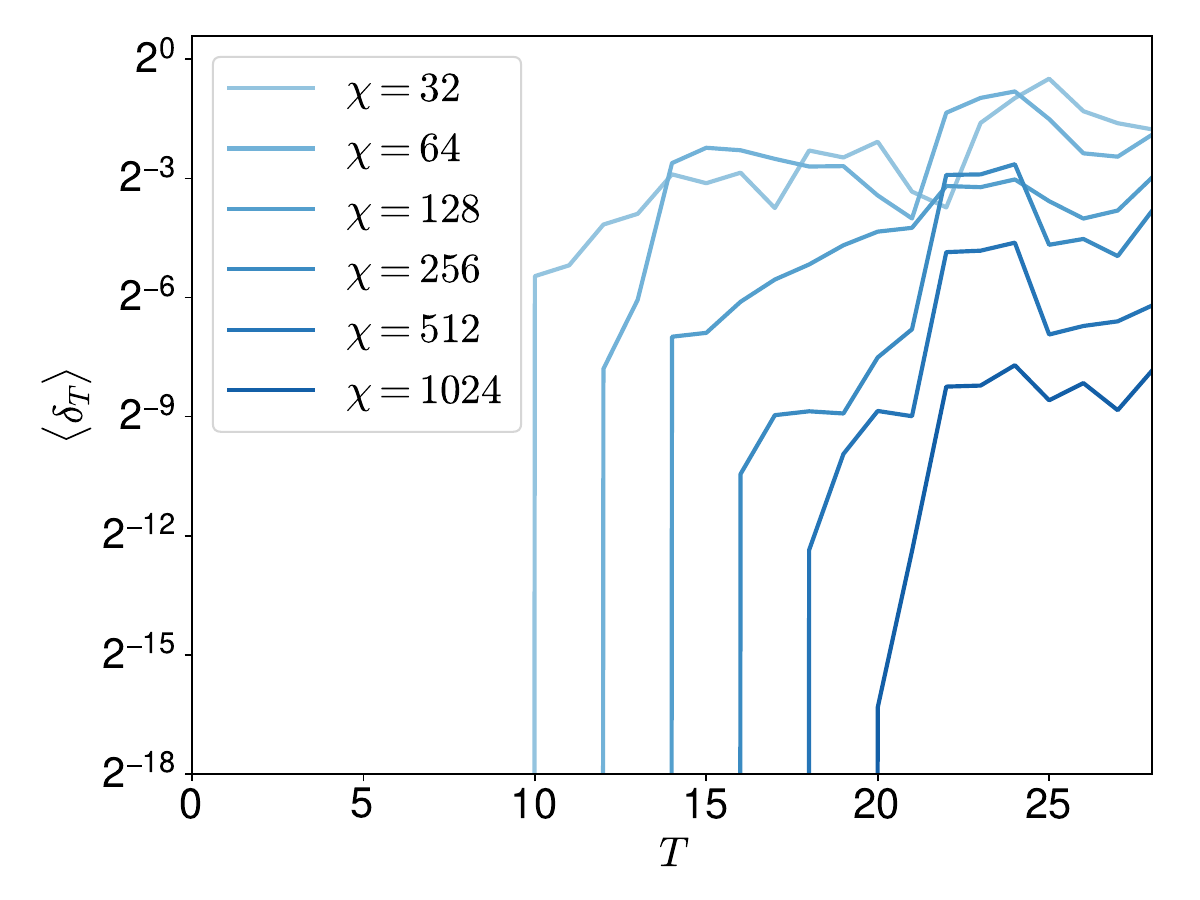}
\caption{\emph{Top pannels:} The local temporal correlator
$C_{xx}(T)$ in the three regimes shows qualitatively similar behavior to their $C_{zz}$ counterparts. \emph{Bottom panels:} relative error for the $C_{xx}$ correlator and different bond dimensions. }\label{fig:C_xx}
\end{figure*}

In this appendix, we provide the numerical results for $C_{xx}$ correlators in the three regimes see Fig.~\ref{fig:C_xx}. While different in magnitude, they demonstrate similar qualitative behavior.

\end{document}